\documentclass[10pt,aps,prd,onecolumn,superscriptaddress,preprintnumbers,floatfix,notitlepage]{revtex4-1}

\usepackage[justification=raggedright,singlelinecheck=false]{caption}

\usepackage{booktabs}
\usepackage{hyperref}
\hypersetup{
    colorlinks=true,
    linkcolor=blue,
    filecolor=magenta,      
    urlcolor=blue,
}
\usepackage{amsmath,amsfonts,amssymb}
\usepackage{fancyhdr}
\usepackage{graphicx}
\usepackage{xspace}
\usepackage{rotating}
\usepackage[normalem]{ulem}
\usepackage{braket}
\usepackage{verbatim}
\usepackage{xcolor}
\usepackage[utf8]{inputenc}
\usepackage{slashed}
\usepackage{subcaption}

\def\lsim{\mathrel{\raise.3ex\hbox{$<$\kern-.75em\lower1ex\hbox{$\sim$}}}}
\def\gsim{\mathrel{\raise.3ex\hbox{$>$\kern-.75em\lower1ex\hbox{$\sim$}}}}

\newcommand{\be}{\begin{equation}}
\newcommand{\ee}{\end{equation}}
\newcommand{\bea}{\begin{equation}\begin{aligned}}
\newcommand{\eea}{\end{aligned}\end{equation}}

\begin{document}

\pagestyle{plain}    

\title{Kinessence, phantom crossing and sign-switching dark energy}

\author{Philippe Brax}
\affiliation{Institut de Physique Théorique, Université Paris-Saclay,CEA, CNRS, F-91191 Gif-sur-Yvette Cedex, France}

\author{Tom Canaguier}
\affiliation{Institut de Physique Théorique, Université Paris-Saclay,CEA, CNRS, F-91191 Gif-sur-Yvette Cedex, France}
\affiliation{Université Paris-Saclay, Université Paris Cité, CEA, CNRS, AIM, 91191 Gif-sur-Yvette, France}

\author{Andrei Lazanu}

\affiliation{Institute of Space Science - INFLPR Subsidiary, Magurele, 077125, Romania}

\begin{abstract}
We introduce Kinessence models where combining  K-mouflage and quintessence leads to phantom-crossing of the dark energy equation of state and sign-switching of the dark energy density. This is realised by joining a dark scalar sector comprising K-essence kinetic terms and a quintessence potential together with a Yukawa coupling to matter. We investigate the dynamics of these models in the Jordan frame both at the background and perturbative levels. We concentrate on the effective equation of state and the matter fraction for the background. The Cosmic Microwave Background (CMB) spectra, the linear power spectrum and the growth of structure via $f\sigma_8$ are also analysed at the perturbative level. The quintessence potential drives the acceleration of the Universe at low redshift with an equation of state greater than minus one. {The Yukawa coupling to matter leads to a phantom crossing behaviour depending on its strength}. The potential fifth force mediated by the scalar field is screened by the K-mouflage mechanism { triggered by the K-essence kinetic terms} and passes Solar System tests, i.e. the Cassini bound and the periastron of the Moon. This allows for relatively large Yukawa couplings to matter $\beta\sim 0.1$, large enough to lead to the crossing of the phantom divide at small redshift $z\sim 0.5$.   Despite the presence of a significant scalar interaction on large scales, we find that {the growth of structure is reduced compared to $\Lambda$CDM.} This is also the case for the matter fraction which deviates from the standard model by a few percent. Similarly the Hubble rate is larger at low redshift than at recombination whilst the CMB spectra show small deviations from $\Lambda$CDM. In these models, the dark energy fraction becomes negative at redshifts greater than two. All these features should alleviate the various tensions such as the larger value of $H_0$ at small redshift, a possible reduction of growth as measured by $S_8$  and the puzzling bound on the sum of the neutrino masses. Deviations from $\Lambda$-CDM of the CMB lensing spectrum may provide a detectable window for these models. 
\end{abstract}
\maketitle 

\section{Introduction}
Probes over the last decades have shown that the current cosmological model, $\Lambda$CDM, is in very good agreement with cosmological observations. This model assumes that the universe is composed of ordinary matter, cold dark matter and a cosmological constant, which accounts for the presence of dark energy. It also assumes that general relativity is valid on cosmological scales. In particular, Cosmic Microwave Background (CMB) probes such as \textit{Planck} \cite{Planck2018params} have measured the parameters of the model with exquisite accuracy. The model can successfully explain the observed accelerated expansion of the Universe \cite{SupernovaSearchTeam:1998fmf, SupernovaCosmologyProject:1998vns}, but the origin of the cosmological constant is currently unknown. Moreover, the model has several theoretical issues, many of which have been known for a long time, e.g. the so-called cosmological constant problem \cite{Carroll:1991mt}.

More recently, an additional discrepancy has been observed. Current data, consisting of combinations of CMB, baryon acoustic
oscillations (BAO) \cite{Shlivko:2026jxa} and supernovae \cite{DESI:2024aqx, DESI:2025zgx, DES:2026jmi} favour a non-constant equation of state, as well as phantom-crossing which cannot be explained by a simple cosmological constant \cite{Li:2026asg,Garcia-Garcia:2026nzy}. Many models have been proposed to explain these observations \cite{Khoury:2025txd, Ozulker:2025ehg, Goldstein:2025epp,Guedezounme:2025wav,Nojiri:2026uvn,Antusch:2026ldp,Abdalla:2026sis,Khoury:2026svx, Gomez-Valent:2026ept,Smith:2025icl,Liu:2025bss, Bedroya:2025fwh,Chanda:2026dmx,Hallam:2026qsk,Yang:2026vkc,Banihashemi:2026twb,Sabogal:2026bsw,Ganguly:2026mgu,Wen:2026xlj,Delaunay:2026fse,Babu:2026tfr}.
{Promising avenues involve higher order operators in the Horndeski  classification \cite{Horndeski:1974wa} of the KGB type \cite{Deffayet:2010qz,Cataneo:2025vae,Wolf:2025acj,Naidoo:2026umv}}.
This behaviour can also be 
easily achieved by coupling the dark energy scalar to matter~  \cite{Martin:2005bp,Das:2005yj,Wolf:2025jed,Yang:2026qmz}.

We will explore a new route to phantom crossing using simple ingredients. The first one is K-essence  \cite{Armendariz-Picon:1999hyi} with a non-linear kinetic term \cite{Chen:2025ywv}.{ In the absence of coupling to matter, they cannot induce phantom crossing \cite{Vikman:2004dc}.} Once coupled to matter, i.e. becoming K-mouflage, they induce modifications to gravity \cite{Babichev:2009ee, Brax:2012jr} where a screening mechanism suppresses fifth forces in regions where the gravitational acceleration is large \cite{Khoury:2013tda,Brax:2021wcv}. These models and their cosmological evolution have been extensively studied \cite{Brax:2014wla, Brax:2014yla}. They cannot explain directly the phantom-crossing, but  show a tendency to lower the equation of state in the past \cite{Brax:2015lra,Benevento:2018xcu}. In this paper we describe a new class of models, Kinessence, that can explain these data by crossing the phantom divide line at low redshift. These models are simply obtained by replacing the  cosmological constant of K-mouflage models, which leads to an equation varying from $-1$ at low redshifts to lower values at larger $z$, by a quintessence potential. At low redshift, the potential drives the equation of state to values larger than $-1$ whilst {the coupling to matter of K-mouflage leads} to a decrease eventually implying the crossing of the phantom divide. This can only be achieved when the coupling to matter is large enough. Obviously for such large couplings of the order $\beta\sim 0.1$ the scalar interaction violates the Cassini bound \cite{Bertotti:2003rm} in the Solar System, i.e. the Shapiro effect for signals emitted between the Cassini satellite and the Earth would have been far too large. 

In the presence of a coupling to matter, a potential for matter and K-essence terms, two screening mechanisms could be at play. The first one is the chameleon mechanism \cite{Khoury:2003aq,Khoury:2003rn} which requires the existence of a minimum for the effective potential combining the potential and the matter coupling, and a thin shell for the scalar profile inside compact objects such as the Sun. Here we will explicitly use a model where the chameleon effect is not efficient enough \cite{Brax:2010gi}. For this we combine a runaway dilaton potential \cite{Gasperini:2001pc} converging exponentially to zero for large field values with the exponentially increasing coupling corresponding to Yukawa interactions. {The exponential decay of the potential could mimic the behaviour of the string dilaton \cite{Gasperini:2001pc} or the one of the Goldstone boson of broken scale invariance \cite{Weinberg:1988cp}.} Although a minimum of the effective potential exists, the chameleon effects cannot operate due to the weak dependence of the effective scalar mass on the matter density \cite{Brax:2026ttj}. This can be remedied as the K-mouflage effect takes over in dense environments \cite{Babichev:2009ee, Desmond:2026lds}. K-mouflage screening requires large scalar gradients which suppress the scalar fifth force. Such models can be easily designed. For instance a cubic correction in the kinetic terms suppressed by the dark energy scale is enough to satisfy the Cassini bound \cite{Barreira:2015aea}. This is not the case of the precision tests associated to the Moon's orbit, i.e. the periastron of the Moon due to potential interactions deviating from $1/r$. More complex $K$-functions are required to satisfy these bounds, an example of which is given by the $\arctan$ model of \cite{Barreira:2015aea}. 
By adding the K-essence terms to the dilaton model with an exponential coupling and a runaway potential, we solve a long-standing puzzle about screening the dilaton. Indeed the non-operative chameleon mechanism is replaced by the K-mouflage effect in the Solar System. So the dilaton is screened due to higher order terms in its kinetic terms. Of course this requires a suppression scale for the kinetic terms of the order of the dark energy scale. From the model building point of view and in particular when applying this setting in a string context \cite{Andriot:2026lac}, this is certainly challenging. Similarly the quantum status of the model needs to be examined. In the absence of potential, K-mouflage models are protected by a non-renormalisation theorem which preserves the classical behaviour of the theory intact \cite{deRham:2014wfa,Brax:2016jjt}. As the potential plays only a role at low redshift where the kinetic terms are essentially canonical, the shift in the vacuum energy due to quantum correction affects the shape of the potential unless the dark energy field is the dilaton of broken scale invariance \cite{Brax:2025ahm}. In this case the amplitude of the exponential potential is only affected and its large magnitude is compensated by the exponential decay of the potential. In this sense, Kinessence models with a runaway potential offer an opportunity to alleviate the quantum sensitivity of quintessence models. 

From a phenomenological point of view, the standard model of cosmology is plagued with tensions. The $H_0$ tension \cite{Schoneberg:2026eys,Schoneberg:2026vaf} is long standing with a local value deemed to be larger than the Planck-normalised value at large redshift. Growth of structure could also be reduced compared to $\Lambda$CDM \cite{Heydenreich:2025xim} although this is highly debated with the recent results from weak lensing experiments \cite{Pantos:2026koc,Wright:2025xka}. Finally, the sum of the neutrino masses seems to be also problematic. They could be in conflict with particle physics and even driven to negative values \cite{Wang:2024hen, Kibris:2026cqq, Green:2024xbb}. Dynamical dark energy could address the neutrino puzzle \cite{daFonseca:2023ury} but seems to require a matter fraction lower than in the standard model calibrated to the Planck satellite values \cite{DESI:2025zgx}. Recently it has been argued that having a period of negative dark energy at redshift larger than two could alleviate the neutrino issue and maybe address the $H_0$ tension \cite{Ibarra-Uriondo:2026zbp,Kibris:2026cqq,Adil:2026gjl}. In our Kinessence model, we will find that the dark energy density in the Jordan frame is driven to negative values for redshift larger than two. Moreover the current matter fraction is lower compared to the standard model. {In a similar vein, increasing the coupling of dark energy to matter does not increase the growth of structure but decreases it \cite{Pookkillath:2026wyg}. This corroborates what is observed in \cite{Smith:2025grk} and corresponds to an increase of the Hubble friction with the coupling strength between matter and dark energy.} Although a thorough study of the model's parameter space and a comparison to data is left for future work, values of the coupling $\beta\sim 0.09$ chosen to give a reasonable fit to the DESI equation of state lead to a slight increase of $H_0$ at $z=0$ and lowers $\Omega_m$. This will not be enough to bridge the gap between the local and cosmological values of $H_0$, although this may help covering parts of the way \cite{Najera:2026odk}.
Moreover we find that the effects of Kinessence models on the CMB lensing spectrum could be substantial. A thorough MCMC study of these models would make it clearer whether  CMB-S4/Simons-Observatory-class lensing reconstruction, targeting sub-percent precision in that $\ell$ range, could distinguish between  Kinessence models and $\Lambda$CDM \cite{CMB-S4:2016ple,LiteBIRD:2023iiy}. {Finally, we also consider the local variation of Newton's constant \cite{Williams:2004qba,Uzan:2024ded,Biskupek:2020fem} $\frac{d\ln G_{\rm eff}}{H_J dt}$ constrained by the Lunar Laser Ranging experiment (LLR) and the potential modification to the exponential shape of the potential that this would require \cite{Albrecht:1999rm}. The change of potential must be effective at small redshift where eventually the field must be nearly stopped. This can certainly be achieved with an Albrecht-Skordis potential, see appendix \ref{app:C} for instance,  although this could be viewed as introducing another coincidence problem. Indeed the DESI data seem to favour the coupling of matter to dark energy and the resulting time evolution of Newton's constant. Why should this behaviour stop at very small redshift for compliance with LLR results?  The detailed matching to the LLR data and the compatibility with the DESI results is left for future work.}

The paper is organised as follows: in Section \ref{sec:km-kin} we introduce the Kinessence models, showing how combining K-mouflage and quintessence models can lead to phantom-crossing, in Section \ref{sec:cosmo} we present the cosmological background and perturbation equations. In Section \ref{sec:results} we discuss the main results of the paper and  conclude in Section \ref{sec:conc}. We have added two appendices where we discuss the coupled dilaton model and the  cosmology for Kinessence for different values of the model parameters.

\section{From K-mouflage to Kinessence}
\label{sec:km-kin}
In this section, we recall what K-essence and K-mouflage models are and introduce Kinessence models where the equation of state in the Jordan frame can cross the phantom divide at low redshift.

\subsection{K-essence}

K-essence models are defined by the Lagrangian density
\begin{equation}
    \mathcal{L}_\phi = M^4 (K(\chi)-1), \qquad 
    \chi \equiv -\frac{1}{2M^4} g^{\mu\nu} \partial_\mu\phi \, \partial_\nu\phi,
    \label{eq:kessence_lagrangian}
\end{equation}
where $M^4$ is an energy scale fixed by dark energy and $K$ is a dimensionless function. In fact we require $M^4 \sim \rho_\Lambda \sim H_0^2 
M_\text{Pl}^2$, which fixes $M$ to be of order the meV scale.
The canonical limit is recovered as $K(\chi) \to \chi$, leading to $\mathcal{L}_\phi = M^4(-1 + \chi) = -M^4 + X$, that is, a cosmological 
constant plus standard kinetic terms {$X = -\frac{1}{2} g^{\mu\nu} \partial_\mu\phi \partial_\nu\phi$}. 
In the low kinetic energy regime ($|\chi| \ll 1$), the function $K$ 
can be expanded as $K(\chi) \approx \chi + \ldots$, and the theory 
reduces to a canonically normalised scalar field with a cosmological constant.  In the high kinetic energy regime ($|\chi| \gg 1$), the non-linear 
terms in $K$ become significant. Although the cut-off scale $M$ is very low and of the order of the energy scales probed cosmologically at late times, quantum corrections for these theories and K-mouflage are under control thanks to non-renormalisation theorems preserving the Lagrangian at large distance.

Cosmologically, 
the energy density and pressure are given by
\begin{equation}
    \rho_\phi = M^4 \left[ 2\chi K_\chi(\chi) - K(\chi)+1 \right], \qquad
    p_\phi = M^4 (K(\chi)-1),
    \label{eq:kessence_rho_p}
\end{equation}
where $K_\chi \equiv \mathrm{d}K/\mathrm{d}\chi$. 
The equation of state 
\be
w_\phi = \frac{K-1}{2\chi K_\chi - K+1}
\ee
depends not only on 
the value of the field but also on its first derivative through $K_\chi$. We will see that when the scalar is coupled to matter, this equation of state is modified and can cross the phantom divide. 

\subsection{K-mouflage}

In the Einstein frame the K-mouflage models correspond to   K-essence with an explicit coupling to matter~\citep{Brax:2014wla,Brax:2014yla}
\begin{equation}
    S = \int \mathrm{d}^4x \sqrt{-g_\mathrm{E}} \left[ \frac{M^2_{\rm Pl}}{2} R_\mathrm{E}
    + \   \underbrace{M^4 (K(\chi_\mathrm{E})-1)}_{\equiv \, \mathcal L_{\phi}}  \right]
    + \int \mathrm{d}^4x \sqrt{-g_\mathrm{J}}\,
    \mathcal{L}_\mathrm{m}\!\left(\psi^{(i)}_\mathrm{m},\, g^\mathrm{J}_{\mu\nu}\right),
    \label{eq:action_einstein}
\end{equation}
where $R_\mathrm{E}$ is the Ricci scalar associated with the Einstein-frame metric
$g^\mathrm{E}_{\mu\nu}$, $M_{\rm Pl} = (8\pi G_{\rm N})^{-1/2}$ is the reduced Planck mass, $G_{\rm N}$ is the gravitational constant,
and $\psi^{(i)}_\mathrm{m}$ denotes matter fields. 
$\chi_\mathrm{E}$ is the kinetic variable as in (\ref{eq:kessence_lagrangian}).
The function $K(\chi_{\mathrm{E}})$ encodes the full dynamics of the
scalar sector with two constraints. 
We impose that 
$
    K'(\chi_{\mathrm{E}}) > 0
$
to avoid ghosts 
where a prime denotes differentiation. 
Matter couples to the Jordan frame where
\begin{equation}
    g^\mathrm{J}_{\mu\nu} = A^2(\phi)\, g^\mathrm{E}_{\mu\nu}.
    \label{eq:conformal_transfo}
\end{equation}
The function $A(\phi)$ is the conformal factor. This introduces a dependence of
the effective gravitational coupling on the local value of the scalar field.
It is convenient to define
\begin{equation}
    \alpha(\phi) \equiv \frac{\mathrm{d}\ln A}{ \mathrm{d}\phi},
    \qquad
    \alpha_{,\phi} \equiv \frac{\mathrm{d}^2 \ln A}{\mathrm{d}\phi^2}.
    \label{eq:alpha_def}
\end{equation}
In the following we will consider that $\alpha$ is constant
\citep{Brax:2014wla,Brax:2014yla}. This corresponds to K-mouflage with a linear Yukawa coupling. { We will also use the dimensionless coupling $\beta= M_{\rm Pl} \alpha$ when comparing to data. }

\subsection{Kinessence}

We generalise the K-mouflage models and introduce Kinessence defined by the Lagrangian
\begin{equation}
    S = \int \mathrm{d}^4x \sqrt{-g_\mathrm{E}} \left[ \frac{M^2_{\rm Pl}}{2} R_\mathrm{E}
    + \   \underbrace{M^4 ( K(\chi_\mathrm{E})-f(\phi))}_{\equiv \, \mathcal L_{\phi}}  \right]
    + \int \mathrm{d}^4x \sqrt{-g_\mathrm{J}}\,
    \mathcal{L}_\mathrm{m}\!\left(\psi^{(i)}_\mathrm{m},\, g^\mathrm{J}_{\mu\nu}\right),
    \label{eq:action_einstein_kin}
\end{equation}
where $f$ is a dimensionless function associated to the potential for the dark energy field. We have normalised $K(0)=0$. This function  corresponds to a potential $V=M^4 f$  and will lead to an effective equation of state greater than $-1$ at small redshift. The { coupling to matter}  will make the equation of state cross the phantom divide. We will focus on models where
\be 
f(\phi)= e^{-\lambda \frac{\phi}{M_{\rm Pl}}}
\label{pott}
\ee
which would lead to quintessence with a constant equation of state \cite{Copeland:1997et} 
\be 
\omega_\phi=-1 + \frac{\lambda^2}{3}
\ee
for canonically normalised fields { and no coupling to matter. Acceleration occurs for $\lambda \le \sqrt 2$ as $\omega_\phi \le -\frac{1}{3}$.}
This type of potential is well-motivated as it could represent the strong coupling limit for the string dilaton \cite{Andriot:2026lac}. This is also the potential for the dilaton of broken scale invariance where a change of $V_0$ by radiative corrections can be compensated by a shift in $\phi$ \cite{Brax:2025ahm}. {We will use values of $\lambda \ge \sqrt 2$ below in section \ref{sec:results} and find that acceleration takes place for kinessence models.}
Indeed, when the K-essence part and {the coupling to matter} are switched on, we will see that the behaviour of the equation of state can mimic the one suggested by the DESI experiment and become negative at high enough redshift. {We will also find that the LLR constraints \cite{Biskupek:2020fem} on the variation of Newton's constant require to change the shape of the potential to an Albrecht-Skordis form
\be
f(\phi)= U(\phi) e^{-\lambda \phi/M_{\rm Pl}}
\label{newpot}
\ee
where $U(\phi)$ has a minimum leading to an attractor with $\dot\phi=0$ at the minimum of the potential. This would drive the equation of state close to $-1$ at $z=0$. It has to be seen if the LLR constraints can be made compatible with the DESI results on the equation of state. This will necessitate a comparison with data taking into account the LLR constraints and is beyond the analysis in this paper. Here we will focus on the cosmological behaviour of the model at non-vanishing redshift, deduce the features of these models and impose that screening takes care of the gravitational tests in the solar system.  We leave the modification of the small $z$ behaviour of the model after using (\ref{newpot}) instead of (\ref{pott}) to future work. We provide an approximate description of the required physics at very small redshift in appendix \ref{app:C}.}

\subsection{Action in the Jordan frame}

Matter couples to the scalar field in the Jordan frame. As most cosmological observations rely on the emission of light by atoms, it is physically motivated  to study the dynamics of the Universe in the Jordan frame where atomic physics is not modified, {i.e. the electron and proton mass are field independent.} This implies that redshifts and distances are not affected by a scalar-induced modification of atomic physics, i.e. spectral lines would be shifted via the scalar dependence of the electron mass on the scalar field in the Einstein frame. In the Jordan frame, atomic physics is not modified and all the new physics effects follow from a change in the geometry of space-time. As observations are interpreted assuming that atomic physics is standard, the Jordan frame becomes the frame in which observations should be understood. We will write all the dynamical equations in the Jordan frame. We use $
    \sqrt{-g_\mathrm{E}} = A^{-4}\sqrt{-g_{\rm J}},\
    R_\mathrm{E} = A^2\!\left[R_\mathrm{J} + 6\,\Box_\mathrm{J} \ln A
    - 6\,(\partial \ln A)^2\right]
$
and obtain
\begin{equation}
    S = \int \mathrm{d}^4x\sqrt{-g_\mathrm{J}}\left[
    \frac{M^2_{\rm Pl}\, A^{-2}(\phi)}{2}\, R_\mathrm{J}
    + 3M^2_{\rm Pl}\, A^{-2}(\phi)\,(\partial \ln A)^2
    + A^{-4}(\phi)\, M^4\left[ K\!\left(A^2(\phi)\,\chi_\mathrm{J}\right)
   {- f(\phi)} \right]\right] + S_\mathrm{m},
    \label{eq:action_jordan}
\end{equation}
where we denote by
    $X \equiv -\frac{1}{2}\,g^{\mu\nu}_{\mathrm{J}}\,\partial_\mu\phi\,\partial_\nu\phi$ the dimensionful Jordan-frame kinetic variable, of mass dimension four. Its dimensionless Jordan-frame
counterpart is $\chi_{\mathrm{J}} = X/M^4$.
    
This expression implies a field-dependent effective Planck
mass
\begin{equation}
    F(\phi) \equiv M^2_{\rm Pl}\, A^{-2}(\phi),
    \label{eq:F_def}
\end{equation}
where $M_{\rm Pl}$ is the reduced Planck mass. The value of $F$ is observationally constrained
by the local measurement of Newton's constant,
\begin{equation}
    G_{\rm eff}(\phi) = G_{\rm N}\, A^2(\phi) = \frac{1}{8\pi F(\phi)}.
    \label{eq:Geff}
\end{equation}
Big Bang Nucleosynthesis (BBN), i.e. the primordial abundances of light 
elements,  is  sensitive to the expansion rate at $T \sim 1\,\mathrm{MeV}$, which 
itself depends on the value of the gravitational coupling at that epoch. Following 
\cite{Uzan:2010pm}, the combined constraints from BBN and other cosmological 
observations require the conservative bound \cite{Yeh:2022heq, Uzan:2024ded}
$-0.04< \Delta G_{\rm eff}/G_{\rm N} < 0.06$, i.e. a maximal variation
\begin{equation}
    \left|\frac{\Delta G_{\rm eff}}{G_{\rm N}}\right| 
    = \Delta (A^2) \lesssim  0.04=4 \%,
    \label{eq:BBNc}
\end{equation}
as an order-of-magnitude constraint between BBN and today. This translates directly into a constraint on the 
conformal factor $A(\phi)$ which must remain close to unity throughout  
the cosmic history.

We will use the effective 
Lagrangian in the Jordan frame
\begin{equation}
    \mathcal{L}^{(\mathrm{J})}_{\phi,\mathrm{eff}}
    \equiv 3F(\phi)\,\alpha^2(\phi)\,g^{\mu\nu}_\mathrm{J}\,\partial_\mu\phi\,
    \partial_\nu\phi
    + A^{-4}(\phi)\,M^4\left[\,K\!\left(A^2(\phi)\,\chi_\mathrm{J}\right) {- f(\phi)} \right]
    \label{eq:Lphi_eff}
\end{equation}
Compared to the Einstein-frame scalar Lagrangian $\mathcal{L}^{(\mathrm{E})}_\phi = 
M^4 K(\chi_\mathrm{E})$, two modifications appear. The second term 
$A^{-4}(\phi)\,M^4\,K(A^2(\phi)\,\chi_\mathrm{J})$ is the direct counterpart of 
$M^4 K(\chi_\mathrm{E})$, with the kinetic variable rescaled in Jordan frame and the amplitude suppressed by $A^{-4}(\phi)$, 
both arising from the Weyl rescaling between the Einstein and Jordan metrics. The first term, i.e.  
$3F(\phi)\,\alpha^2(\phi)\,g^{\mu\nu}_\mathrm{J}\,\partial_\mu\phi\,\partial_\nu
\phi$, is an additional kinetic contribution with no counterpart in the 
Einstein frame. 
The complete action reads
\begin{equation}
    {
    S = \int \mathrm{d}^4x\sqrt{-g_\mathrm{J}}\left[
    \frac{F(\phi)}{2}\, R_\mathrm{J}
    + \mathcal{L}^{(\mathrm{J})}_{\phi,\mathrm{eff}}
    \right] + S_\mathrm{m}.
    }
    \label{eq:action_jordan_compact}
\end{equation}
This can be decomposed in three sectors 
\begin{equation}
    S = S_{\mathrm{GK}} + S_{\phi} + S_\mathrm{m},
\end{equation}
where $S_{\mathrm{GK}} = \int \mathrm{d}^4x\sqrt{-g_\mathrm{J}}\, \frac{F(\phi)}{2} R_\mathrm{J}$ 
is the gravitational sector  modified
by the field-dependent Planck mass $F(\phi)$, $S_{\phi} = \int 
\mathrm{d}^4x\sqrt{-g_\mathrm{J}}\, \mathcal{L}^{(\mathrm{J})}_{\phi,\mathrm{eff}}$ encodes the 
scalar field dynamics, and $S_\mathrm{m}$ is the matter action, minimally 
coupled to the Jordan-frame metric $g^\mathrm{J}_{\mu\nu}$.

\subsection{Equations of motion: covariant formulation in the Jordan frame}

Starting from the action~\eqref{eq:action_jordan_compact}, we now derive the
equations of motion directly in the Jordan frame. 

\subsubsection{Stress-energy tensors}

We first define the stress-energy tensor of the scalar field by variation of
$S^{(\mathrm{J})}_\phi$ with respect to $g^{\mu\nu}_\mathrm{J}$ at fixed $\phi$ by the usual definition
\begin{equation}
    T^{(\phi)}_{\mu\nu}
    \equiv -\frac{2}{\sqrt{-g_\mathrm{J}}}
    \frac{\delta S^{(\mathrm{J})}_\phi}{\delta g^{\mu\nu}_\mathrm{J}}
    \label{eq:Tphi_def}
\end{equation}
which reads
\begin{equation}
    T^{(\phi)}_{\mu\nu}
    = g^\mathrm{J}_{\mu\nu}\,\mathcal{L}^{(\mathrm{J})}_{\phi,\mathrm{eff}}
    + Z(\phi,X)\,\partial_\mu\phi\,\partial_\nu\phi \,,
    \label{eq:Tphi}
\end{equation}
We have introduced the composite coefficient
\begin{equation}
    Z(\phi,X) \equiv A^{-2}(\phi)\,K_\chi\!\left(A^2(\phi)\,\chi_\mathrm{J}
    \right) - 6F(\phi)\,\alpha^2(\phi)\,.
    \label{eq:Z_def}
\end{equation}
The matter stress-energy tensor is also defined as
\begin{equation}
    T^{(\mathrm{m})}_{\mu\nu}
    \equiv -\frac{2}{\sqrt{-g_\mathrm{J}}}
    \frac{\delta S_\mathrm{m}}{\delta g^{\mu\nu}_\mathrm{J}}\,.
    \label{eq:Tm_def}
\end{equation}
Since matter is already minimally coupled to $g^\mathrm{J}_{\mu\nu}$, this 
expression is identical to its General Relativity counterpart.

\subsubsection{Modified Einstein equations}

The non-minimal coupling $F(\phi)$ generates additional geometric terms in the Einstein equation 
involving the operator
\begin{equation}
    \mathcal{D}^{(\mathrm{J})}_{\mu\nu}\,
    \equiv \left(g^\mathrm{J}_{\mu\nu}\,\Box^{(\mathrm{J})}
    - \nabla^{(\mathrm{J})}_\mu\nabla^{(\mathrm{J})}_\nu\right)\,.
    \label{eq:D_def}
\end{equation}
As $F(\phi)$ is a scalar, the commutator of covariant 
derivatives vanishes identically, $[\nabla^{(\mathrm{J})}_\mu, \nabla^{(\mathrm{J})}_\nu]F = 0$, 
so that $\mathcal{D}^{(\mathrm{J})}_{\mu\nu}F$ is symmetric in $(\mu,\nu)$. The modified Einstein equation reads
\begin{equation}
    {
    F(\phi)\,G^\mathrm{J}_{\mu\nu}
    + \mathcal{D}^{(\mathrm{J})}_{\mu\nu}\,F
    = T^{(\mathrm{m})}_{\mu\nu}
    + g^\mathrm{J}_{\mu\nu}\,\mathcal{L}^{(\mathrm{J})}_{\phi,\mathrm{eff}}
    + Z(\phi,X)\,\partial_\mu\phi\,\partial_\nu\phi.
    }
    \label{eq:einstein_jordan}
\end{equation}
The term $\mathcal{D}^{(\mathrm{J})}_{\mu\nu}\,F$ has no counterpart in General
Relativity, it comes uniquely from the variation of the field-dependent Planck
mass.
We can write this equation in a convenient manner
\begin{equation}
     M^2_{\rm Pl}G^\mathrm{J}_{\mu\nu} =  T_{\mu\nu}^{(m)} + T_{\mu\nu}^{(\rm DE, eff)},
\end{equation}
where $T_{\mu\nu}^{(\rm DE,eff)}$ is the effective energy-tensor of dark energy
\begin{equation}
    T_{\mu\nu}^{(\rm DE,eff)} = \left(\frac{M_{\rm Pl}^2}{F} - 1 \right)T^{(\mathrm{m})}_{\mu\nu} + \frac{M^2_{\rm Pl}}{F}\left(g^\mathrm{J}_{\mu\nu}\,\mathcal{L}^{(\mathrm{J})}_{\phi,\mathrm{eff}}
    + Z(\phi,X)\,\partial_\mu\phi\,\partial_\nu\phi - \mathcal{D}^{(\mathrm{J})}_{\mu\nu}\,F \right),
    \label{eq:T_DE_covariant}
\end{equation}
We can see that, in the Jordan frame, dark energy mixes the scalar and matter parts of the action. 
This will directly lead to a non-standard expression for the dark energy equation of state. 

\subsubsection{Klein-Gordon equation}

The scalar field is subject to the Klein-Gordon equation 
\begin{equation}
    {
    \nabla^{(\mathrm{J})}_\mu\!\left(Z(\phi,X)\,
    \nabla^{(\mathrm{J})\mu}\phi\right)
    + \frac{\partial\mathcal{L}^{(\mathrm{J})}_{\phi,\mathrm{eff}}}{\partial\phi}
    - \alpha(\phi)\,F(\phi)\,R_\mathrm{J} = 0\,.
    }
    \label{eq:KG}
\end{equation}
Introducing the scalar current $J^\mu$ and the effective source term
$\mathcal{S}_{\phi,\mathrm{eff}}$
\begin{align}
    J^\mu &\equiv Z(\phi,X)\,\nabla^{(\mathrm{J})\mu}\phi\,,
    \label{eq:current}\\[4pt]
    \mathcal{S}_{\phi,\mathrm{eff}}
    &\equiv
    \frac{\partial\mathcal{L}^{(\mathrm{J})}_{\phi,\mathrm{eff}}}{\partial\phi}
    - \alpha(\phi)\,F(\phi)\,R_\mathrm{J}\,,
    \label{eq:source}
\end{align}
we obtain $\mathcal{S}_{\phi,\mathrm{eff}} = \,\partial_\phi\mathcal{L}_{\rm Kin}$, where we have collected the whole Lagrangian
\be 
\mathcal{L}_{\rm Kin}=\frac{F(\phi)}{2}\, R_\mathrm{J}
    + \mathcal{L}^{(\mathrm{J})}_{\phi,\mathrm{eff}} \,,
\ee
which will reappear when we deal with perturbations.
The Klein-Gordon equation becomes a conservation law
\begin{equation}
    {\nabla^{(\mathrm{J})}_\mu J^\mu + \mathcal{S}_{\phi,\mathrm{eff}} = 0 }.
    \label{eq:KG_compact}
\end{equation}
The term $-\alpha F R_J$ in $\mathcal{S}_{\phi,\mathrm{eff}}$ arises
from the variation of the non-minimal gravitational coupling
$F(\phi)R_\mathrm{J}/2$ with respect to $\phi$, and is a direct signature of the
conformal structure of the theory, it has no counterpart in a minimally
coupled theory and vanishes identically when $\alpha = 0$.

{This  dependence can be removed using the trace of the modified Einstein equation \eqref{eq:einstein_jordan},
with $G^{J\,\mu}{}_{\mu}=-R_J$, $\mathcal D^{(J)\mu}{}_{\mu}F=3\Box^{(J)}F$, we obtain
\begin{equation}
F R_J = 3\,\Box^{(J)}F - T^{(m)} - 4\mathcal L^{(J)}_{\phi,\rm eff} + 2XZ\, ,
\qquad
\Box^{(J)}F=-2\alpha F\,\Box^{(J)}\phi-2XF\big(4\alpha^2-2\alpha_{,\phi}\big)\,.
\label{eq:traceR}
\end{equation}
Inserting (\ref{eq:traceR}) into (\ref{eq:KG}), the $\Box^{(J)}\phi$ terms combine as
$\big(Z+6F\alpha^2\big)\Box^{(J)}\phi=A^{-2}K_\chi\,\Box^{(J)}\phi$.
The Klein-Gordon equation takes the unmixed form 
\begin{equation}
\nabla^{(J)}_\mu\!\Big(A^{-2}(\phi)\,K_\chi(\chi)\,\nabla^{(J)\mu}\phi\Big)
= A^{-4}(\phi)\,M^4 f_{,\phi}(\phi)-\alpha(\phi)\,T^{(m)}
\label{eq:KGdemixed}
\end{equation}}
where the role of the potential and the coupling to matter are transparent.

\subsubsection{Conservation of matter }

In the Jordan frame, matter is minimally coupled to $g^\mathrm{J}_{\mu\nu}$ and  as a result the matter
stress-energy tensor is covariantly conserved
\begin{equation}
    \nabla^{(\mathrm{J})}_\mu T^{(\mathrm{m})\mu\nu} = 0\,.
    \label{eq:conservation_matter}
\end{equation}

For a perfect fluid with energy density $\rho$, pressure $p$, and
four-velocity $u^\mu$, the stress-energy tensor reads
\begin{equation}
    T^{(\mathrm{m})}_{\mu\nu} = (\rho + p)\,u_\mu u_\nu + p\,g^\mathrm{J}_{\mu\nu}\,.
    \label{eq:Tm_fluid}
\end{equation}
Inserting this into~\eqref{eq:conservation_matter} and projecting along
$u^\nu$ and orthogonally via the projector
$h^{\mu\nu} \equiv g^{(\mathrm{J})\mu\nu} + u^\mu u^\nu$ leads to the
continuity equation
\begin{equation}
    u^\mu\nabla_\mu\rho + (\rho + p)\,\nabla_\mu u^\mu = 0\,,
    \label{eq:continuity_cov}
\end{equation}
and the Euler equation
\begin{equation}
    (\rho + p)\,u^\mu\nabla_\mu u^\alpha
    + h^{\alpha\nu}\nabla_\nu p = 0 \,.
    \label{eq:euler_cov}
\end{equation}
These are the standard perfect-fluid equations of General Relativity,
unmodified explicitly by the Kinessence dynamics as  a direct consequence of the
Jordan-frame matter conservation~\eqref{eq:conservation_matter}.
We can also use the Bianchi identity 
\begin{equation}
    \nabla^{(\mathrm{J})}_\mu T^{(\mathrm{DE,eff})\mu\nu} = 0.
    \label{eq:conservation_DE}
\end{equation}
corresponding to the conservation of the effective dark energy energy momentum tensor. This conservation law allows us to deduce that the effective dark energy density and pressure satisfy a standard conservation equation with their ratio being the effective equation of state.

\subsection{Stability condition and  K-mouflage screening}
\label{subsec:stability_screening}
The scalar field is 
universally coupled to matter through the conformal factor $A(\phi)$.
This coupling can  mediate an additional gravitational-strength
interaction which could be seen as a fifth force whose strength is controlled
by $\alpha(\phi)$. 
The Kinessence models can in principle be subject to two screening effects. 
First of all, in the Einstein frame in the presence of matter, the scalar dynamics are governed by the effective potential \cite{Khoury:2003aq,Khoury:2003rn}
\be
V_{\rm eff}(\phi)= M^4 f(\phi) + \rho_m A(\phi) \,,
\ee
where $\rho_m$ is the conserved matter density in the Einstein frame. When this potential has a minimum depending on $\rho_m$ and the effective mass of the scalar at this minimum is large enough in matter, chameleon screening can be effective. For the exponential potential that we consider, chameleon screening is not efficient enough to screen the Solar System.
Fortunately the K-mouflage screening mechanism can be present in the models described in this paper. The possibility of having two operating screening mechanisms, i.e. chameleon and K-mouflage, is  intriguing and   would have to be considered if $f(\phi)$ were of the inverse power law type for instance. This is left for future work. {For recent works along this direction, see \cite{Sirera:2026klo}}.

\subsubsection{Quadratic Lagrangian and stability conditions}
\label{subsubsec:stability}

We decompose $\phi = \phi_0 + \delta\phi $ around a background value $\phi_0$  and expand the Lagrangian of Kinessence to second order. 
The first-order term vanishes as $\phi_0$ is an exact 
solution of the equations of motion.  
We find that the quadratic Lagrangian takes the form:
{\begin{align}
\mathcal L^{(2)} &= -\tfrac12\Big[\partial_X\mathcal{L}_{\rm Kin}\big|_0\, g_J^{\mu\nu}
 - \partial_X^2\mathcal{L}_{\rm Kin}\big|_0\,\nabla^\mu\phi_0\nabla^\nu\phi_0\Big]
 \partial_\mu\delta\phi\,\partial_\nu\delta\phi
 \nonumber\\
 &\quad + \tfrac12\Big[\partial_\phi^2\mathcal{L}_{\rm Kin}\big|_0
 + \nabla^{(J)}_\mu\!\Big(\partial_\phi\partial_X\mathcal{L}_{\rm Kin}\big|_0\,\nabla^{(J)\mu}\phi_0\Big)\Big]\delta\phi^2\, .
\label{eq:L2_raw}
\end{align}}
{Specialized in the FLRW background $\phi_0=\phi_0(\tau)$, where $\nabla^\mu\phi_0$ has only a time
component, this becomes}
\begin{equation}
    \mathcal{L}^{(2)}
    =
    -\frac{1}{2}\,Z_{\rm eff}(\phi_0,\chi_0)\,
    g^{00}_0\,(\partial_0\delta\phi)^2
    \;-\;
    \frac{1}{2}\,Z(\phi_0,\chi_0)\,
    g^{ij}_0\,\partial_i\delta\phi\,\partial_j\delta\phi - \tfrac{1}{2}m^2_{\rm
eff}(\phi_0,\chi_0)(\delta\phi)^2
    \label{eq:L2_FLRW}
\end{equation}
where we have
\begin{align}
    Z(\phi_0,X_0) &= \,\partial_X\mathcal{L}_{\rm Kin}\big|_0\,,
    \label{eq:Z_id}\\
    Z_{\rm eff}(\phi_0,X_0) &= \left[\partial_X\mathcal{L}_{\rm Kin}\big|_0
        + 2X_0\,\partial^2_X\mathcal{L}_{\rm Kin}\big|_0\right]\,,
    \label{eq:Zeff_id}\\
    m^2_{\rm eff}(\phi_0,X_0)
    &= -\,\partial_\phi^2 \mathcal{L}_{\rm Kin}\big|_0
      \;-\;\nabla^{(J)}_\mu\!\Big(\partial_\phi Z\,\nabla^{(J)\mu}\phi_0\Big)\, .
    \label{eq:meff_id2}
\end{align}
The $Z$ factor is the composite coefficient already introduced in
eq. ~\eqref{eq:Z_def}, and
\begin{equation}
    {
    Z_{\rm eff}(\phi,X) \equiv A^{-2}(\phi)\Big[K_\chi(\chi) + 2\chi\,K_{\chi \chi}(\chi)\Big] - 6F(\phi)\,\alpha^2(\phi)\,.
    } 
    \label{eq:Zeff_def}
\end{equation}

Finally the mass term becomes
{
\begin{equation}
m^2_{\rm eff} = -\,\partial^2_\phi\mathcal L^{(J)}_{\phi,\rm eff}
+\big(\alpha_{,\phi}-2\alpha^2\big)F R_J
-\nabla^{(J)}_\mu\Big(\big[\alpha\,(Z_{\rm eff}-3Z)-12F\alpha\,\alpha_{,\phi}\big]\nabla^{(J)\mu}\phi\Big)\,,
\label{eq:meff_compact}
\end{equation}
for constant $\alpha$, i.e.\ $\alpha_{,\phi}=0$, the equation reduces to}
\begin{equation}
m^2_{\rm eff} = V^{(\rm J)}_{,\phi\phi} + 2\alpha V^{(\rm J)}_{,\phi}\;-\; 4\alpha^2 A^{-4}(\phi)\,M^4\Big[2K(\chi) - 2\chi K_\chi(\chi) + \chi^2 K_{\chi \chi}(\chi)\Big]
 \;-\; 2\alpha\,\nabla^{(J)}_\mu\!\Big[A^{-2}(\phi)\,\chi K_{\chi \chi}(\chi)\,\nabla^{(J)\mu}\phi\Big]\, ,
\label{eq:meff_explicit}
\end{equation}
\noindent where $V^{(\rm J)}(\phi)\equiv A^{-4}M^4 f(\phi)$ is the potential associated to the dark energy field in the Jordan frame.

\noindent The two quantities $Z$ and $Z_{\rm eff}$  control the amplitude of the spatial 
propagation of the perturbation and its temporal kinetic energy. They 
differ precisely by the term $2A^{-2}\chi_0 K_{\chi \chi}(\chi_0)$, which vanishes for 
a canonical kinetic function ($K = \chi$, $K_{\chi \chi} = 0$) but is 
generically non-zero in Kinessence. 
The absence of ghosts
requires the coefficient of $(\partial_0\delta\phi)^2$ in~\eqref{eq:L2_FLRW}
to be positive
\begin{equation}
    Z_{\rm eff}(\phi_0, \chi_0) > 0\,.
    \label{eq:no_ghost}
\end{equation}
The condition for gradient stability is 
\begin{equation}
    Z(\phi_0, \chi_0) > 0\,,
    \label{eq:no_gradient}
\end{equation}
i.e.\ $A^{-2}K_{\chi}(\chi_0) > 6F\alpha^2$, which in the regime $\alpha^2 \ll
A^{-2}K_{X}/6F$ reduces simply to $K_{X}(\chi_0) > 0$.
The ratio of the two coefficients defines the propagation speed of the
scalar perturbation
\begin{equation}
    c_\mathrm{s}^2
    = \frac{Z}{Z_{\rm eff}}
    = \frac{A^{-2}K_{\chi}(\chi_0) - 6F\alpha^2}
           {A^{-2}\left[K_{\chi}(\chi_0) + 2\chi_0 K_{\chi \chi}(\chi_0)\right] - 6F\alpha^2}\,.
    \label{eq:cs2}
\end{equation}
For $\alpha = 0$ this reduces to the standard K-essence result
$c_\mathrm{s}^2 = K_{\chi}/(K_{\chi}+2\chi K_{\chi \chi})$, and
sub-luminality $c_\mathrm{s}^2 \leq 1$ is ensured whenever $K_{\chi \chi}(X_0) \geq 0$.
Note that the effective mass term $m^2_{\rm eff}$ appearing in
$\mathcal{L}^{(2)}$ does not enter the stability conditions. At large $k$, the
dispersion relation $\omega^2 = c_\mathrm{s}^2 k^2 + m^2_{\rm eff}/Z_{\rm
eff}$ is dominated by $c_\mathrm{s}^2 k^2$, so the ultraviolet stability is
controlled by the signs of $Z$ and $Z_{\rm eff}$. At small $k$, a 
negative $m^2_{\rm eff}$ could in principle source an infrared 
instability, but since $m_{\rm eff} \sim \mathcal{O}(H)$, this term is 
only relevant near the Hubble horizon.  It is worth noting that in the screened regime, the scalar 
field perturbations can propagate superluminally, with a 
speed $c_\mathrm{s}^2 > 1$ when $K_{\chi \chi} < 0$. As shown 
in~\cite{Brax:2014gra}, this superluminality is accompanied 
by the formation of transient shocks during the dynamical 
relaxation of the scalar field towards its static screened 
configuration in high-density environments.

\subsubsection{Screening}
\label{subsubsec:screening_kmouflage}

The two stability conditions have a direct physical consequence for
screening. In the linear, low-kinetic-energy regime
$|\chi_0| \ll 1$, one has $K_{X}(X_0) \approx 1$ and $K_{\chi \chi} \approx 0$, so
that $Z \approx Z_{\rm eff} \approx 1 - 6F\alpha^2$. The scalar
current~\eqref{eq:current} is of order unity, and the field propagates
freely, mediating a fifth force of gravitational strength proportional
to $\alpha^2$.
In the high-kinetic-energy regime $|\chi_0| \gg 1$, the
no-ghost and no-gradient conditions require $K_{\chi}(\chi_0) \gg 1$, so that
\begin{equation}
    Z \approx Z_{\rm eff} \approx A^{-2}K_{\chi}(\chi_0) \gg 1\,.
    \label{eq:Z_screened}
\end{equation}
{We can now discuss the effects of screening. 
K-mouflage screening is due to the large value of $Z$ in dense environments. This is a dominant effect as long as
\be 
k\sim \frac{1}{R}\gg \frac{m_{\rm eff}}{\sqrt Z}\,,
\ee
where $R$ is the typical size of the screened object. At large density the quintessence potential is irrelevant and
\be 
m_{\rm eff}\sim \beta \frac{\sqrt \rho}{M_{\rm Pl}}
\ee
implying that K-mouflage effects dominate compared to the mass term, which would lead to chameleon screening if $m_{\rm eff}R \gg 1$, if we have 
\be 
m_{\rm eff}R \ll \sqrt{Z}\,.
\label{bound1}
\ee
The K-mouflage dominance must be checked in the Solar System with $R=R_\odot$ and the density $\rho_\odot \sim 150 \ {\rm g/cm^3}$. The kinetic terms are given by \cite{Barreira:2015aea}
\be 
\sqrt{-\chi_{ss}}K'_{ss} \sim 10^6 \,,
\ee
where we take $\beta\sim 0.1$ as a benchmark value which satisfies the BBN constraints. This gives the value of $\chi_{\rm ss}$ for each model. In the ones that we consider below, this bound (\ref{bound1}) is always satisfied and we have verified that $m_{\rm eff} R\lesssim 1$.
}
From the Klein-Gordon equation~\eqref{eq:KG_compact}, a large $Z$
suppresses the spatial gradients of $\delta\phi$ relative to its source
$\mathcal{S}_{\phi,\mathrm{eff}}$: the scalar field becomes kinetically
rigid. The fifth force is therefore suppressed, i.e.  this is the
K-mouflage screening mechanism.

In the following, we will present two models of K-mouflage where screening takes place. In particular, Solar System tests such as the one provided by the Cassini satellite will be satisfied. On the other hand, the stronger constraint provided by the periastron of the Moon will only be satisfied by one of these two examples. 
{
More precisely we focus on models defined by \cite{Barreira:2015aea}
\bea 
K(\chi)&=\chi + K_0 \chi^m\, ,\nonumber \\ 
K(\chi)&=\chi + K_* \left[\chi-\chi_* \arctan\left(\frac{\chi}{\chi_*} \right) \right] \,. \nonumber \\
\eea}
Benchmark values are such that $\beta=0.1$, $K_0=1$, $m=3$, $K_*=1000$, $\chi_*=100$. With these values the Cassini \cite{Bertotti:2003rm} and periastron bounds \cite{Williams:2004qba} are satisfied for the $\arctan$ model only, i.e
\be
\frac{\beta^2}{K'_{ss}}\lesssim 10^{-5}
\ee
for Cassini and
\be 
\frac{\beta^2}{K'_{ss}}\frac{\vert \chi_{ss}K''_{ss}\vert}{K'_{ss}+2\chi_{ss}K_{ss}''}\lesssim 8\times 10^{-13}
\label{eq:periastron}
\ee
for the periastron. In Table \ref{bigtable}, the numerical values for each of the considered models are given and show that the chameleon effect is not present and K-mouflage dominates in the Solar System.

\begin{table}[h]
\centering
\caption{Solar-system screening criteria for the benchmark models ($\beta=0.1$, $M^4={(H_0^{\Lambda\rm CDM})^2}M_{\rm Pl}^2${, with $H_0^{\Lambda\rm CDM}=67.32\,{\rm km\,s^{-1}\,Mpc^{-1}}$}).
$\chi_{ss}$ and $K_{\chi,ss}$ are the exact solutions of $\sqrt{-\chi}\,K_\chi(\chi)=\sqrt2\,\beta\,M_{\rm Pl}\,g_\odot/M^2$
in the solar system \cite{Barreira:2015aea}. The Cassini column gives $\beta^2/K_{\chi,ss}$ and the periastron column
the combination of \eqref{eq:periastron}. The last two columns use $m_{\rm eff}\simeq\beta\sqrt{\rho_\odot}/M_{\rm Pl}$ with
$\rho_\odot=150\,{\rm g/cm^3}$, $R_\odot=6.96\times10^{8}\,{\rm m}${, and $Z \approx  K_{\chi,ss}$}.}
\label{bigtable}
\renewcommand{\arraystretch}{1.1}
\setlength{\tabcolsep}{4pt}
\begin{tabular}{lcccccc}
\hline
Model & $\chi_{ss}$ & $K_{\chi,ss}$ & Cassini & Periastron & $m_{\rm eff}R_\odot$ & $m_{\rm eff}R_\odot/\sqrt{Z}$ \\
\hline
Bound & -- & -- & {$\lesssim10^{-5}$} & $\lesssim8\times10^{-13}$ & $\lesssim1$ & $\ll1$ \\
\hline
arctan ($K_*=1000$, $\chi_*=100$) & {$-1.64\times10^{6}$} & $1.001\times10^{3}$ & $9.99\times10^{-6}$  & {$7.41\times10^{-14}$} & $3.68\times10^{-3}$ & $1.16\times10^{-4}$ \\
polynomial ($K_0=1$, $m=3$)       & {$-1.79\times10^{2}$} & {$9.59\times10^{4}$}  & {$1.04\times10^{-7}$} & {$4.17\times10^{-8}$}  & $3.68\times10^{-3}$ & {$1.19\times10^{-5}$} \\
\hline
\end{tabular}
\end{table}
{ Finally, the time drift of Newton's constant depends on  the cosmological variation of the  scalar field $\phi$ which penetrates inside the solar system and is unscreened~ \cite{Brax:2018zvh}.  This crucially depends on the shape of the potential for the values of the scalar field in the recent past of the Universe. In a nutshell, potentials like the runaway dilaton (\ref{pott}) chosen in this paper do not have the stopping power to slow the scalar field down to the level required by the LLR constraint. The dynamics at very small redshift must be altered. This can be achieved by modifying the exponential potential to an Albrecht-Skordis shape \cite{Albrecht:1999rm} where the local minimum of the potential (\ref{newpot}) can slow down the rolling scalar field. This would leave an imprint on the low $z$ evolution of the equation of state which would become close to $-1$. For larger values of $z$ the exponential behaviour dominates and leads to the results presented in this paper. In particular the distance to the last scattering surface hardly depends on the very small $z$ behaviour of the equation of state and the CMB results in section \ref{sec:results} stand. }

\section{Cosmological equations: background and perturbations}
\label{sec:cosmo}

\subsection{The FLRW metric}
\label{subsec:FLRW}

We consider a spatially flat Friedmann–Lemaître–Robertson–Walker (FLRW)
spacetime. In the Jordan frame, the background line element in conformal time reads
\begin{equation}
    \mathrm{d}s^2_\mathrm{J} = a_\mathrm{J}^2(\eta)
    \left(-\mathrm{d}\eta^2 + \delta_{ij}\,\mathrm{d}x^i\mathrm{d}x^j\right),
    \label{eq:FLRW_Jordan}
\end{equation}
where $a_\mathrm{J}(\eta)$ is the Jordan-frame scale factor. The conformal time $\eta$ is 
shared by both frames \footnote{verifying that $a_\mathrm{J} = A\,a_\mathrm{E}$ implies 
$\mathrm{d}\eta_\mathrm{J} = \mathrm{d}\eta_\mathrm{E}$}, the transformation $g^\mathrm{J}_{\mu\nu} = A^2(\phi)\,g^\mathrm{E}_{\mu\nu}$ 
acts as a global rescaling of the metric at fixed coordinates, leaving the  global coordinate $(\eta , x)$  chart 
unchanged. 
The
conformal Hubble parameter as usual is defined by :
\begin{equation}
    \mathcal{H}_\mathrm{J} \equiv \frac{a_\mathrm{J}'}{a_\mathrm{J}}\,,
    \label{eq:calH_J}
\end{equation}
where a prime denotes differentiation with respect to $\eta$.
We also define the
cosmic-time Hubble rate $H_\mathrm{J} \equiv \dot{a}_\mathrm{J}/a_\mathrm{J}$, related to $\mathcal{H}_\mathrm{J}$ by
$\mathcal{H}_\mathrm{J} = a_\mathrm{J} H_\mathrm{J}$.\\

\subsection{Background equations}
\label{subsec:transfo_J_E}
\label{subsec:background}

We now specialise to  the Jordan frame equations. The scalar
field reduces to $\phi = \bar{\phi}(\eta)$. We work throughout in conformal time $\eta$ and
denote background quantities with an over bar.

\subsubsection{Modified Friedmann-Lemaître equations}
The Friedmann-Lemaître equations are obtained by  inserting the FLRW metric~\eqref{eq:FLRW_Jordan} into the modified Einstein equations \eqref{eq:einstein_jordan}. 
The matter stress-energy tensor for a perfect fluid takes the standard form
$
    \bar{T}^{(\mathrm{m})}_{00} = a^2\left(\rho_\mathrm{m} + \rho_\mathrm{r}\right),
    \bar{T}^{(\mathrm{m})}_{ij} = a^2\left(p_\mathrm{m} + p_\mathrm{r}\right)\delta_{ij} ,
    \label{eq:Tm_background}
$
where $\rho_\mathrm{m}$, $\rho_\mathrm{r}$, $p_\mathrm{m}$, $p_\mathrm{r}$ are the energy densities and pressures of
matter and radiation.
On the homogeneous background, the
non-minimal coupling operator \eqref{eq:D_def} becomes
$
    \bar{\mathcal{D}}_{00}F = 3\mathcal{H}\bar{F}',
    \bar{\mathcal{D}}_{ij}F = -\left(\bar{F}'' + \mathcal{H}\bar{F}'\right)\delta_{ij}.
    \label{eq:D_background_1}
$
For the scalar stress-energy tensor, since $\partial_i\bar\phi = 0$, the term $Z\,\partial_\mu\phi\,\partial_\nu\phi$ in
equation~\eqref{eq:Tphi} contributes only through its $(00)$ component. The background value of
$\mathcal{L}^{(\mathrm{J})}_{\phi,\text{eff}}$ is
\begin{equation}
    \bar{\mathcal{L}}^{(\mathrm{J})}_{\phi,\text{eff}}
    = -\frac{3\bar{F}\bar{\alpha}^2\bar{\phi}'^{\,2}}{a^2}
    + \bar{A}^{-4}M^4 \Big[ K(\bar{\chi}) - f(\bar{\phi}) \Big] \,,
    \label{eq:Leff_background}
\end{equation}
where $\bar{\chi} \equiv \bar{A}^2\bar{\phi}'^{\,2}/(2M^4 a^2)$ is the
background value of the Einstein-frame kinetic variable.
A direct computation gives
\begin{align}
    \bar{T}^{(\phi)}_{00} &= -a^2 \bar{\mathcal{L}}^{(\mathrm{J})}_{\phi,\text{eff}} + \bar{Z}\bar{\phi}'^{\,2}
    = \left(\bar{Z} + 3\bar{F}\bar{\alpha}^2\right)\bar{\phi}'^{\,2}
    - a^2 \bar{A}^{-4}M^4\Big[K(\bar{\chi}) - f(\bar{\phi})\Big] \,,
    \label{eq:T00_phi} \\
    \bar{T}^{(\phi)}_{ij}
    &= a_\mathrm{J}^2\,\bar{\mathcal{L}}^{(\mathrm{J})}_{\phi,\text{eff}}\,\delta_{ij} \,.
    \label{eq:Tij_phi}
\end{align}
Inserting equations \eqref{eq:Leff_background}, \eqref{eq:T00_phi} and \eqref{eq:Tij_phi} into the
$(00)$ and $(ij)$ projections of equation~\eqref{eq:einstein_jordan}, and
identifying the scalar energy density and pressure as
$\rho_\phi \equiv \bar{T}^{(\phi)}_{00}/a^2$ and
$p_\phi \equiv \bar{T}^{(\phi)}_{ij}/(a^2\delta_{ij})$, leads to  the
two Friedmann equations
\begin{equation}
    3\left(\mathcal{H}^2\,\bar{F} + \mathcal{H}\bar{F}'\right)
    = a^2\left(\rho_\mathrm{m} + \rho_\mathrm{r} + \rho_\phi\right),
    \label{eq:Friedmann1}
\end{equation}
\begin{equation}
    \bar{F}'' + \mathcal{H}\bar{F}'
    + \bar{F}\left(2\mathcal{H}' + \mathcal{H}^2\right)
    = -a^2\left(p_\mathrm{m} + p_\mathrm{r} + p_\phi\right) \,,
    \label{eq:Friedmann2}
\end{equation}
with
\begin{align}
    \rho_\phi &= \frac{\left(\bar{Z} + 3\bar{F}\bar{\alpha}^2\right)
                \bar{\phi}'^{\,2}}{a^2}
                - \bar{A}^{-4}M^4 \left[ K(\bar{\chi}) {- f(\phi)} \right]\,,
    \label{eq:rho_phi} \\
    p_\phi &= -\frac{3\bar{F}\bar{\alpha}^2\,\bar{\phi}'^{\,2}}{a^2}
              + \bar{A}^{-4}M^4 \left[ K(\bar{\chi}) {- f(\phi)} \right]\,,
    \label{eq:p_phi}
\end{align}
and the equation of state parameter for $\phi$
\begin{equation}
    w_\phi = \frac{p_\phi}{\rho_\phi} = 
    \frac{-3\bar{F}\bar{\alpha}^2 + \mathcal{K}}
         {\bar{Z} + 3\bar{F}\bar{\alpha}^2 - \mathcal{K}} \,,
\end{equation}
where $\mathcal{K} \equiv \dfrac{a^2\bar{A}^{-4}M^4 \left[ K(\bar{\chi}) {- f(\phi)} \right]}{\bar{\phi}'^{\,2}}$.
Using
\begin{equation}
    \rho_\phi + p_\phi = \frac{\bar{Z}\,\bar{\phi}'^{\,2}}{a^2} 
    \label{eq:rho_plus_p}
\end{equation}
we can solve for ${\cal H}$ in the first Friedmann equation to find 
\begin{equation}
    \mathcal{H} = \frac{-3\bar{F}' + 
    \sqrt{9\bar{F}'^{\,2} + 12\bar{F}\,a^2\bar{\rho}_\mathrm{tot}}}
    {6\bar{F}} \,,
    \label{eq:H_quadratic}
\end{equation}
where $\bar{\rho}_\mathrm{tot} = \rho_\mathrm{m} + \rho_\mathrm{r} + \rho_\phi$. 
Compared to the standard Friedmann equation, there are  terms  proportional 
to $\bar{F}' = -2\bar{\alpha}\bar{F}\bar{\phi}'$, i.e. the expansion rate depends not only on the energy content 
$\rho_{\rm tot}$ but also on the velocity of the scalar field 
$\bar{\phi}'$.

\subsubsection{Effective dark energy}

An observer would simply analyse the expansion of the Universe using the standard Friedmann equations and assign to a dark energy component anything which does not correspond to matter and radiation. 
We obtain  the first Friedmann equation in standard form
\begin{equation}
    3M_{\mathrm{Pl}}^2\,\mathcal{H}^2
    = a^2\!\left(\rho_\mathrm{m} + \rho_\mathrm{r} + \rho_{\mathrm{DE}}^{\mathrm{eff}}\right),
    \label{eq:Friedmann1_std}
\end{equation}
with the effective dark energy density written as
\begin{equation}
    \rho_{\mathrm{DE}}^{\mathrm{eff}} \equiv
    \left(\frac{M_{\rm Pl}^2}{\bar{F}} - 1\right)
    (\rho_\mathrm{m} + \rho_\mathrm{r})
    + \frac{M_{\rm Pl}^2}{\bar{F}}
    \left(\rho_\phi - \frac{3\mathcal{H}\bar{F}'}{a^2}\right).
    \label{eq:rhoDE_eff}
\end{equation}
Applying the same procedure to equation~\eqref{eq:Friedmann2} yields the
second equation in standard form
\begin{equation}
    -M_{\mathrm{Pl}}^2\!\left(2\mathcal{H}' + \mathcal{H}^2\right)
    = a^2\!\left(p_\mathrm{m} + p_\mathrm{r} + p_{\mathrm{DE}}^{\mathrm{eff}}\right),
    \label{eq:Friedmann2_std}
\end{equation}
with the effective dark energy pressure
\begin{equation}
    p_{\mathrm{DE}}^{\mathrm{eff}} \equiv
    \left(\frac{M_{\rm Pl}^2}{\bar{F}} - 1\right)
    (p_\mathrm{m} + p_\mathrm{r})
    + \frac{M_{\rm Pl}^2}{\bar{F}}
    \left(p_\phi + \frac{\bar{F}'' + \mathcal{H}\bar{F}'}{a^2}\right).
    \label{eq:pDE_eff}
\end{equation}
The term $(\bar{F}'' + \mathcal{H}\bar{F}')/a^2$ in
equation~\eqref{eq:pDE_eff}, together with $-3\mathcal{H}\bar{F}'/a^2$ in
equation~\eqref{eq:rhoDE_eff}, originate again from the time variation of
$F(\phi)$.
Both equations are exactly recovered by using the covariant definition
\eqref{eq:T_DE_covariant} with the FLRW background quantities as it should, i.e. they follow from writing the Einstein equation like in General Relativity in the absence of coupling to matter.
We can then identify  the effective equation of state
\begin{align}
    w_{\mathrm{DE}} &\equiv \frac{p_{\mathrm{DE}}^{\mathrm{eff}}}{\rho_{\mathrm{DE}}^{\mathrm{eff}}}
    \nonumber \\
    &= 
    \frac{\epsilon_F\left(p_\mathrm{m} + p_\mathrm{r}\right) 
    + (\epsilon_F +1) \left (p_\phi + \dfrac{\bar{F}'' + \mathcal{H}\bar{F}'}{a^2}\right)}
    {\epsilon_F\left(\rho_\mathrm{m} + \rho_\mathrm{r}\right) 
    + (\epsilon_F +1) \left (\rho_\phi - \dfrac{3\mathcal{H}\bar{F}'}{a^2}\right)}\,,
    \label{eq:wDE_details}
\end{align}
and the dimensionless deviation of the effective gravitational coupling from its General Relativity value
\begin{align}
    \epsilon_F &\equiv \frac{M_{\rm Pl}^2}{\bar{F}} - 1 \,.
    \label{eq:epsilonF}
\end{align}
Note that $\epsilon_F = 0$ when $A = 1$, i.e., in the absence of scalar coupling. 

The first term in equation~\eqref{eq:rhoDE_eff}, proportional to $\epsilon_F$, 
is a feature of K-mouflage. The effective dark energy density receives 
a contribution from matter and radiation, weighted by the deviation of the 
gravitational coupling from its General Relativity value. 
The BBN constraint imposes that  $|A^2(\phi) - 1| \lesssim 0.1$ throughout the cosmic history  and therefore $|\epsilon_F| \lesssim 0.1$, so that the 
matter and radiation contributions to $\rho_{\rm DE}^{\rm eff}$ are suppressed 
by at least one order of magnitude relative to $\rho_\mathrm{r}$ and $\rho_\mathrm{m}$ 
respectively. In their respective epochs, one therefore recovers 
$\rho_{\rm DE}^{\rm eff} \ll \rho_\mathrm{r}$ and $\rho_{\rm DE}^{\rm eff} \ll \rho_\mathrm{m}$. Here the BBN constraint serves two purposes, i.e it bounds the variation of $G_{\rm eff}$, and 
simultaneously guarantees that the modified gravity sector does not 
spoil the standard cosmological history with radiation-dominated and matter-dominated epochs.

A consequence of the Kinessence structure is that $w_{\rm DE}$ can cross the phantom
divide $w_{\rm DE} = -1$. From equations~\eqref{eq:rhoDE_eff}
and~\eqref{eq:pDE_eff} $ w_{\rm DE}=-1$, i.e. the crossing condition
$\rho_{\rm DE}^{\rm eff} + p_{\rm DE}^{\rm eff} = 0$,  reads
\begin{equation}
    \epsilon_F\left( \rho_\mathrm{m} + p_\mathrm{m} + \rho_\mathrm{r} + p_\mathrm{r}\right)+ \left( \epsilon_F +1 \right)\left[\frac{\bar{Z}\,\bar{\phi}'^{\,2}}{a^2}
    + \frac{\bar{F}'' -2\mathcal{H}\bar{F}'}{a^2}\right] = 0\,,
    \label{eq:full_phantom_crossing_condition}
\end{equation}
which becomes, when $\rho_\mathrm{m},\rho_\mathrm{r} \ll \rho_{\mathrm{DE}}^{\mathrm{eff}} $, and thus when the first term is negligible
\begin{equation}
    \bar{Z}\,\bar{\phi}'^{\,2}
    + \bar{F}'' -2\mathcal{H}\bar{F}' = 0 \,.
    \label{eq:phantom_crossing_condition}
\end{equation}
This is a dynamical condition on the field trajectory rather than a sign of
instability, the no-ghost and no-gradient-instability conditions of
Section~\ref{subsec:stability_screening} remain satisfied throughout, and the crossing
is a genuine prediction of the model that can in principle be
tested with supernova and BAO data~\citep{Brax:2014wla}.

\subsubsection{Klein-Gordon equation}

The modified Einstein equations determine the geometry once the scalar-field configuration is known.
The background value of the scalar field is determined by  the Klein-Gordon equation.
The scalar field equation on the FLRW background follows from the covariant
equation~\eqref{eq:KG} evaluated on the homogeneous solution. Using
$\nabla^{(\mathrm{J})}_\mu J^\mu = -(a^2\bar{Z}\bar{\phi}')'/a^4$ in conformal time, we find 
\begin{equation}
    \bar{Z}\,\bar{\phi}'' + \left(2\mathcal{H}\bar{Z} + \bar{Z}'\right)\bar{\phi}'
    =  a^2\,\mathcal{S}_{\phi,\mathrm{eff}} \,,
    \label{eq:KG_background}
\end{equation}
where the effective source term is given by \eqref{eq:source}.
Expanding each contribution explicitly,
we find
\begin{align}
    \mathcal{S}_{\phi,\mathrm{eff}} =\;&
    \frac{1}{a^2}\Big[
        6\bar{F}\alpha\left(\alpha^2 - \alpha_{,\phi}\right)\bar{\phi}'^{\,2}
        + \alpha\bar{A}^{-2}K_{\chi}(\bar{\chi})\,\bar{\phi}'^{\,2}
    \Big] \notag \\
    &- 4\alpha\bar{A}^{-4}M^4 \Big[K(\bar{\chi}){- f(\bar\phi)}\Big]
    - \bar{A}^{-4}M^4 f_{,\phi}(\bar\phi)
    - \frac{6\alpha\bar{F}}{a^2}
      \left(\mathcal{H}' + \mathcal{H}^2\right) \,,
    \label{eq:S_eff_explicit}
    \end{align}
where the last term originates from $\alpha\bar{F}R$ with the background Ricci
scalar $R = 6(\mathcal{H}' + \mathcal{H}^2)/a^2$.

\subsubsection{Conservation equations for matter and radiation}

In the Jordan frame, matter is minimally coupled to $g^\mathrm{J}_{\mu\nu}$ and its
energy tensor is independently conserved,
$\nabla^{(\mathrm{J})}_\mu T^{(\mathrm{m})\mu\nu} = 0$. The temporal
component $\nu = 0$ of this equation yields the continuity equation for each
fluid species
\begin{equation}
    \bar{\rho}_\mathrm{s}' + 3\mathcal{H}\left(\bar{\rho}_\mathrm{s} + \bar{p}_\mathrm{s}\right) = 0,
    \qquad \mathrm{s} \in \{\mathrm{m},\, \mathrm{r}\}.
    \label{eq:continuity_background}
\end{equation}
Inserting the equations of state $p_\mathrm{m} = 0$ and $p_\mathrm{r} = \rho_\mathrm{r}/3$ respectively, ~\eqref{eq:continuity_background}  gives 
$
    \bar{\rho}_\mathrm{m} \propto a^{-3},
    \bar{\rho}_\mathrm{r} \propto a^{-4}.
$
These dilution laws are identical to their $\Lambda$CDM counterparts, which is
a direct  consequence of working in the Jordan frame.
In the
Einstein frame, the matter conservation equation acquires an
explicit source term proportional to the field velocity \citep{Brax:2014yla} and reads 
\begin{equation}
    \rho_\mathrm{E}' + 3 \mathcal H_\mathrm{E}\left(\rho_\mathrm{E} + p_\mathrm{E}\right)
    = \alpha\,\phi'\left(\rho_\mathrm{E} - 3p_\mathrm{E}\right) \,.
    \label{eq:conservation_Einstein_conformal}
\end{equation}
This is the usual non-conservation of matter in the Einstein frame.

\subsection{Linear perturbations in the Newtonian gauge}
\label{subsec:perturbations}

\subsubsection{Perturbed metric}
\label{subsubsec:perturbed_metric}

 In the Newtonian gauge, the perturbed
line element in conformal time reads
\begin{equation}
    \mathrm{d}s^2 = a^2(\eta)\left[
    -(1+2\Phi_\mathrm{J})\,\mathrm{d}\eta^2
    + (1-2\Psi_\mathrm{J})\,\delta_{ij}\,\mathrm{d}x^i\mathrm{d}x^j
    \right]\,,
    \label{eq:perturbed_metric}
\end{equation}
where $\Phi_\mathrm{J}(\eta,\mathbf{x})$ and $\Psi_\mathrm{J}(\eta,\mathbf{x})$\footnote{Note that there is a convention inconsistency in the literature. Theoretical works preferred the notation $(\Psi,\Phi)$, whereas in parts of the cosmology and numerical-simulation literature the opposite convention for the metric potentials is often adopted}
 are the
Bardeen potentials \citep{Bardeen:1980kt} in the Jordan frame. The inverse metric components
at first order are
$
    g^{00}_\mathrm{J} = -a^{-2}(1-2\Phi_\mathrm{J}), \qquad
    g^{ij}_\mathrm{J} = a^{-2}(1+2\Psi_\mathrm{J})\delta^{ij}, \qquad
    g^{0i}_\mathrm{J} = 0.
    \label{eq:inverse_metric}
$
The scalar field is decomposed as
$
    \phi(\eta,\mathbf{x}) = \bar{\phi}(\eta) + \delta\phi(\eta,\mathbf{x}),
$
where the perturbation $\delta\phi$ depends on both conformal
time and spatial coordinates. This spatial dependence propagates to all composite
quantities derived from $\phi$, and in particular generates spatial
gradients $\partial_i\delta\phi \neq 0$. All derived quantities are then expanded as
$X = \bar{X} + \delta X$ at first order, with background values carrying
an overbar and evaluated at $\bar{\phi}(\eta)$.
Since $\delta F$ depends on
$\phi$ through $\bar{\alpha}$ and $\bar{F}$, its spatial and temporal
derivatives differ
\begin{equation}
    \partial_\mu\,\delta F = -2\bar{F}\!\left[
    \bar{\alpha}\,\partial_\mu\delta\phi
    + \left(\bar{\alpha}_{,\phi} - 2\bar{\alpha}^2\right)
    \bar{\phi}'\,\delta^0_\mu\,\delta\phi
    \right] \,,
    \label{eq:grad_deltaF}
\end{equation}
which gives
\begin{equation}
    \partial_i\,\delta F = -2\bar{\alpha}\bar{F}\,\partial_i\delta\phi,
    \qquad
    \delta F' = -2\bar{\alpha}\bar{F}\,\delta\phi'
    - 2\bar{F}\bar{\phi}'
    \!\left(\bar{\alpha}_{,\phi} - 2\bar{\alpha}^2\right)\delta\phi \,.
    \label{eq:deltaF_components}
\end{equation}

\subsubsection{Perturbed Einstein equations}
\label{subsubsec:einstein_perturbed}

We now linearise the modified Einstein equation~\eqref{eq:einstein_jordan}
around the FLRW background. At first order, this leads to
\begin{equation}
    \bar{F}\,\delta G^\mathrm{J}_{\mu\nu}
    + \bar{G}^\mathrm{J}_{\mu\nu}\,\delta F
    + \delta\mathcal{D}_{\mu\nu}F
    = \delta T^{(\mathrm{m})}_{\mu\nu} + \delta T^{(\phi)}_{\mu\nu}\,,
    \label{eq:einstein_perturbed}
\end{equation}
where the perturbed geometric correction operator is
\begin{align}
    \delta\mathcal{D}_{\mu\nu}F
    &= \delta g_{\mu\nu}\,\bar{\Box}\bar{F}
    + \bar{g}_{\mu\nu}\,\delta g^{\rho\sigma}
      \bar{\nabla}_\rho\bar{\nabla}_\sigma\bar{F}
    + \bar{g}_{\mu\nu}\,\bar{\Box}\,\delta F
    - \bar{\nabla}_\mu\bar{\nabla}_\nu\delta F
    \nonumber\\
    &\quad
    - \bar{g}_{\mu\nu}\,\bar{g}^{\rho\sigma}
      \delta\Gamma^\lambda_{\rho\sigma}\,\partial_\lambda\bar{F}
    + \delta\Gamma^\lambda_{\mu\nu}\,\partial_\lambda\bar{F} \,,
    \label{eq:deltaD}
\end{align}
whose components in the Newtonian gauge evaluate to
\begin{subequations}
\label{eq:deltaD_components}
\begin{align}
    \delta\mathcal{D}_{00}F &=
    -\nabla^2\delta F + 3\mathcal{H}_\mathrm{J}\delta F'
    - 3\bar{F}'\Psi_\mathrm{J}' \,,
    \label{eq:deltaD00} \\
    \delta\mathcal{D}_{0i}F &=
    -\partial_i\!\left(
        \delta F' - \mathcal{H}_\mathrm{J}\delta F - \bar{F}'\Phi_\mathrm{J}
    \right) \,,
    \label{eq:deltaD0i} \\
    \delta\mathcal{D}_{ij}F &=
    -\partial_i\partial_j\delta F
    + \delta_{ij}\Big[
        -\delta F'' - \mathcal{H}_\mathrm{J}\delta F'
        + \nabla^2\delta F \nonumber\\
    &\quad + \bar{F}'(\Phi_\mathrm{J}' + 2\Psi_\mathrm{J}')
        + 2(\Phi_\mathrm{J}+\Psi_\mathrm{J})
          \!\left(\bar{F}'' + \mathcal{H}_\mathrm{J}\bar{F}'\right)
    \Big] \,.
    \label{eq:deltaDij}
\end{align}
\end{subequations}
Specialising to the various components of the Einstein equation we find 
\\

\paragraph{$\bf{(00)}$:}
Using $\delta G^\mathrm{J}_{00} = 2\nabla^2\Psi_\mathrm{J} - 6\mathcal{H}_\mathrm{J}\Psi_\mathrm{J}'$ and
$\bar{G}^\mathrm{J}_{00} = 3\mathcal{H}_\mathrm{J}^2$, the $(00)$ projection of
equation~\eqref{eq:einstein_perturbed} gives
\begin{equation}
    2\bar{F}\!\left(\nabla^2\Psi_\mathrm{J} - 3\mathcal{H}_\mathrm{J}\Psi_\mathrm{J}'\right)
    + 3\mathcal{H}_\mathrm{J}^2\delta F
    + 3\mathcal{H}_\mathrm{J}\delta F'
    - \nabla^2\delta F
    - 3\bar{F}'\Psi_\mathrm{J}'
    = a_\mathrm{J}^2\!\left(\delta\rho_\mathrm{m} + \delta\rho_\mathrm{r} + \delta\rho_\phi\right) \,,
    \label{eq:einstein00_perturb}
\end{equation}
where the perturbation of the scalar energy and pressure are
\begin{align}
    \delta\rho_\phi &\equiv
    \frac{\bar{\phi}'^{\,2}\,\delta Z
    + 2\bar{Z}\bar{\phi}'\delta\phi'}{a_\mathrm{J}^2}
    - \delta\mathcal{L}^{(\mathrm{J})}_{\phi,\mathrm{eff}}
    - 2\Phi_\mathrm{J}\bar{\mathcal{L}}^{(\mathrm{J})}_{\phi,\mathrm{eff}}\,,
    \label{eq:delta_rho_phi}\\
    \delta p_\phi & \equiv \delta \mathcal{L}^{(\mathrm{J})}_{\phi,\mathrm{eff}}- 2\Psi_J\bar{\mathcal{L}}^{(\mathrm{J})}_{\phi,\mathrm{eff}}\,.
\end{align}
{In general we have $\delta\rho_s\equiv\delta T^{(s)}_{00}/a_J^2$ and $\delta p_s\equiv\delta T^{(s)}_{ii}/a_J^2$ for every species
$s\in\{m,r,\phi\}$.}

\paragraph{$\bf{(0i)}$:}
Using $\delta G^\mathrm{J}_{0i} = 2\partial_i(\Psi_\mathrm{J}' + \mathcal{H}_\mathrm{J}\Phi_\mathrm{J})$
and $\bar{G}^\mathrm{J}_{0i} = 0$, and noting that all terms are spatial
gradients 
one obtains by neglecting any vorticity
\begin{equation}
    2\bar{F}\!\left(\Psi_\mathrm{J}' + \mathcal{H}_\mathrm{J}\Phi_\mathrm{J}\right)
    - \delta F' + \mathcal{H}_\mathrm{J}\delta F + \bar{F}'\Phi_\mathrm{J}
    = - a_\mathrm{J}^2\!\left(\bar{\rho}_\mathrm{m} + \bar{\rho}_\mathrm{r}
    + \bar{p}_\mathrm{r}\right)v
    +\bar{Z}\bar{\phi}'\delta\phi \,,
    \label{eq:einstein0i}
\end{equation}
where $v$ is the fluid velocity potential defined by $\partial_i v \equiv v_i$ and we also introduce for future reference $\theta= \partial_i v^i$. In Fourier space $v=-\frac{\theta}{k^2}$.

\paragraph{$\bf{(ij)}$:}
For the spatial components, it is useful to separate the
off-diagonal ($i \neq j$) and diagonal parts. The traceless and off-diagonal part gives
\begin{equation}
    \partial_i\partial_j\!\left[
        \bar{F}(\Psi_\mathrm{J} - \Phi_\mathrm{J}) - \delta F
    \right] = 0 \qquad (i \neq j) \,.
    \label{eq:offdiag}
\end{equation}
Since all functions must have their Fourier transformation defined for all wave vectors $k_ik_j$ in Fourier space,
the derivatives must vanish, yielding the equation
\begin{equation}
{
    \Psi_\mathrm{J} - \Phi_\mathrm{J} = \frac{\delta F}{\bar{F}}
    = -2\bar{\alpha}\,\delta\phi\,.}
    \label{eq:slip}
\end{equation}
This discrepancy between both Bardeen potentials  is  a signature of the non-minimal coupling.
The diagonal ($i = j$) part gives the propagation equation for the potentials
\begin{align}
    \bar{F}\Big[2\Psi_\mathrm{J}'' &+ 2\mathcal{H}_\mathrm{J}(\Phi_\mathrm{J}' + 2\Psi_\mathrm{J}')
    + 2(2\mathcal{H}_\mathrm{J}' + \mathcal{H}_\mathrm{J}^2)(\Phi_\mathrm{J} + \Psi_\mathrm{J})
    + \nabla^2(\Phi_\mathrm{J} - \Psi_\mathrm{J})\Big]
    \nonumber\\
    &- (2\mathcal{H}_\mathrm{J}' + \mathcal{H}_\mathrm{J}^2)\delta F
    - \delta F'' - \mathcal{H}_\mathrm{J}\delta F'
    + \nabla^2\delta F \nonumber\\
    &+ \bar{F}'(\Phi_\mathrm{J}' + 2\Psi_\mathrm{J}')
    + 2(\Phi_\mathrm{J} + \Psi_\mathrm{J})(\bar{F}'' + \mathcal{H}_\mathrm{J}\bar{F}')
    = a_\mathrm{J}^2(\delta p_\mathrm{m} + \delta p_\mathrm{r} + \delta p_\phi)
    \label{eq:einsteinij}
\end{align}
which we will analyse below.

\subsubsection{Modified Poisson equations}

We now rewrite the perturbed $(00)$ Einstein
equation~\eqref{eq:einstein00_perturb} as a Poisson equation for
$\Psi_\mathrm{J}$. This gives us 
\begin{align}
{
    \nabla^2\Psi_\mathrm{J}
    = \frac{a^2}{2M_{\mathrm{Pl}}^2}
        (\delta\rho_\mathrm{m} + \delta\rho_\mathrm{r})   - \bar{\alpha}\,\nabla^2\delta\phi \
    +
    \frac{a^2}{2M_{\mathrm{Pl}}^2}\left[
         \epsilon_F
          (\delta\rho_\mathrm{m} + \delta\rho_\mathrm{r})
        + \frac{M_{\mathrm{Pl}}^2}{\bar{F}}\,\delta\rho_\phi\right]
    + \mathcal{R}_{\mathrm{vel}}}
    \label{eq:poisson_Psi_full}
\end{align}
where  $\nabla^2\delta F = -2\bar{\alpha}\bar{F}\nabla^2\delta\phi$
from equation~\eqref{eq:deltaF_components}, and the
velocity-type terms are 
\begin{equation}
    \mathcal{R}_{\mathrm{vel}} \equiv
    \frac{1}{2\bar{F}}\!\left[
        6\bar{F}\mathcal{H}_\mathrm{J}\Psi_\mathrm{J}'
        + 3\bar{F}'\Psi_\mathrm{J}'
        - 3\mathcal{H}_\mathrm{J}^2\delta F
        - 3\mathcal{H}_\mathrm{J}\delta F'
    \right] \,.
    \label{eq:Rvel}
\end{equation}
The Poisson equation for $\Phi_\mathrm{J}$ follows from the slip
relation~\eqref{eq:slip}, $\Phi_\mathrm{J} = \Psi_\mathrm{J} + 2\bar{\alpha}\delta\phi$,
\begin{align}
    {\nabla^2\Phi_\mathrm{J}
    = \frac{a_\mathrm{J}^2}{2M_{\mathrm{Pl}}^2}
        (\delta\rho_\mathrm{m} + \delta\rho_\mathrm{r})   + \bar{\alpha}\,\nabla^2\delta\phi \
    +
    \frac{a_\mathrm{J}^2}{2M_{\mathrm{Pl}}^2}\left[
         \epsilon_F
          (\delta\rho_\mathrm{m} + \delta\rho_\mathrm{r})
        + \frac{M_{\mathrm{Pl}}^2}{\bar{F}}\,\delta\rho_\phi\right]
    + \mathcal{R}_{\mathrm{vel}}} \,.
    \label{eq:poisson_Phi_full}
\end{align}
This is particularly useful when studying structure formation. 

\subsubsection{Perturbed Klein-Gordon equation}
\label{sec:perturbed_KG}

We now linearise the scalar field equation of motion~\eqref{eq:KG_compact} around
the FLRW background of Section~\ref{subsec:background}, just as the background equation governed the evolution of $\phi$, 
the linearised equation will determine the evolution of the 
scalar field perturbation $\delta\phi$. Recall that the covariant
Klein-Gordon equation reads
\begin{equation}
    \nabla^{(\mathrm{J})}_\mu J^\mu = - \mathcal{S}_{\phi,\mathrm{eff}} \,.
    \label{eq:KG_compact_bis}
\end{equation}
Using the identity
$\nabla^{(\mathrm{J})}_\mu J^\mu = (\sqrt{-g_\mathrm{J}})^{-1}\,
\partial_\mu\!\left(\sqrt{-g_\mathrm{J}}\,J^\mu\right)$,
and expanding to first order
$\sqrt{-g_\mathrm{J}} = a_\mathrm{J}^4(1+\Phi_\mathrm{J}-3\Psi_\mathrm{J})$,
$g^{00}_J = -a_\mathrm{J}^{-2}(1-2\Phi_\mathrm{J})$ and
$g^{ij}_J = a_\mathrm{J}^{-2}(1+2\Psi_\mathrm{J})\delta^{ij}$,
the temporal and spatial components of $\sqrt{-g_\mathrm{J}}\,J^\mu$ read to first order 
\begin{align}
    \sqrt{-g_\mathrm{J}}\,J^0\,\big|^{(1)}
    &= -a_\mathrm{J}^2\!\left[
        \bar{Z}\bar\phi'
        + \delta J^0_{\mathrm{eff}}
    \right] \,, \\[4pt]
    \partial_i\!\left(\sqrt{-g_\mathrm{J}}\,J^i\right)^{(1)}
    &= a_\mathrm{J}^2\,\bar{Z}\,\nabla^2\delta\phi \,,
\end{align}
where we have introduced the notation $J^0_{\mathrm{eff}}$,
\begin{equation}
    \delta J^0_{\mathrm{eff}}
    \equiv
    \bar{Z}\,\delta\phi'
    + \delta Z\,\bar\phi'
    - \bar{Z}\,\bar\phi'\!\left(\Phi_\mathrm{J} + 3\Psi_\mathrm{J}\right) \,.
    \label{eq:deltaJ0eff}
\end{equation}
Rewriting this term as
\begin{equation}
    \delta J^0_{\mathrm{eff}}
    =
    \bar{Z}_{\mathrm{eff}}\!\left(\delta\phi' - \bar\phi'\,\Phi_\mathrm{J}\right)
    - 3\bar{Z}\,\bar\phi'\,\Psi_\mathrm{J}
    + \bar\alpha\,\bar\phi'\!\left(\bar{Z}_{\mathrm{eff}} - 3\bar{Z} - 12\bar{F}\,\bar\alpha_{,\phi}\right)\delta\phi \,,
    \label{eq:deltaJ0eff_explicit}
\end{equation}
the first-order perturbation of the
divergence is 
\begin{equation}
    \left[\nabla^{(\mathrm{J})}_\mu J^\mu\right]^{(1)}
    = \frac{1}{a_\mathrm{J}^4}\!\left[
        -\!\left(a_\mathrm{J}^2\,\delta J^0_{\mathrm{eff}}\right)'
        + a_\mathrm{J}^2\,\bar{Z}\,\nabla^2\delta\phi
        + \left(\Phi_\mathrm{J} - 3\Psi_\mathrm{J}\right)\!\left(a_\mathrm{J}^2\bar{Z}\bar\phi'\right)'
    \right].
    \label{eq:div_J_pert}
\end{equation}
The last term involving $(a_\mathrm{J}^2\bar{Z}\bar\phi')'$ is directly related to
$a_\mathrm{J}^4\,\bar{\mathcal{S}}_{\phi,\mathrm{eff}}$.
The effective source $\mathcal{S}_{\phi,\mathrm{eff}}$ receives first-order
contributions from both the scalar Lagrangian derivative and the curvature
coupling. The perturbation of the curvature coupling term splits 
\begin{equation}
    \left[\alpha(\phi)\,F(\phi)\,R_J\right]^{(1)}
    =
    \bar\alpha\,\bar{F}\,R_J^{(1)}
    + \bar{F}\,\bar{R}_J
      \!\left(\bar\alpha_{,\phi} - 2\bar\alpha^2\right)\delta\phi \,,
    \label{eq:curv_coupling_pert}
\end{equation}
where the two terms come from the perturbation of the Ricci scalar and from the $\phi$-dependence of $\alpha F$ evaluated
at $\bar\phi$.
This gives the perturbed Klein-Gordon equation 
\begin{equation}
    \frac{1}{a_\mathrm{J}^4}\!\left[
        -\!\left(a_\mathrm{J}^2\,\delta J^0_{\mathrm{eff}}\right)'
        + a_\mathrm{J}^2\,\bar{Z}\,\nabla^2\delta\phi
        + \left(\Phi_\mathrm{J} - 3\Psi_\mathrm{J}\right)\!\left(a_\mathrm{J}^2\bar{Z}\bar\phi'\right)'
    \right] =\bar\alpha\,\bar{F}\,R_J^{(1)}
    + \bar{F}\,\bar{R}_J
      \!\left(\bar\alpha_{,\phi} - 2\bar\alpha^2\right)\delta\phi \,.
    \label{eq:KG_perturbated}
\end{equation}
The complete perturbed Klein-Gordon equation follows from inserting both
\eqref{eq:div_J_pert} and the perturbation of $\mathcal{S}_{\phi,\mathrm{eff}}$
into~\eqref{eq:KG_compact_bis}.

\subsubsection{Matter and radiation conservation equations at first order}
\label{sec:matter_perturbations}

Starting from the covariant conservation law $\nabla^{(\mathrm{J})}_\mu T^{(\mathrm{m})\mu\nu} = 0$
we have 
\begin{equation}
    \bar{\nabla}^{(J)}_\mu \delta T^{(\mathrm{s})\mu\nu}
    + \delta\Gamma^\mu_{\mu\lambda}\,\bar{T}^{(\mathrm{s})\lambda\nu}
    - \delta\Gamma^\nu_{\mu\lambda}\,\bar{T}^{(\mathrm{s})\mu\lambda} = 0 \,,
    \label{eq:conservation_pert}
\end{equation}
whose $\nu = 0$ and $\nu = i$ projections yield respectively the continuity
and Euler equations for each fluid species $s \in \{m,r\}$.
We obtain from the time components
\begin{equation}
    \delta_\mathrm{s}' + (1+w_\mathrm{s})\,\theta_\mathrm{s} - 3(1+w_\mathrm{s})\Psi_\mathrm{J}' = 0 \,,
    \label{eq:continuity}
\end{equation}
where $w_\mathrm{s} \equiv \bar p_\mathrm{s}/\bar\rho_\mathrm{s}$ is the equation-of-state parameter.
Likewise, the $\nu = i$ component gives
\begin{equation}
    \theta_\mathrm{s}' + \mathcal{H}_\mathrm{J}(1-3c^2_{a,\mathrm{s}})\,\theta_\mathrm{s}
    + \frac{c^2_{a,\mathrm{s}}}{1+w_\mathrm{s}}\,\nabla^2\delta_\mathrm{s}
    + \nabla^2\Phi_\mathrm{J} = 0 \,,
    \label{eq:euler}
\end{equation}
where $c^2_{a,\mathrm{s}} \equiv \bar p_\mathrm{s}'/\bar\rho_\mathrm{s}'$ is the adiabatic sound speed.
In Fourier space for pressureless matter ($w_\mathrm{m} = 0$, $c^2_{a,\mathrm{m}} = 0$)
and radiation ($w_\mathrm{r} = 1/3$, $c^2_{a,\mathrm{r}} = 1/3$),
equations~\eqref{eq:continuity} and \eqref{eq:euler} become respectively 
\begin{align}
    \delta_\mathrm{m}' + \theta_\mathrm{m} - 3\Psi_\mathrm{J}' &= 0 \,,
    \label{eq:continuity_matter}\\
    \theta_\mathrm{m}' + \mathcal{H}_\mathrm{J}\,\theta_\mathrm{m} - k^2\Phi_\mathrm{J} &= 0\,,
    \label{eq:euler_matter}\\[6pt]
    \delta_\mathrm{r}' + \frac{4}{3}\,\theta_\mathrm{r} - 4\Psi_\mathrm{J}' &= 0\,,
    \label{eq:continuity_rad}\\
    \theta_\mathrm{r}' - k^2\!\left(\frac{1}{4}\delta_\mathrm{r} + \Phi_\mathrm{J}\right) &= 0 \,.
    \label{eq:euler_rad}
\end{align}
These equations are formally identical to their $\Lambda$CDM counterparts ~\citep{Ma:1995ey}.

\subsubsection{Subhorizon and quasi-static approximation}

The full system of linear perturbations derived above is exact at first order in metric perturbations. We will simplify it before presenting exact results based on numerical calculations in the following section. 
The metric potentials $\Phi_\mathrm{J}$ and $\Psi_\mathrm{J}$, the scalar perturbation 
$\delta\phi$, and the matter perturbations are all dynamically coupled. Here we will be interested in the evolution of matter perturbations on scales well inside the 
Hubble radius. We therefore consider the subhorizon approximation, which in 
Fourier space translates to
$
    k^2 \gg \mathcal{H}_\mathrm{J}^2 ,
$
where $k$ is the comoving wavenumber and $\mathcal{H}_\mathrm{J}$ is the conformal Hubble 
rate defined above. The quasi-static approximation (QSA) then assumes that, in 
this regime, spatial gradients dominate over time derivatives of perturbative 
quantities such as the gravitational potentials and the scalar perturbation, i.e. 
$ |\Phi_\mathrm{J}'|,\ |\Psi_\mathrm{J}'|,\ |\delta\phi'|
    \ll
    k\,|\Phi_\mathrm{J}|,\ k\,|\Psi_\mathrm{J}|,\ k\,|\delta\phi| .
$
All terms suppressed by powers of $\mathcal{H}_\mathrm{J}^2/k^2$ are neglected.

We now apply this approximation to the perturbation equations. For pressureless 
matter, the conservation equations in Newtonian gauge read
\begin{align}
    &\delta_\mathrm{m}' = -\theta_\mathrm{m} + 3\Psi_\mathrm{J}' \,, \\
    &\theta_\mathrm{m}' + \mathcal{H}_\mathrm{J}\,\theta_\mathrm{m} = k^2\Phi_\mathrm{J} \,,
\end{align}
where $\theta_\mathrm{m} \equiv \partial_i v_\mathrm{m}^i$. In the subhorizon limit, the term 
$3\Psi_\mathrm{J}'$ is suppressed by $\mathcal{H}_\mathrm{J}^2/k^2$ relative to $k\theta_\mathrm{m}$ and 
is therefore neglected, giving
\begin{equation}
    \delta_\mathrm{m}' \simeq -\theta_\mathrm{m}\, .
\end{equation}
Taking one derivative with respect to conformal time and substituting into the Euler 
equation yields
\begin{equation}
    \delta_\mathrm{m}'' + \mathcal{H}_\mathrm{J}\,\delta_\mathrm{m}' + k^2\Phi_\mathrm{J} \simeq 0 \,.
    \label{eq:delta_m_growth_phi_subhorizon}
\end{equation}
The growth of matter perturbations is therefore controlled by the potential 
$\Phi_\mathrm{J}$, which is the Bardeen potential entering the non-relativistic geodesic 
equation in the Jordan frame.

The modified Poisson equation follows from the $(00)$ component of the perturbed 
Einstein equations. Keeping only the dominant $k^2$ terms gives
\begin{equation}
    -k^2\Psi_\mathrm{J}
    \simeq
    \frac{a_\mathrm{J}^2}{2\bar{F}}\,\bar{\rho}_\mathrm{m}\,\delta_\mathrm{m}
    + \bar{\alpha}\,k^2\,\delta\phi \,,
    \label{eq:poisson_psi_qs}
\end{equation}

The scalar perturbation is obtained from the perturbed Klein-Gordon equation. 
In the same quasi-static limit, all time-derivative terms are suppressed relative 
to the $k^2\delta\phi$ contribution, and the equation reduces to the algebraic 
relation
\begin{equation}
    \bar{Z}\,k^2\,\delta\phi
    \simeq
    2\bar{\alpha}\bar{F}\,k^2\!\left(2\Psi_\mathrm{J} - \Phi_\mathrm{J}\right) \,,
    \label{eq:delta_phi_qs}
\end{equation}
or equivalently, using the slip relation \eqref{eq:slip},
\begin{equation}
    \delta\phi
    =
    \frac{2\bar{\alpha}\bar{F}}{\bar{Z} + 4\bar{\alpha}^2\bar{F}}\,\Psi_\mathrm{J} \,.
    \label{eq:deltaphi_qs_explicit}
\end{equation}
This expression shows explicitly the physical role of K-mouflage screening when 
$\bar{Z}$ becomes large, the scalar field perturbation responds weakly to matter 
perturbations, and the fifth-force contribution is suppressed.

\begin{figure}[!h]
\centering
\includegraphics[width=0.98\linewidth]{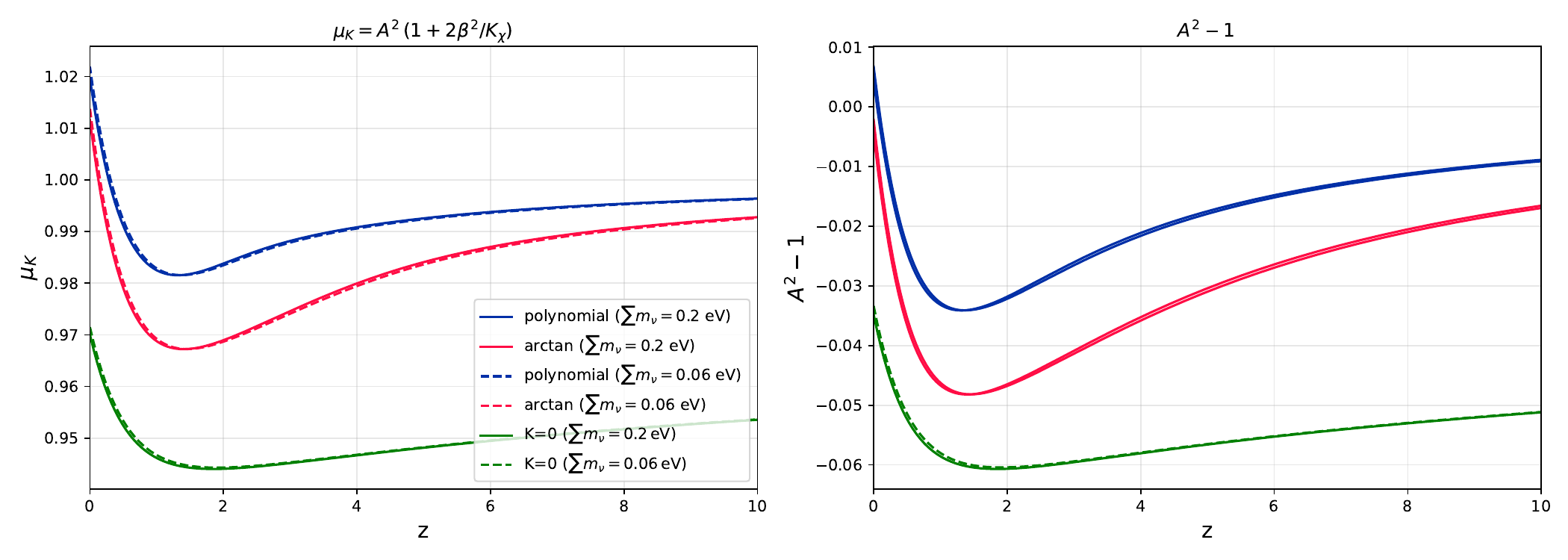} 
\caption{$\mu_{\rm K}$ (left) and $A^2(\phi)-1$ (right) for the models considered: $\sum m_{\nu}=0.2$ (solid lines) and $\sum m_{\nu}=0.06$ (dashed lines). For the coupled dilation model, we consider in both cases the model with $\beta=0.05$.}
\label{fig:muk}
\end{figure}

Combining equations \eqref{eq:poisson_psi_qs}, 
\eqref{eq:slip}, and \eqref{eq:deltaphi_qs_explicit}, 
one obtains a modified Poisson equation for the physical potential $\Phi_\mathrm{J}$:
\begin{equation}
    -k^2\Phi_\mathrm{J}
    \simeq
    \frac{a^2}{2\bar{F}}
    \left(1 + \frac{2\bar{F}\,\bar{\alpha}^2}{\bar A^{-2} K_\chi(\bar{\chi})}\right)
    \bar{\rho}_\mathrm{m}\,\delta_\mathrm{m} \,.
    \label{eq:poisson_phi_qs_raw}
\end{equation}
It is convenient to rewrite this in the standard form
\begin{equation}
    -k^2\Phi_\mathrm{J}
    =
    \frac{a^2}{2M_{\rm Pl}^2}\,\mu_{\rm K}(a)\,\bar{\rho}_\mathrm{m}\,\delta_\mathrm{m} \,,
    \label{eq:poisson_phi_mu_qs}
\end{equation}
where the dimensionless effective Newton coupling is defined as
\begin{align}
    &{
    \mu_{\rm K}(a)
    \equiv
    \frac{M_{\rm Pl}^2}{\bar{F}}
    \left(1 + \frac{2\bar{F}\,\bar{\alpha}^2}{\bar A^{-2} K_\chi(\bar{\chi})}\right)
    } \,. 
     \label{eq:mu_K}
 \end{align}
 The first factor $M_{\rm Pl}^2/\bar{F}$
accounts for the variation of the effective Planck mass, while the second 
term is the genuine fifth-force contribution mediated by the scalar field.
By substituting equation \eqref{eq:poisson_phi_mu_qs} into 
\eqref{eq:delta_m_growth_phi_subhorizon} gives the subhorizon growth equation in 
conformal time:
\begin{equation}
    \delta_\mathrm{m}''
    + \mathcal{H}\,\delta_\mathrm{m}'
    - \frac{a^2}{2M_{\rm Pl}^2}\,\mu_{\rm K}\,\bar{\rho}_\mathrm{m}\,\delta_\mathrm{m}
    = 0 \,.
    \label{eq:delta_growth_conformal_qs}
\end{equation}
{In this equation, two terms are relevant. The Hubble friction reduces the growth when the Hubble rate increases as a function of the coupling strength $\beta$. The fifth force contribution in $\mu_{\rm K}$ increases the growth of structure with $\beta$. These are two competing effects. {Finally the amplitude of $\mu_K$ is controlled by $\bar A^2$ which is smaller than unity for the models we consider, see Fig. \ref{fig:muk}. In our models, the Hubble friction is relevant and reduces growth with increasing $\beta$ whilst growth is reduced as $\mu_K<1$ during most of the cosmological history.}}

The linear growth factor $D_+(a)$ is defined by $\delta_\mathrm{m}(k,a) = 
D_+(a)\,\delta_\mathrm{m}(k,a_{\rm ini})$ which is scale-independent in the quasi-static 
approximation. It satisfies
\begin{equation}
    {
    \frac{d^2 D_+}{d\ln^2 a}
    +
    \left(1 + \frac{d\ln\mathcal{H}_\mathrm{J}}{d\ln a}\right)
    \frac{dD_+}{d\ln a}
    -
    \frac{3}{2}\,\mu_{\rm K}(a)\,\Omega_\mathrm{m}(a)\,D_+
    = 0
    } \,.
    \label{eq:Dplus_qs}
\end{equation}
We will see how the growth of structure is affected in the Kinessence models below. 

\subsection{Embedding in Horndeski}

A full numerical analysis of the model comprising the Cosmic Microwave Background and the low-redshift growth of structure necessitates to go beyond the quasi-static approximation. For this we employ the $\mathcal{H}$-EFTCAMB code \cite{Ye:2026qqf}, which requires writing the Lagrangian in terms of the well-known Horndeski terms. The Horndeski Lagrangian \cite{Horndeski1974} is given by
\begin{align}
    \mathcal{S}_{\mathcal{H}} = \int d^4x \sqrt{-g} \left( L_2 + L_3 + L_4 + L_5 \right),
\end{align}
with
\begin{align}
L_2 &= G_2(\phi, X)\,, \nonumber \\
L_3 &= G_3(\phi, X)\Box\phi\,, \nonumber \\
L_4 &= G_4(\phi, X)R - 2G_{4X}(\phi, X)\left[ (\Box\phi)^2 - \phi_{;\mu\nu}\phi^{;\mu\nu} \right]\,, \nonumber \\
L_5 &= G_5(\phi, X)G_{\mu\nu}\phi^{;\mu\nu} + \frac{1}{3}G_{5X}(\phi, X) \left[ (\Box\phi)^3 - 3\Box\phi \, \phi^{;\mu\nu} \phi_{;\mu\nu} + 2\phi_{;\mu}^{\;\;\nu} \phi_{;\nu}^{\;\;\sigma} \phi_{;\sigma}^{\;\;\mu} \right]\,.
\end{align}
Rewriting our starting-point Lagrangian here in the Jordan frame
$\mathcal{L}_{\phi,\mathrm{eff}}^{(J)} = -6F(\phi)\,\alpha^2(\phi)\,X
    + A^{-4}(\phi)\,M^4\Big[K(\mathcal{X}) - f(\phi)\Big],
   $ where $ \mathcal{X} \equiv \frac{A^2(\phi)\,X}{M^4}$. 
We can identify the Horndeski functions
\begin{align}
    G_2(\phi,X) &= -6F(\phi)\,\alpha^2(\phi)\,X \;+\; A^{-4}(\phi)\,M^4\Big[K(\mathcal{X}) - f(\phi)\Big], \\
    G_3(\phi,X) &= 0, \\
    G_4(\phi,X) &= \frac{F(\phi)}{2} = \frac{M_{\rm Pl}^2}{2}A^{-2}(\phi), \\
    G_5(\phi,X) &= 0 \, ,
\end{align}
where there is no {explicit} braiding \cite{Bellini:2014fua} and the constraints on the speed of gravitational waves are automatically satisfied. The perturbation equations require a long list of derivatives which is given below for $G_2$,
\begin{align}
    G_{2X} &= A^{-2}(\phi)\,K_X(\mathcal{X}) - 6F(\phi)\,\alpha^2(\phi) \;=\; Z(\phi,\chi_J)\,, \\
    G_{2XX} &= \frac{1}{M^4}\,K_{XX}(\mathcal{X})\,, \\
    G_{2XXX} &= \frac{A^2(\phi)}{M^8}\,K_{XXX}(\mathcal{X})\,, \\[6pt]
    G_{2\phi} &= 12F(\phi)\,\alpha(\phi)\big(\alpha^2(\phi)-\alpha_{,\phi}(\phi)\big)\,X
    \;+\; A^{-4}(\phi)\,M^4\Big[2\alpha(\phi)\,\mathcal{X}\,K_X(\mathcal{X})
    - 4\alpha(\phi)\,K(\mathcal{X}) + 4\alpha(\phi)\,f(\phi) - f'(\phi)\Big]\,, \\[6pt]
    G_{2\phi X} &= 2\alpha(\phi)\,A^{-2}(\phi)\Big[\mathcal{X}\,K_{XX}(\mathcal{X}) - K_X(\mathcal{X})\Big]
    + 12F(\phi)\,\alpha(\phi)\big(\alpha^2(\phi)-\alpha_{,\phi}(\phi)\big) \, ,
\end{align}
and also for $G_4$
\begin{align}
    G_{4\phi} &= -\alpha(\phi)\,F(\phi)\,, \\
    G_{4\phi\phi} &= F(\phi)\big(2\alpha^2(\phi)-\alpha_{,\phi}(\phi)\big)\,, \\
    G_{4X} &= G_{4\phi X} = G_{4XX} = 0\,.
\end{align}
{With these we can compute the CMB spectra,  and observables related to the growth of structure,  which are then compared to the standard model results.}

\section{Cosmological Results and constraints from the variation of $G_N$}
\subsection{Cosmological models and CMB constraints}
\label{sec:results}
In order to compare with observations, one must choose explicit models for the functions $K(\chi)$ and $A(\phi)$. These are based on Ref. \cite{Brax:2014wla}, and the functions take the following form {
\begin{align}
A(\phi)&=e^{\beta \phi/M_{\rm Pl}}  \, , \\
K(\chi)&=\begin{cases} 
\chi + K_0 \chi^m \;\;\text{or} \\ 
\chi + K_* \left[\chi-\chi_* \arctan\left(\frac{\chi}{\chi_*} \right) \right] . 
\end{cases}
\end{align}}
The overall parameter $M$ defining the overall scale of the model will be numerically implemented as $M= M_{\rm num} (H_0^{\rm \Lambda CDM} M_{\rm Pl})^{1/2}$ where $M_{\rm num}$ is of order unity. {The integration is performed from an initial redshift $z_{\rm ini}=10^5$.} {We choose initial $\phi_{\rm ini}=0$ and $\dot \phi_{\rm ini}=0$ but the late time behaviour of the field evolution is almost independent of this choice. }
The parameters are free to vary, with default ones taken to be $\beta=0.1$, $K_0=1$, $m=3$, $K_*=1000$, $\chi_*=100$. These values are chosen as they allow for screening in the Solar System. In the appendix \ref{app:plots}, we will show how the cosmological results vary when these values are slightly changed. 

Using the connection with the Horndeski models described in the previous Section, we employ the $\mathcal{H}$-EFTCAMB code to determine the angular and the matter power spectra for the K-mouflage models. These spectra are compared to the $\Lambda$CDM model, with best fit parameters obtained by the \textit{Planck} Consortium \cite{Planck2018params}. In particular, we fix the parameter initially as  $\Omega_b h^2=0.022383$, $\Omega_c h^2=0.12011$, $\tau=0.0543$, $n_s=0.96605$ and $A_s=2.100582962 \times 10^{-9}$. For $\Lambda$CDM we have 3.044 {effective} neutrinos, and a single massive one of mass 0.06 eV. {We fix for $\Lambda$CDM $H_0=67.32$, and we impose that the angular scale of the sound horizon at the time of photon decoupling $\theta_*$ remains within the error bars of the value measured by \textit{Planck} $100 \theta_* = 1.0411  \pm 0.0003$. This is achieved by fine-tuning $M_{\rm num}$.} In addition, we will obtain a dark-energy equation of state compatible to the one measured by DESI \cite{DESI:2025zgx, DESI:2025fii}. We will not perform a detailed fit as this is left for future work but compare qualitative features such as the redshift for the crossing of the phantom divide and the slope of the curve.
We investigate both cases for the kinetic function $K(\chi)$ and we analyse the impact of each of the parameters of the models. 

\begin{figure}[!h]
\centering
\includegraphics[width=0.98\linewidth]{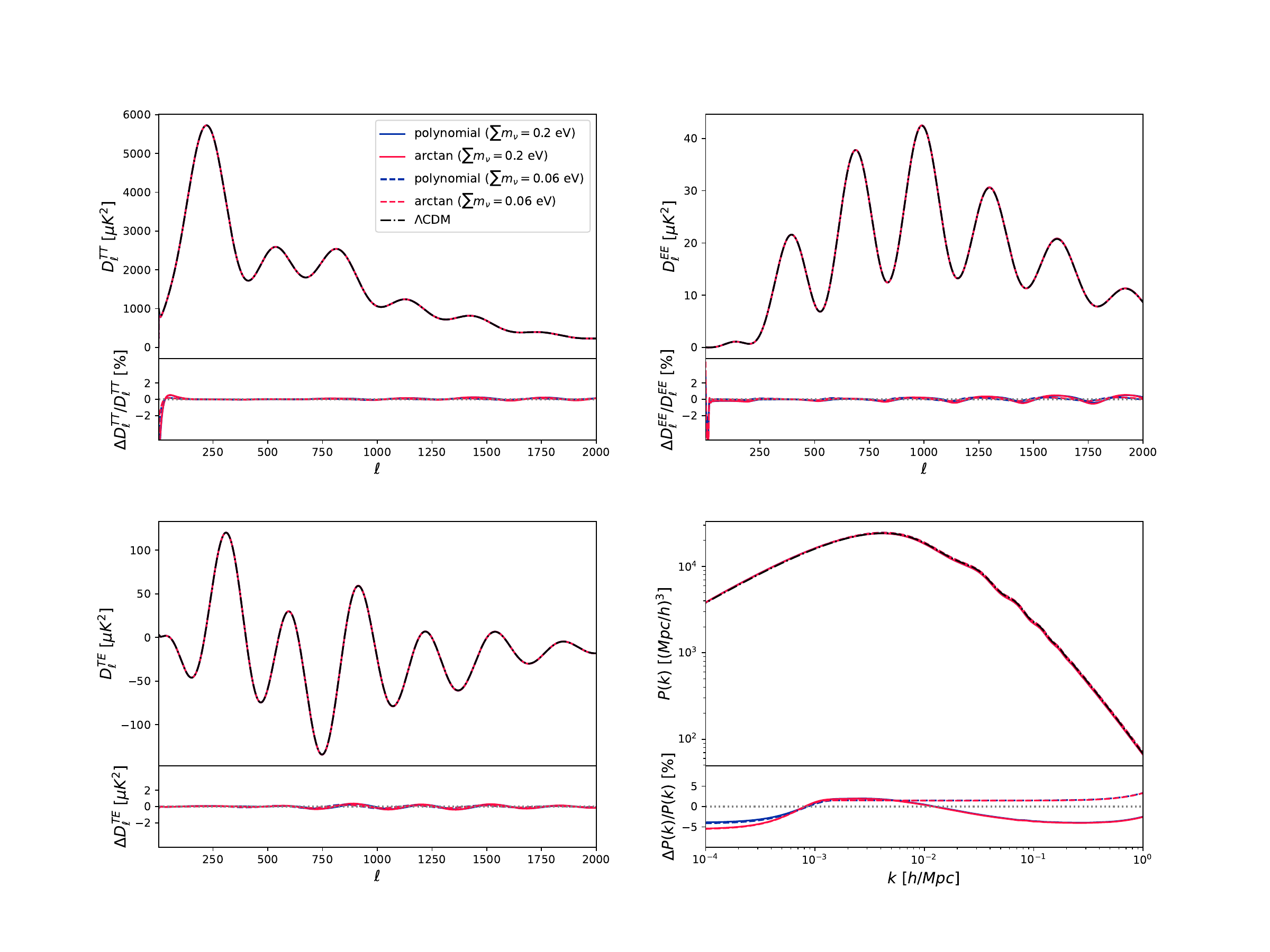} 
\caption{Angular power spectra (from top to bottom and left to right: temperature, E-polarisation and TE cross-spectrum) and matter power spectra for a reasonable fit of Kinessence power-law and arctan models. Solid lines represent Kinessence model with 2 massive neutrinos with a total mass of 0.2 eV, with the dashed lines the model with one massive neutrino of mass 0.06 eV.}
\label{fig:powersp}
\end{figure}

We have obtained a good fit of the observed data (power spectrum from \textit{Planck}, together with the observed DESI equation of state) using parameters close to  the default values for $K_0$ and $M_{\rm num}$  and $K_*$, and $\chi_*$ respectively.  A reasonable  fit corresponds to $\beta=0.09$ and 2 massive neutrinos with mass 0.2 eV for both models, while $\lambda=2.4$ and mass $M_{\rm num}=0.955$ for the polynomial model and $\lambda=2.3$ and mass $M_{\rm num}=0.945$ for the arctan one. {We have chosen large neutrino masses to illustrate the fact that our results can be made compatible with particle physics bounds.} {These models yield relative variations of $\theta_*$ with respect to $\Lambda$CDM of $-8.6 \times 10^{-3} \,\%$ and $2.4 \times 10^{-4}\,\%$ respectively.} {In addition, we consider a model having the same parameters, but with just one single massive neutrino, with the same mass as in the $\Lambda$CDM case. In this case, the calibrated values for the polynomial and arctan models are 0.9695 and 0.9585 respectively, with relative deviations of $\theta_*$ with respect to $\Lambda$CDM of $3 \times 10^{-5}$ and $-7\times 10^{-6}$; both models yield $\sigma_8=0.818$. The model with a non-standard neutrino mass is studied here in order to show that such a model can be compatible with current particle physics bounds.}
The models presented  satisfy the \textit{Planck} CMB constraints, with subpercent variations of the $C_\ell$'s {in all the scenarios considered} (Fig. \ref{fig:powersp}). {When representing the angular power spectra, we plot $D_\ell^{AB} \equiv \ell (\ell+1) C_\ell^{AB} /(2\pi)$.}

\begin{figure}[!h]
\centering
\includegraphics[width=0.98\linewidth]{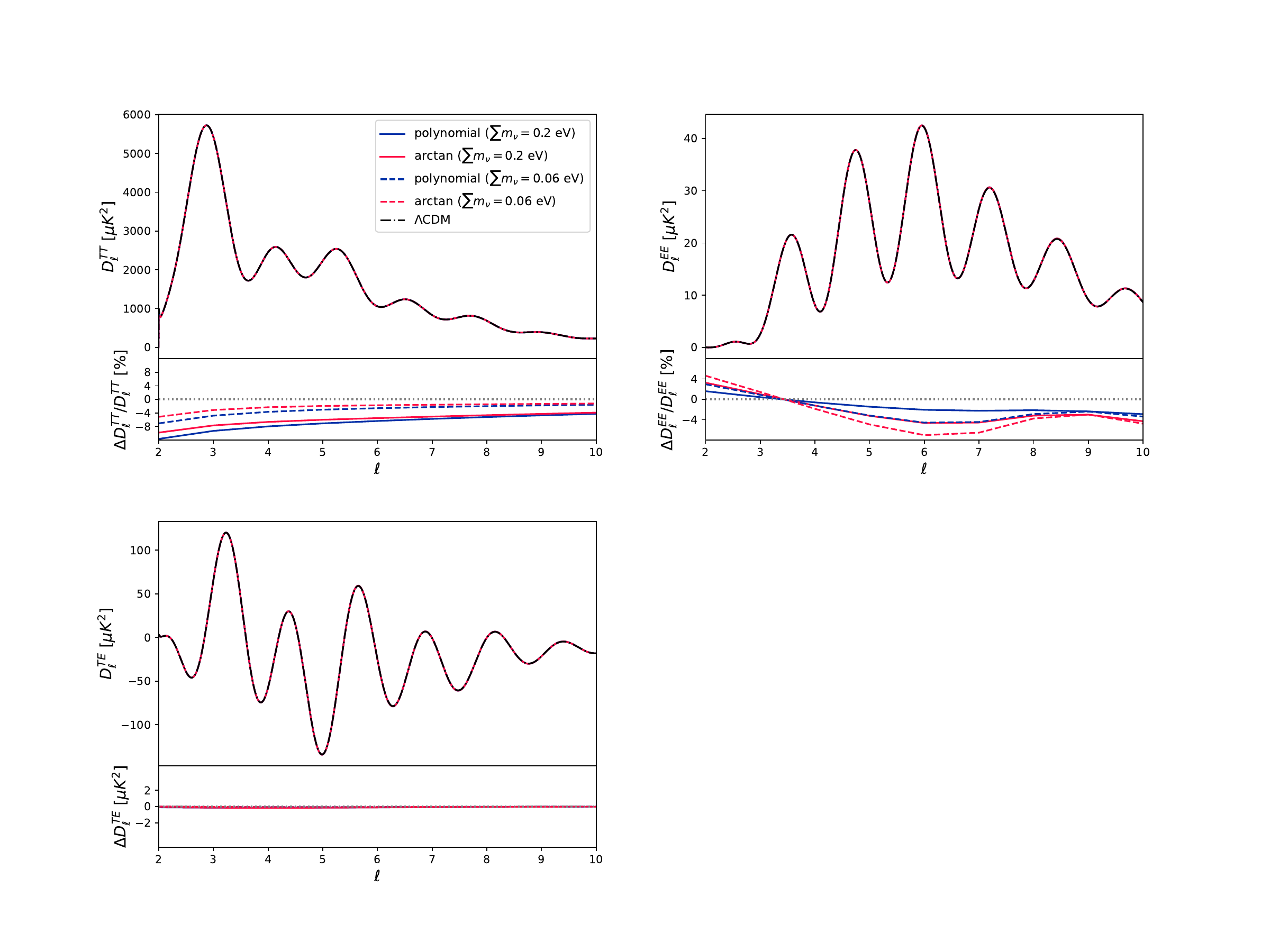} 
\caption{{Angular power spectra (from top to bottom and left to right: temperature, E-polarisation and TE cross-spectrum)  for a reasonable fit of Kinessence power-law and arctan models for $\ell \le 10$.}}
\label{fig:powersp_low_ell}
\end{figure}

We note that the deviation of the spectra is larger at low multipole values, and in Fig. \ref{fig:powersp_low_ell}, we plot the spectra for $\ell \le 10$. We note that the relative differences with respect to $\Lambda$CDM peak at around $\ell=6$, and these are always less than 8\%, making them undetectable due to cosmic variance. {We have also investigated the behaviour of the angular spectra up to $\ell=4000$, as measured by the ACT \cite{ACT:2020frw} and the Kinessence models do not deviate further from $\Lambda$CDM as in the plots shown here.}

{We study the deviation of the power spectrum with the standard model. We find, see Fig. \ref{fig:powersp}, that the reduction in power is effective on large scales whilst being non-operative on short scales. We find that the integrated effect on $\sigma_8$ makes the difference with {respect to} the standard model for one neutrino of mass 0.06 eV {negligible} whilst reducing it for a larger total mass of 0.2 eV. This matters as we will see below that $f\sigma_8$ is reduced even with a neutrino of mass 0.06 corresponding to the reduction $\mu_K<1$. }

\begin{figure}[!h]
\centering
\includegraphics[width=0.98\linewidth]{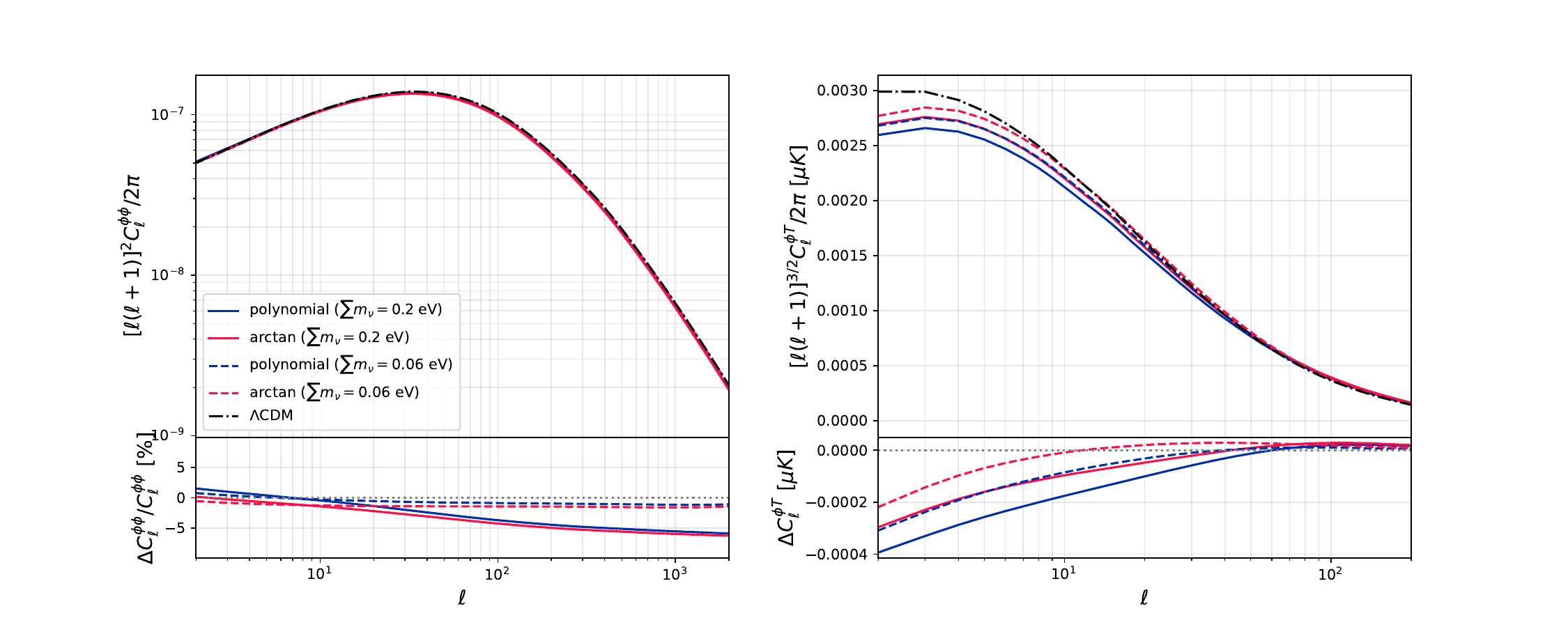} 
\caption{{Power spectrum of the lensing potential $C_\ell^{\phi\phi}$ (left) and lensing potential cross temperature $C_\ell^{\phi T}$ (right) for the two Kinessence models and $\Lambda$CDM. The bottom panels present the relative difference in percentage of $C_\ell^{\phi\phi}$ for the two models with respect to $\Lambda$CDM (left) and the difference between the $C_\ell^{\phi T}$ for the models and the corresponding $\Lambda$CDM spectrum (right). }}
\label{fig:lensing}
\end{figure}

We also derive the lensing power spectrum $C_\ell^{\phi\phi}$ together with the lensing cross temperature spectrum (Fig. \ref{fig:lensing}) left and right respectively. The plots show a relative difference of less than 7\% for all $\ell$ considered (bottom panels) {for the models with two massive neutrinos, while the models with one massive neutrino presents subpercent variations}. These differences are currently undetectable by the CMB. The  $\ell \lesssim 10$ is cosmic-variance dominated, making the modifications undetectable, while for $100 \lesssim \ell \lesssim 2000$, the \textit{Planck} reconstruction noise is in the regime of less than 10\% uncertainty, making the models marginally detectable \cite{Planck:2018lbu}. CMB-S4/Simons-Observatory-class lensing reconstruction, targeting sub-percent precision in that $\ell$ range, would be however able to distinguish between these Kinessence models {with larger neutrino masses} and $\Lambda$CDM \cite{CMB-S4:2016ple,LiteBIRD:2023iiy}.

\begin{figure}[!h]
\centering
\includegraphics[width=0.55\linewidth]{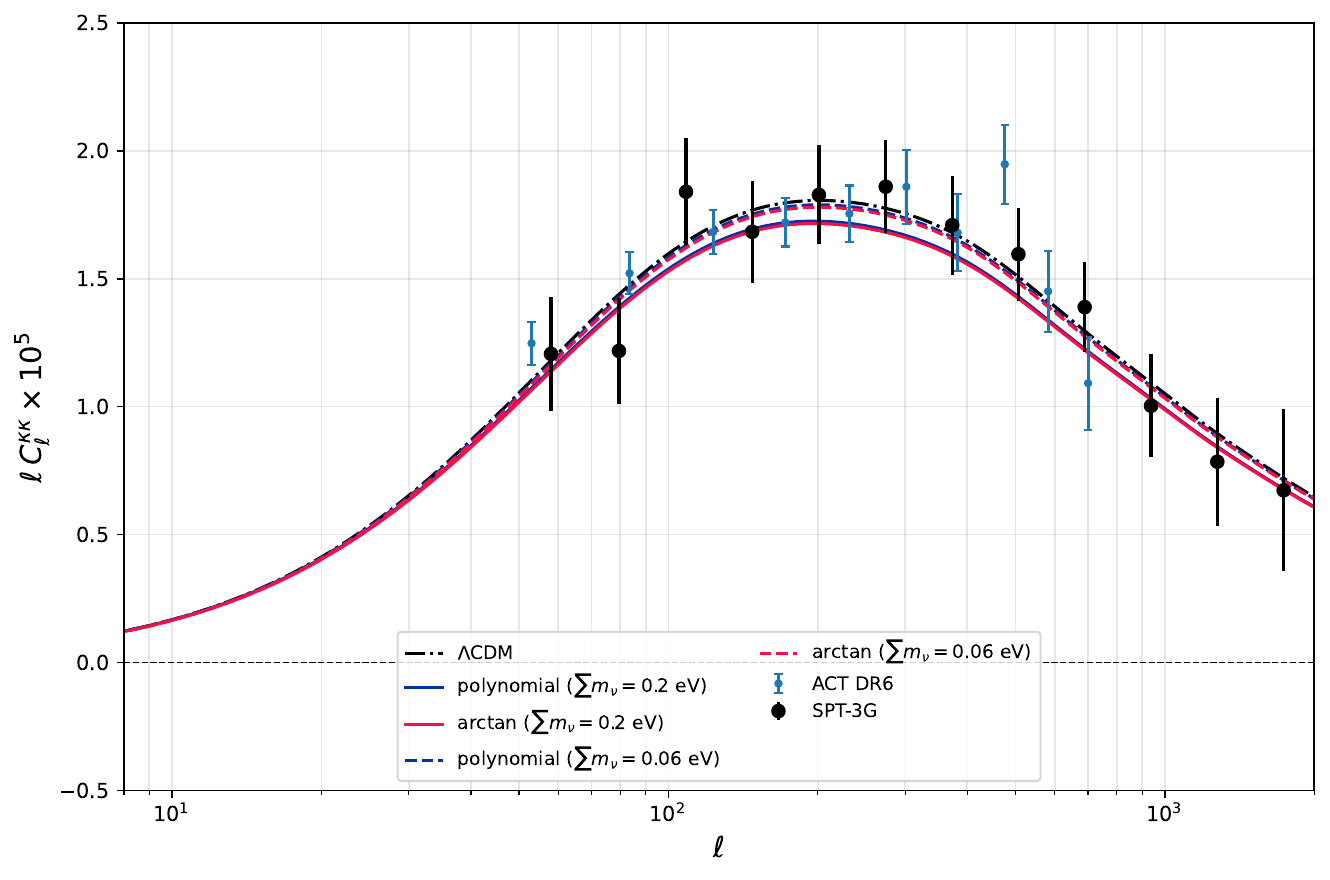} 
\caption{Lensing power spectrum from $\Lambda$CDM, the two Kinessence models and the ACT DR6 and SPT-3G measurements from \cite{SPT-3G:2026kbu}}
\label{fig:lensing_ms}
\end{figure}

The Kinessence models discussed satisfy the lensing constraints from 
\cite{SPT-3G:2026kbu}. In Fig. \ref{fig:lensing_ms} we present the lensing power spectra from $\Lambda$CDM and  the two Kinessence models discussed in this work, together with the experimental measurements, showing that the models presented in this work satisfy the latest lensing constraints. {The kinessence models with one massive neutrino have a lensing spectrum similar to the $\Lambda$CDM one.}

\subsection{Phantom crossing and the Hubble rate}

\begin{figure}[!h]
\centering
\includegraphics[width=0.49\linewidth]{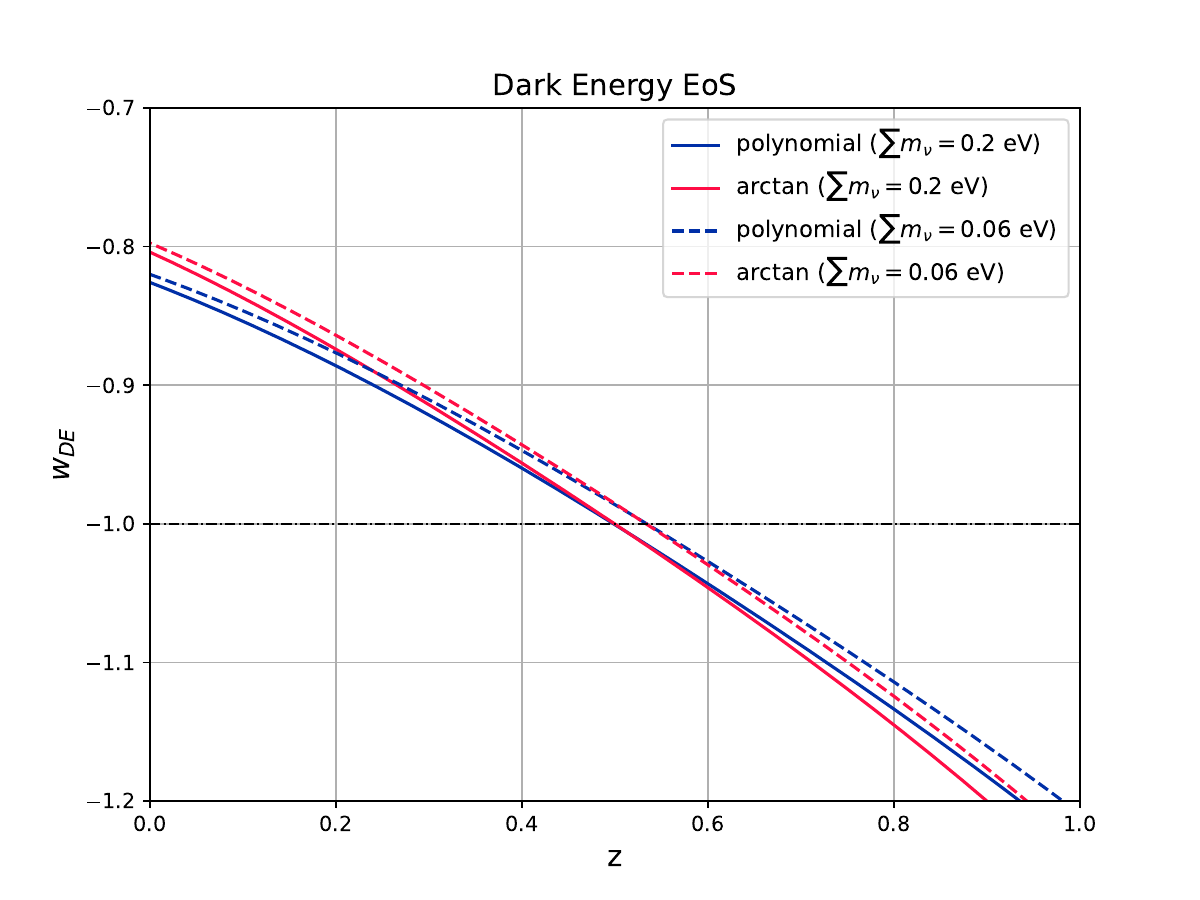} 
\includegraphics[width=0.49\linewidth]{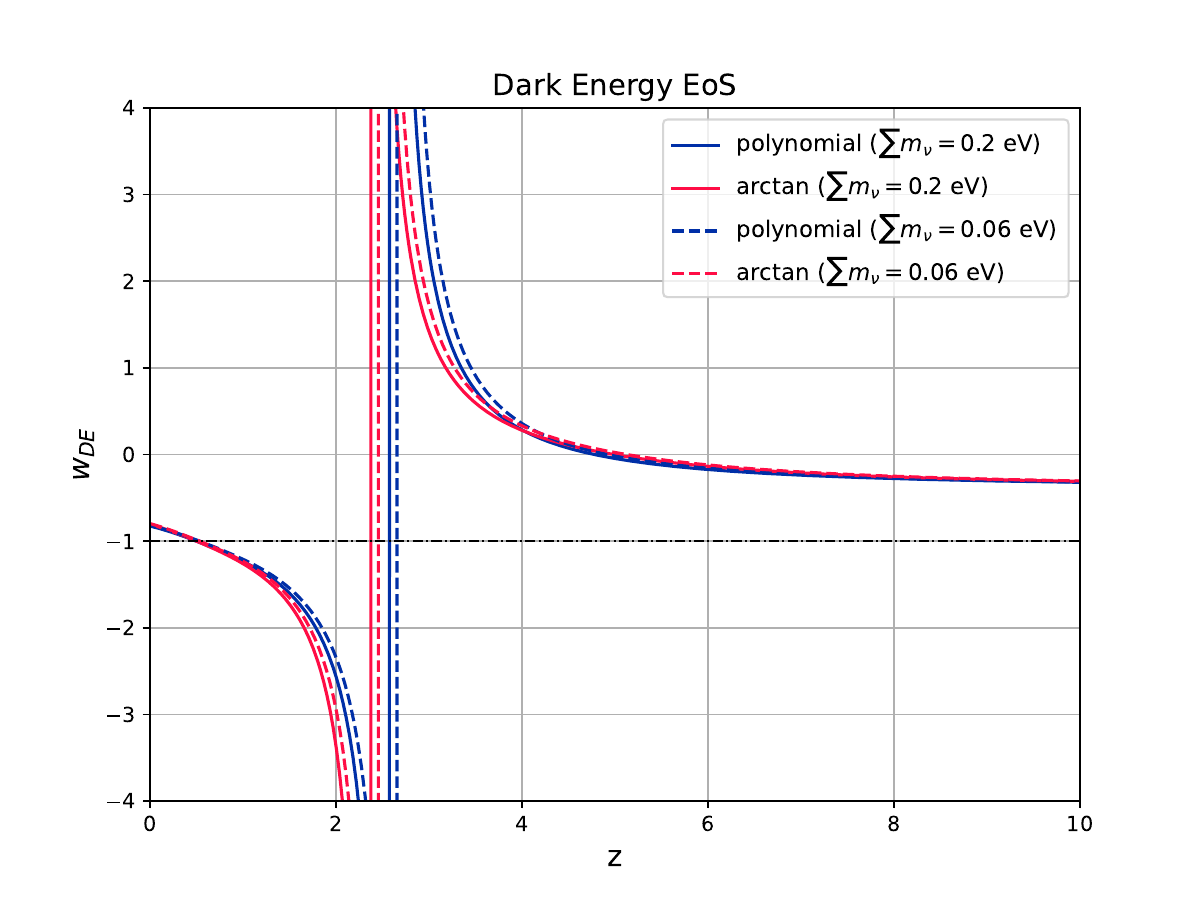} 
\caption{Dark energy equation of state $w_{DE}$ for the  Kinessence polynomial and arctan models. Both models exhibit the phantom crossing behaviour.}
\label{fig:DE_EOS}
\end{figure}

The phantom crossing is represented in Figures \ref{fig:DE_EOS}, where we show the dark energy equation of state for the two models. For both models, the phantom crossing happens at redshift $z \approx 0.5$. {In order to compare directly with the standard parameterisations of dynamical dark energy constrained by late-time data,  we investigate numerically the behaviour of the equation of state around the phantom crossing redshift $z_c$, and we compare it with the same expansion of the parametrisation  $\omega = \omega_0 + \omega_a(1-a)$ \cite{Brax:2025ahm}, which reads}
\begin{equation}
    \omega(z)=-1+\frac{\omega_a}{(1+z_c)^2}(z-z_c) \, ,
\end{equation}
corresponding to the phantom crossing at
\begin{equation}
    \omega_0=-1-\frac{z_c}{1+z_c}\omega_a \,.
\end{equation}
For the models in Fig. \ref{fig:DE_EOS}, 
\begin{equation}
    \omega(z)=-1+C(z-z_c) \, ,
\end{equation}
and hence $\omega_a=C(1+z_c)^2$. In the models presented here $z_{c, \rm{poly}}=0.4881$, $C_{\rm{poly}}=-0.4202$ $z_{c, \rm{arctan}}=0.5101$, $C_{\rm{arctan}}= -0.4457$, yielding $\omega_a=-0.93, -1.02$ and $\omega_0=-0.69, -0.66$ respectively. These values are compatible with the DESI measurements, showing that the chosen value of $\beta$ is reasonable.

\begin{figure}[!h]
\centering
\includegraphics[width=0.49\linewidth]{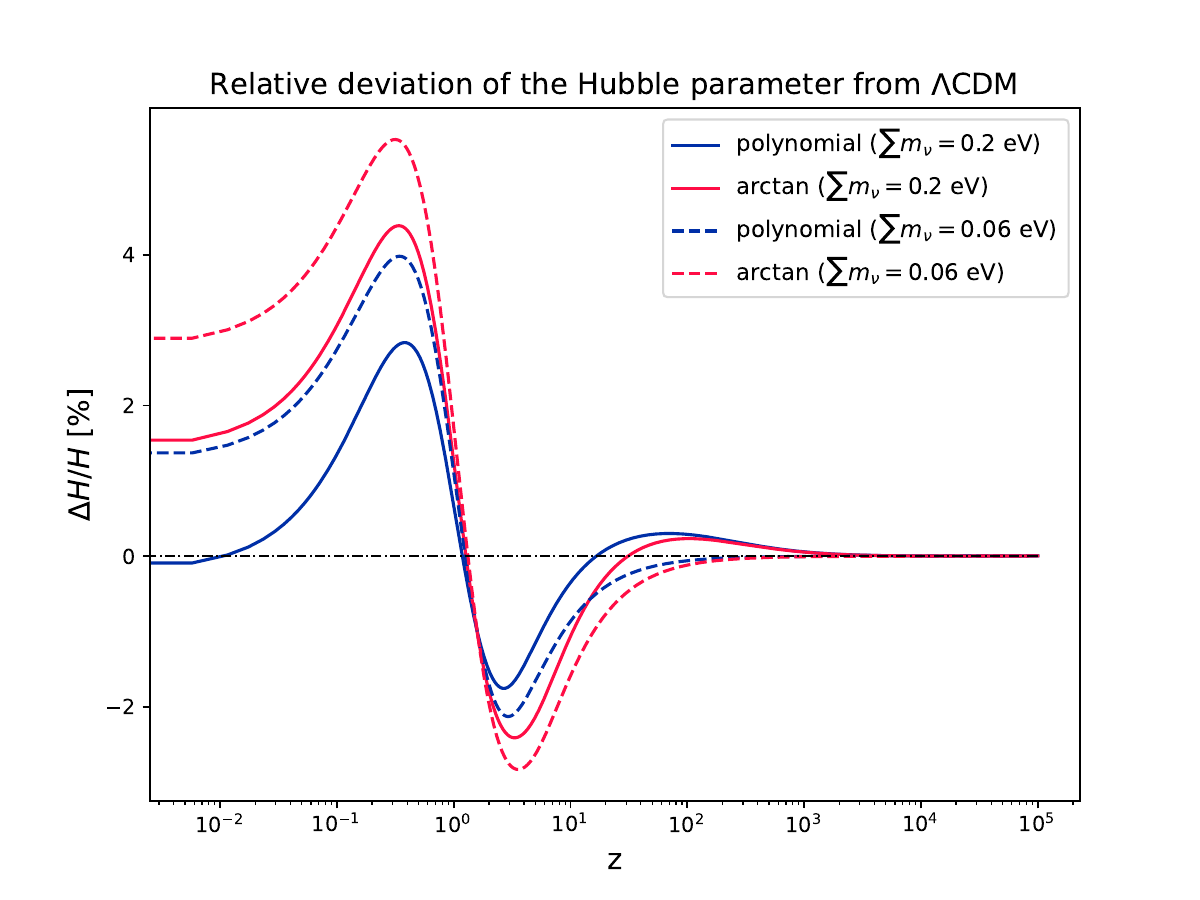} 
\includegraphics[width=0.49\linewidth]{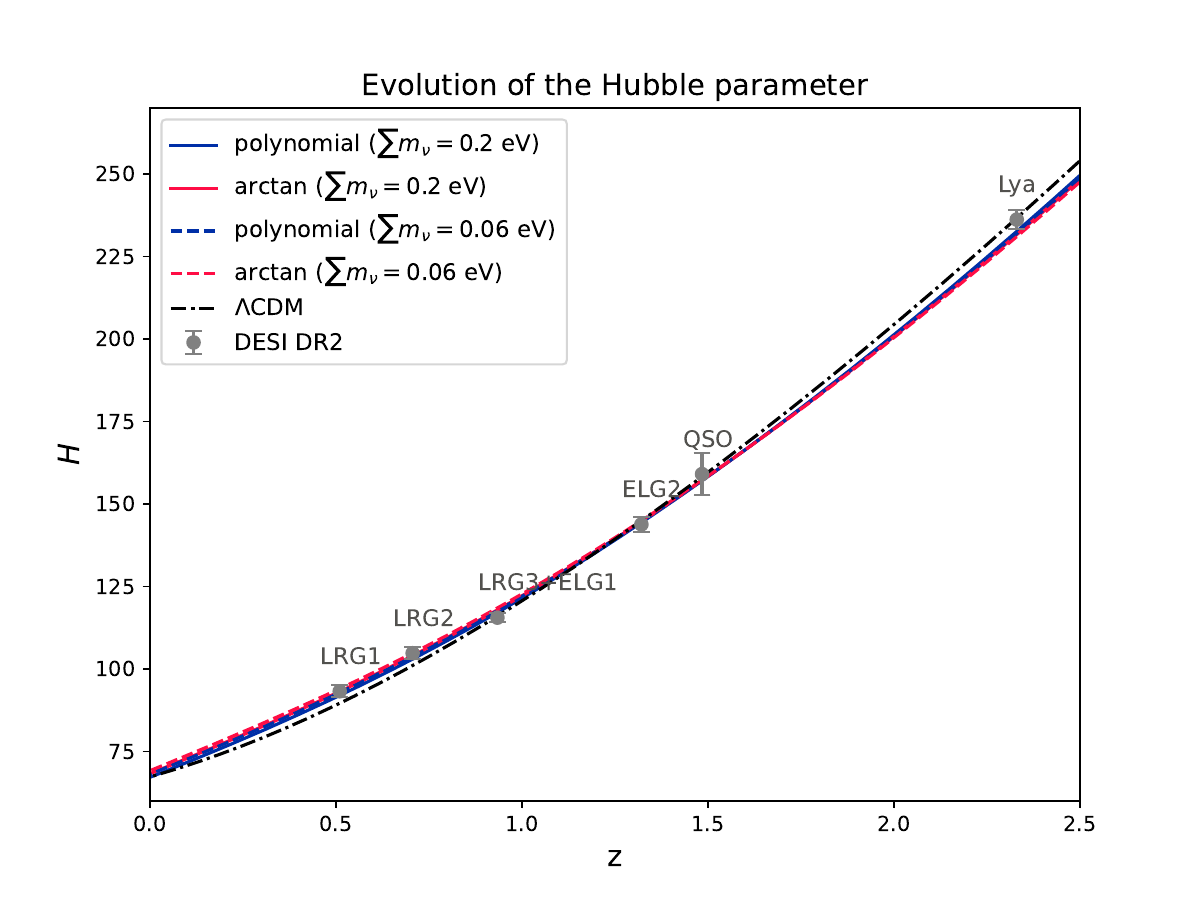} 
\caption{Hubble deviation with respect to $\Lambda$CDM for the two Kinessence models (left) and the Hubble parameter for the two models and $\Lambda$CDM, with the points measured by DESI (right).}
\label{fig:H}
\end{figure}

The Kinessence model also leads to a modification of the Hubble parameter, which varies with redshift. In Figure \ref{fig:H}, we represent the variation of the Hubble parameter with respect to $\Lambda$CDM (left), showing a less than 3\% and 5\% variation for the polynomial and arctan model respectively {for the models with two massive neutrinos. The models with a less massive neutrino enhance the Hubble parameter further, to values of 4\% and 6\%}. The value of $H_0$  decreases by {0.1\%} for the polynomial model and is enhanced by around {1.5\%} for the arctan one {in the 2-massive neutrino scenario and are enhanced by around 1.5\% and 3\% respectively in the 1-massive neutrino scenario}. The  plot on the right shows the Hubble parameter together with the seven tracers (galaxies, quasars and Ly$\alpha$ forest) and their corresponding error bars considered by DESI \cite{DESI:2025zgx}. These show a good agreement between the proposed models and the measurements {in all the cases considered.}

\subsection{Growth of structure}

\begin{figure}[!h]
\centering 
\includegraphics[width=0.49\linewidth]{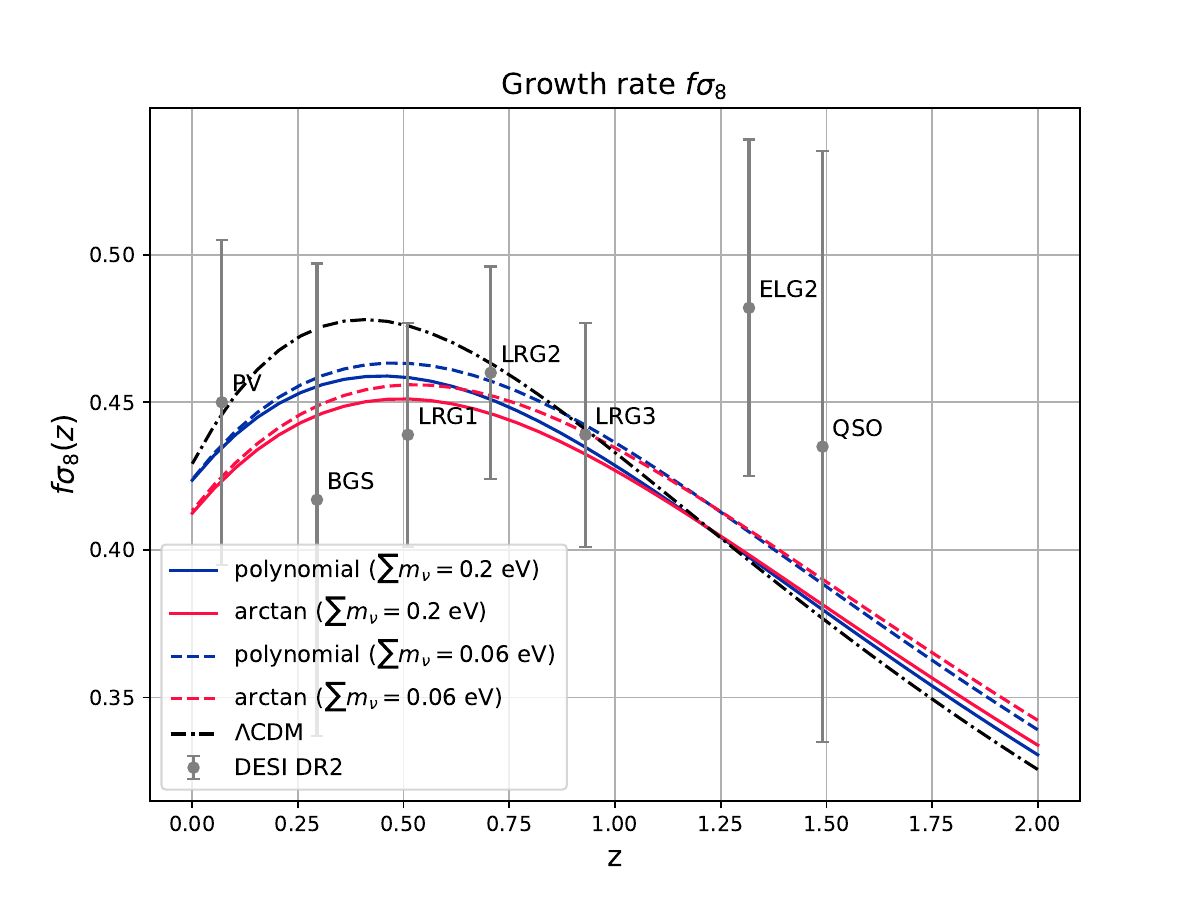} 

\caption{The  growth rate $f\sigma_8$ for the  two best-fit Kinessence models (polynomial and arctan) and $\Lambda$CDM with DESI DR2 measurements.}
\label{fig:fs8}
\end{figure}

We also investigate the modifications of the structure growth rate $f\sigma_8$ with respect to $\Lambda$CDM (Fig. \ref{fig:fs8}). $\sigma_8$ is defined as the amplitude of density fluctuations in the universe on a scale of 8 Mpc/h, and is output by $\mathcal{H}$-EFTCAMB. The K-mouflage models lead to a decrease in the amplitude of $f\sigma_8$ for $z<1.25$, within the error bars of the DESI measurements. 
{This follows from $\mu_K<1$ compared to $\Lambda$-CDM at low redshift, see fig. \ref{fig:H} and \ref{fig:muk}.}
{The current constraints on the $\mu$ and $\Sigma= A^2$ parameters \cite{Ferte:2017bpf} from DESI \cite{Ishak:2024jhs} are deviations from unity at small redshift $\mu-1 = 0.05\pm 0.22$ and $\Sigma-1=0.008\pm 0.045$ and Euclid forecasts \cite{Euclid:2025tpw}. They are satisfied by our models.}

The matter fraction of the Universe is also modified by  K-mouflage, as can be seen in Fig. \ref{fig:omegam}. In the left subfigure, we have represented the relative deviation of $\Omega_m$ for the K-mouflage models with respect to $\Lambda$CDM. At $z=0$, there is an enhancement of around 2.5\% in the polynomial case and 2.5\% decrease for the arctan model. For both models, this first decreases with redshift and then increases again at around $z \approx 0.5$. In the right subfigure, we represent $\Omega_m$ for the two K-mouflage models, $\Lambda$CDM and the seven points corresponding the  DESI DR2 measurements. {We note that $\Omega_m$ differs from the Kinessence value at high $z$ due to the {different  number of massive species (one in $\Lambda$CDM, two in Kinessence).} 

\begin{figure}[!h]
\centering
\includegraphics[width=0.49\linewidth]{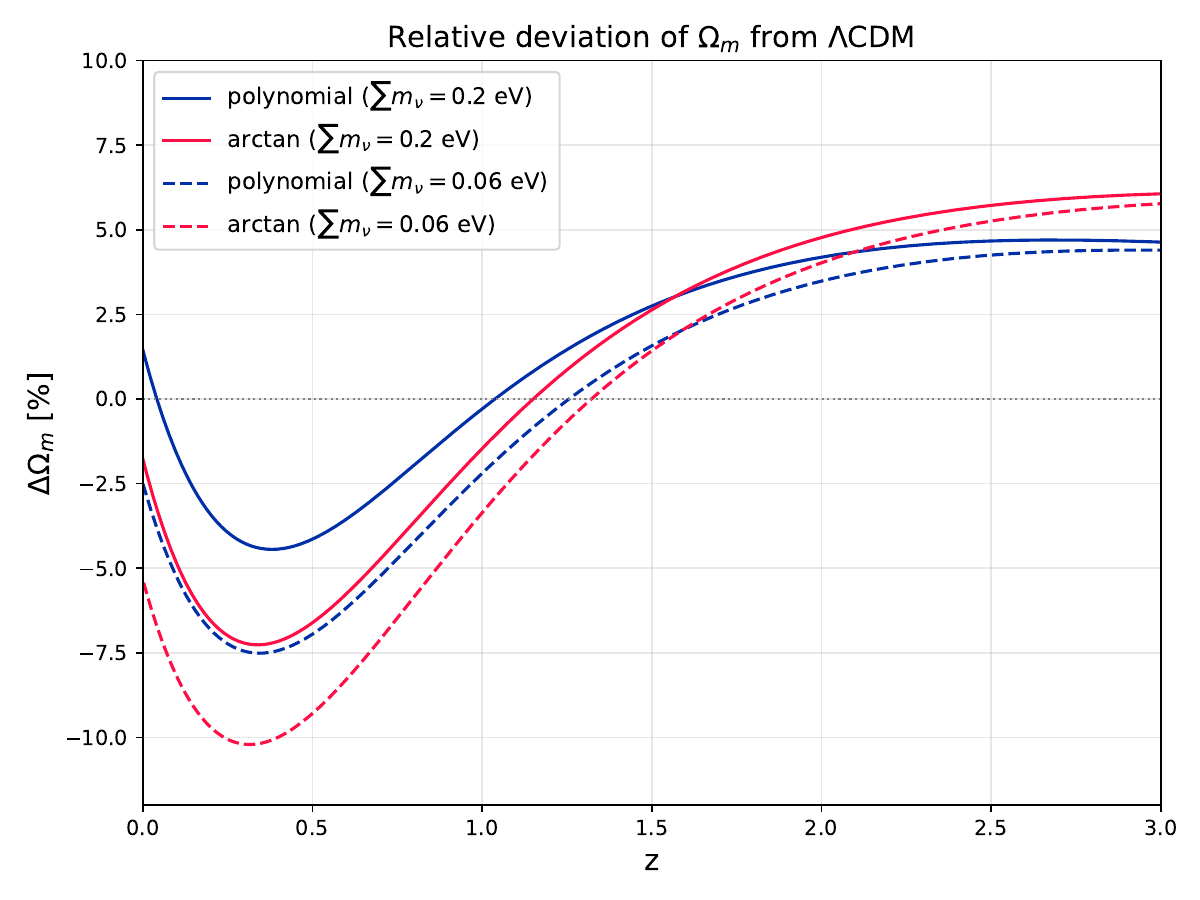} 
\includegraphics[width=0.49\linewidth]{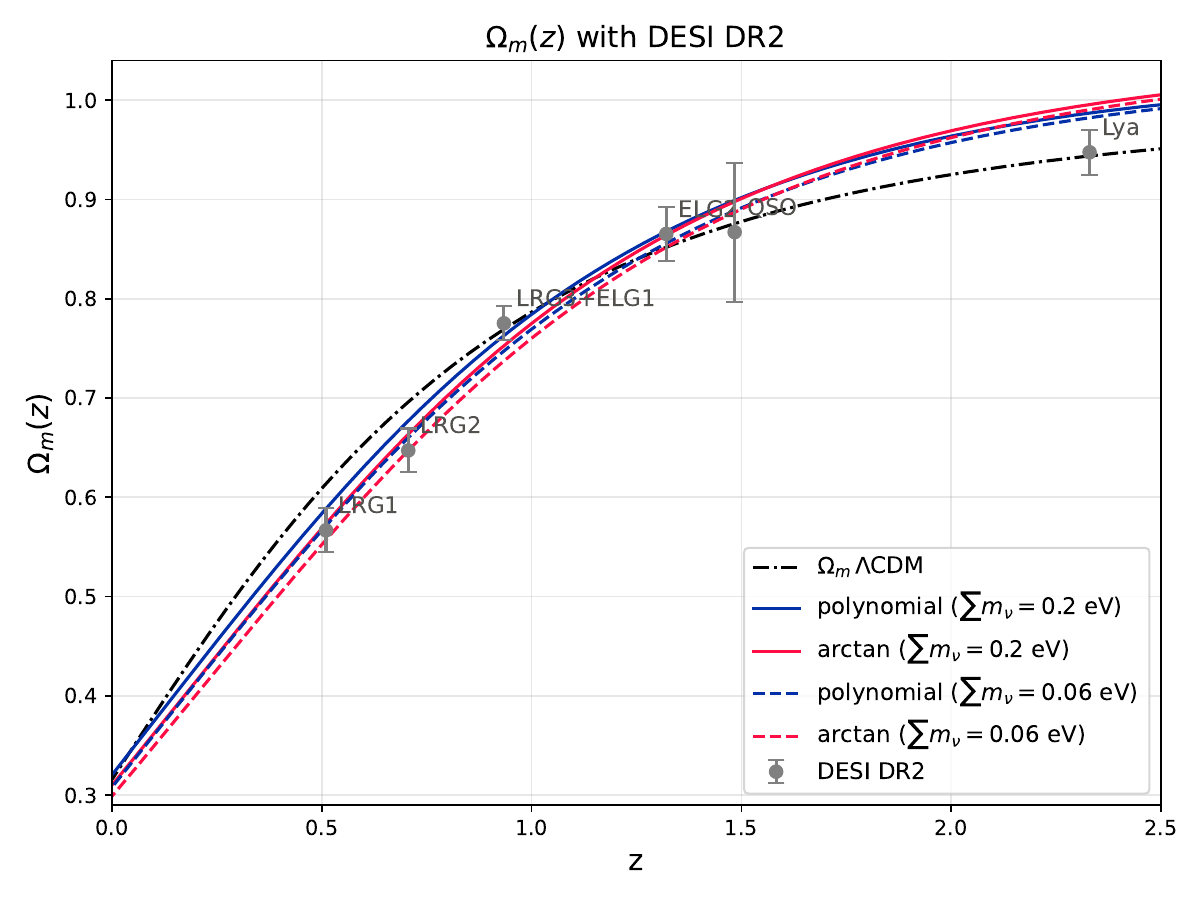} 
\caption{The relative deviation of $\Omega_m(z)$  with respect to $\Lambda$CDM (in percentage) for the two Kinessence models (left) and $\Omega_m(z)$ the two models and $\Lambda$CDM, with the points measured by DESI (right).}
\label{fig:omegam}
\end{figure}

{In Fig. \ref{fig:muk} We represent the function $\mu_{\rm K}$ from Eq. \eqref{eq:mu_K} and $A^2(\phi)-1$ for the models considered ($\sum m_{\nu}=0.2$ and $\sum m_{\nu}=0.06$), in the case of the polynomial and arctan couplings, together with the coupled dilaton discussed in appendix \ref{app:dil}. For the coupled dilaton models we plotted the models having $\beta=0.05$ to satisfy the BBN constraint. We observe that the constraints from \eqref{eq:BBNc} are satisfied by all the models considered.}

The redshift evolution of the densities of the different components $\Omega$ of the universe in  the two scenarios are represented in Fig. \ref{fig:omega}. 
{We find that in both cases the dark energy density switches sign at redshift $z\gtrsim 2$. This is an unusual feature of the models presented here which has an impact on the dynamics of the Hubble rate and affects its value at low redshift. }
\begin{figure}[!h]
\centering
\includegraphics[width=0.98\linewidth]{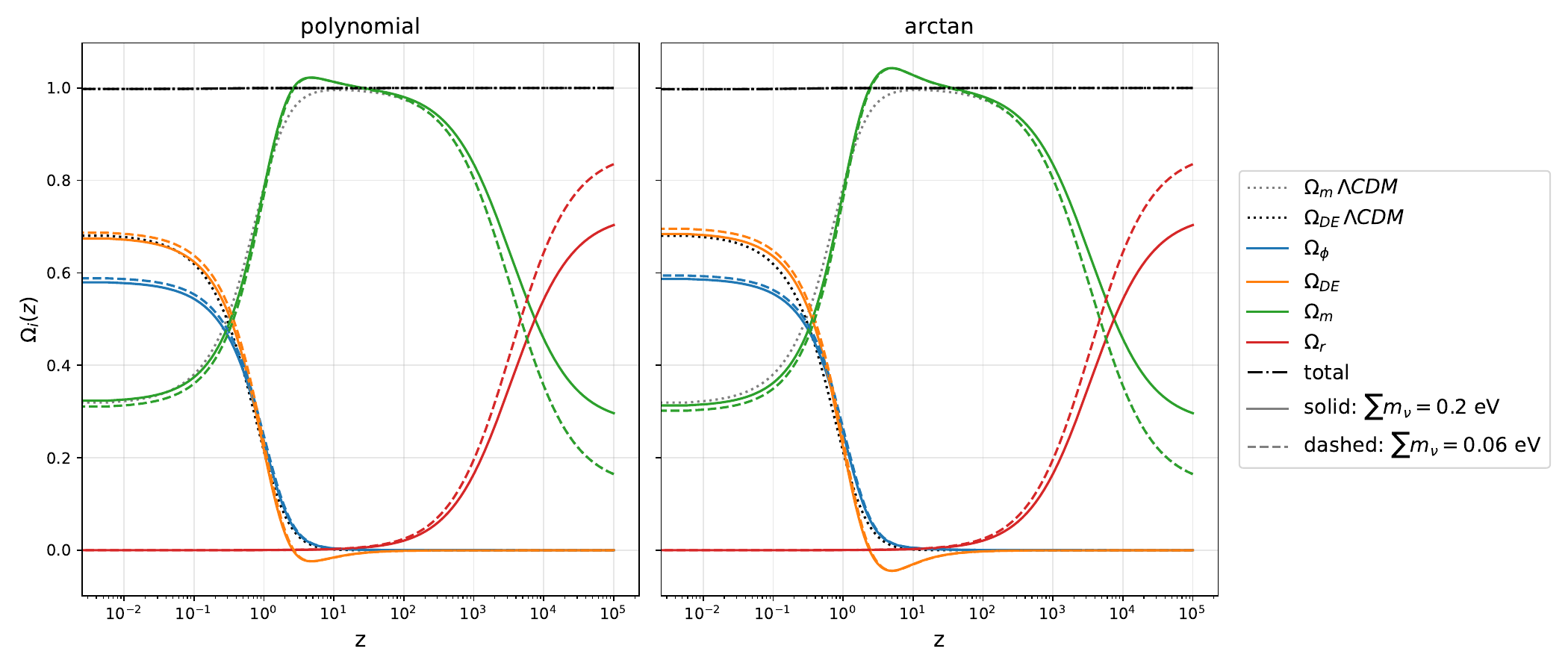} 
\caption{{The different components of} $\Omega(z)$ for $\Lambda$CDM (matter and dark energy) and for the two K-mouflage model with polynomial coupling (left) and arctan coupling (right):  $\Omega_{DE}$, $\Omega_\phi$, $\Omega_m$, $\Omega_r$. {The difference in $\Omega_m$ at early times in the case of the solid lines is due to the convention being used, massive neutrinos being counted as matter while still relativistic (there is one massive species in $\Lambda$CDM, two in Kinessence). }}
\label{fig:omega}
\end{figure}

We have investigated the impact of varying the parameters of the models on the quantities considered, around the best parameters that we obtained. This is detailed in the appendix.
The values of $\sigma_8$ for the models considered are presented in Tables \ref{tab:poly_scan} and \ref{tab:arctan_scan}, while the corresponding plots for the dark energy equation of state, Hubble deviation, $f\sigma_8$, $\Omega$ and the angular power spectra are presented in appendix \ref{app:plots}.

{The peak positions of the CMB power spectrum {scale as  $\ell \approx n \pi/\theta_*$, where $\theta_* = r_s(z_*)/D_A(z_*)$. The sound horizon \(r_s\) is determined by pre-recombination physics, which is essentially unchanged by the scalar sector, as its contribution is negligible at \(z_*\). There is, however, a change in  $D_A(z_*) = \int_0^{z_*} dz/H(z)$. $H(z)$ is systematically more enhanced than depleted relative to $\Lambda$CDM over the range $z\sim 0.01-10$. A larger $H(z)$ reduces the integral $\int dz/H(z)$, and therefore reduces $D_A(z_*)$, which in turn increases $\theta_* = r_s/D_A$, {as the other cosmological parameters are kept fixed.} A larger $\theta_*$ means that the peaks are shifted towards smaller $\ell$. }

\subsection{Varying the model parameters}

\paragraph{${\bf{\sum m_\nu}}.$}
We have considered massive neutrinos, with a sum of masses of 0.1, 0.2 and 0.3 eV. This change has a very small influence on the dark energy equation of state, as well as $\Omega$'s.

The Hubble deviation with respect to $\Lambda$CDM tends to be shifted up as the neutrino mass is increased, while the growth rate $f\sigma_8$ is decreased \cite{Lesgourgues:2006nd,Pan:2015bgi,Bayer:2021kwg}.  
The increase of the neutrino mass leads to a reduction in the amplitude of the matter power spectrum, and hence to a reduction in $\sigma_8$. There is also a slight shift towards smaller multipoles with respect to $\Lambda$CDM of the acoustic peaks in the angular power spectrum \cite{Lesgourgues:2006nd,Lesgourgues:2012uu}. {Increasing neutrino masses increases  $\Omega_\nu h^2$, adding to the total non-relativistic matter density at late times. The CMB gives an extremely precise measurement of the angular size of the sound horizon at recombination. Adding more massive neutrinos increases $\Omega_\nu h^2$, leading to a deceleration of the expansion during that epoch, which then leads to an increase in  $H(z)$, which stretches distances back out to preserve the correct angular scale. Massive neutrinos free-stream out of small-scale overdensities, damping matter clustering on those scales.  This directly suppresses the growth rate of structure $f$ and the amplitude $\sigma_8$, hence their product. 
Raising the sum of the neutrino masses alters the angular diameter distance to the last-scattering surface and the matter-radiation equality epoch (via the neutrino-to-matter transition), slightly changing the sound horizon angle, shifting  the peak positions to lower $\ell$.}  The plots for the polynomial and arctan model are given in Figs. \ref{fig:varnu_poly} and \ref{fig:varnu_arctan} respectively.

\paragraph{$\beta$}

In the case of the equation of state, increasing $\beta$ increases the slope of the curve, and also increases the value of the equation of state at redshift 0.
The behaviour of the variation of the Hubble parameter with respect to $\Lambda$CDM changes significantly when $\beta$ is modified.
Increasing $\beta$ decreases the magnitude of $f\sigma_8$,  the matter power spectrum and $\sigma_8$. {Indeed as can be seen in Figs. \ref{fig:varbeta_arctan} and \ref{fig:varbeta_poly} increasing $\beta$ increases the Hubble rate at low $z$ where structures form. This implies that the Hubble friction in (\ref{eq:delta_growth_conformal_qs}) overcomes the effects of the fifth force and reduces the growth of structure.} The acoustic peaks also move to the left.

\paragraph{$\lambda$}
The increase of $\lambda$ leads to an increase of the intercept of the $w$ curve and to a higher amplitude of the Hubble parameter deviation. It also decreases the growth rate and the amplitude of the matter power spectrum, while the acoustic peaks are shifted to the left. 

\paragraph{$M_{\rm num}$}
The increase of $M$ leads to an increase of the value of $w(z=0)$ and a slight lowering of the magnitude of the slope of the $w(z)$. The Hubble parameter is enhanced, while the growth rate $f\sigma_8$ is reduced, as is the amplitude of the matter power spectrum and $\sigma_8$. The acoustic peaks are shifted to the left.

\paragraph{$K_0$}
The quantities considered vary weakly with $K_0$. Increasing $K_0$ lowers the redshift of the phantom crossing and reduces the Hubble parameters. The growth rate is increased, and so is $\sigma_8$ and the matter power spectrum. The acoustic peaks are shifted towards higher multipoles.

\paragraph{$m$}
The value of the $m$ parameter has a very mild influence on the quantities considered.  Increasing $m$ reduces the  redshift of the phantom crossing and increases the Hubble parameter. $\sigma_8$ is reduced, as is the growth rate and the matter power spectrum.

\paragraph{$K_*$}
In this case, the influence of $K_*$ is again mild on the quantities considered. The increase of $K_*$ decreases the redshift of the phantom crossing and the Hubble parameter. $\sigma_8$, the growth rate and the matter power spectrum are increased and the acoustic peaks are shifted towards higher multipoles.

\paragraph{$\chi_*$}
In this case, increasing $\chi_*$ increases the redshift of the phantom crossing and the Hubble parameter. $\sigma_8$, the growth rate and the matter power spectrum are reduced and the acoustic peaks are shifted towards higher multipoles.

\begin{table}[htbp]
\centering
\begin{tabular}{cccccccc}
\toprule
$K_0$ & $m$ & $\beta$ & $M_{\rm num}$ & $\lambda$ & $\Sigma m_\nu$ & $\sigma_8$ & $\Delta \theta_*/\theta_*$\\
\hline
$\Lambda$CDM & & & & & & 0.8118 & 0\\
\hline
1   & 3 & 0.09 & 0.955 & 2.4 & 0.1 & 0.8187 & $-0.002945$ \\
    &   &      &      &     & 0.2 & 0.7968 & $-8.6099 \times 10^{-5}$ \\
    &   &      &      &     & 0.3 & 0.7559 & $0.005404$ \\
\hline
1   & 3 & 0.01 & 0.955 & 2.4 & 0.2 & 0.8112 & $-0.01145$ \\
    &   & 0.05 &      &     &     &  0.7989 & $-0.00354$\\
    &   & 0.09 &      &     &     &  0.7968 & $-8.60993 \times 10^{-5}$ \\
\hline
1   & 3 & 0.09 & 0.955 & 2.0 & 0.2 & 0.8010 & $-0.002328$ \\
    &   &      &      & 2.4 &     &  0.7968 & $-8.6099 \times 10^{-5}$ \\
    &   &      &      & 2.8 &     &  0.7939 & 0.001411\\
\hline
1   & 3 & 0.09 & 0.905 & 2.4 & 0.2 & 0.8233 & $-0.01360$ \\
    &   &      & 0.955 &     &     & 0.7968 & $-8.6099 \times 10^{-5}$\\
    &   &      & 1.005  &     &     & 0.7705 & 0.01392\\
\hline
0.1 & 3 & 0.09 & 0.955 & 2.4 & 0.2 & 0.7921 & 0.002337 \\
1   &   &      &      &     &     &  0.7968 & -8.6099e-05\\
10  &   &      &      &     &     &  0.8006 & $-0.002331$\\
\hline
1   & 3 & 0.09 & 0.955 & 2.4 & 0.2 & 0.7968 & $-8.6099\times 10^{-5}$ \\
    & 4 &      &      &     &      & 0.7966 & 0.0001477 \\
    & 5 &      &      &     &      &  0.7964 & 0.0002872\\
\bottomrule
\end{tabular}
\caption{Kinessence (polynomial model) $\sigma_8$ {and $\Delta \theta_*/\theta_*$} values.}
\label{tab:poly_scan}
\end{table}

\begin{table}[htbp]
\centering
\begin{tabular}{cccccccc}
\toprule
$K_*$ & $\chi_*$ & $\beta$ & $M_{\rm num}$ & $\lambda$ & $\Sigma m_\nu$ & $\sigma_8$ & $\Delta \theta_*/\theta_*$\\
\hline
 $\Lambda$CDM   &     &      &      &     &     & 0.81182 \\
\hline
1000 & 100 & 0.09 & 0.945 & 2.3 & 0.1 & 0.8183 & $-0.002845$ \\
     &     &      &      &     & 0.2 & 0.7964 & $2.3228 \times 10^{-6}$\\
     &     &      &      &     & 0.3 & 0.7556 & 0.005468\\
\hline
1000 & 100 & 0.01 & 0.945 & 2.3 & 0.2 & 0.8142 & $-0.012951$ \\
     &     & 0.05 &      &     &     & 0.7997 & $-0.003997$ \\
     &     & 0.09 &      &     &     & 0.7965 & $2.3228 \times 10^{-6}$ \\
\hline
1000 & 100 & 0.09 & 0.945 & 1.9 & 0.2 & 0.8032 & $-0.003593$ \\
     &     &      &      & 2.3 &      & 0.7964 & $2.3228 \times 10^{-6}$ \\
     &     &      &      & 2.7 &      & 0.7916 & 0.002527 \\
\hline
1000 & 100 & 0.09 & 0.895 & 2.3 & 0.2 & 0.8244 & $-0.01404$ \\
     &     &      & 0.945 &     &     & 0.7964 & $2.3228 \times 10^{-6}$ \\
     &     &      & 0.995 &     &     & 0.7689 & 0.01449 \\
\hline
100   & 100 & 0.09 & 0.95 & 2.3 & 0.2 & 0.7906 & 0.002746 \\
1000  &     &      &      &     &     & 0.7964 & $2.3228 \times 10^{-6}$ \\
10000 &     &      &      &     &     & 0.8011 & $-0.002307$ \\
\hline
1000 & 50  & 0.09 & 0.95 & 2.3 & 0.2 & 0.7993 & $-0.001407$ \\
     & 100 &      &      &     &     & 0.7964  & $2.3228 \times 10^{-6}$ \\
     & 200 &      &      &     &     & 0.7932 & 0.001497 \\
\bottomrule
\end{tabular}
\caption{Kinessence (arctan model): $\sigma_8$ {and $\Delta \theta_*/\theta_*$} values.}
\label{tab:arctan_scan}
\end{table}
A full Markov-Chain-Monte-Carlo analysis of the Kinessence model is required to derive the best fitting parameters. This will be the subject of a subsequent publication. {It will be particularly important to determine whether higher neutrino masses and lower growth are viable when comparing to data. }

\begin{figure}[!h]
\centering 
\includegraphics[width=0.7\linewidth]
{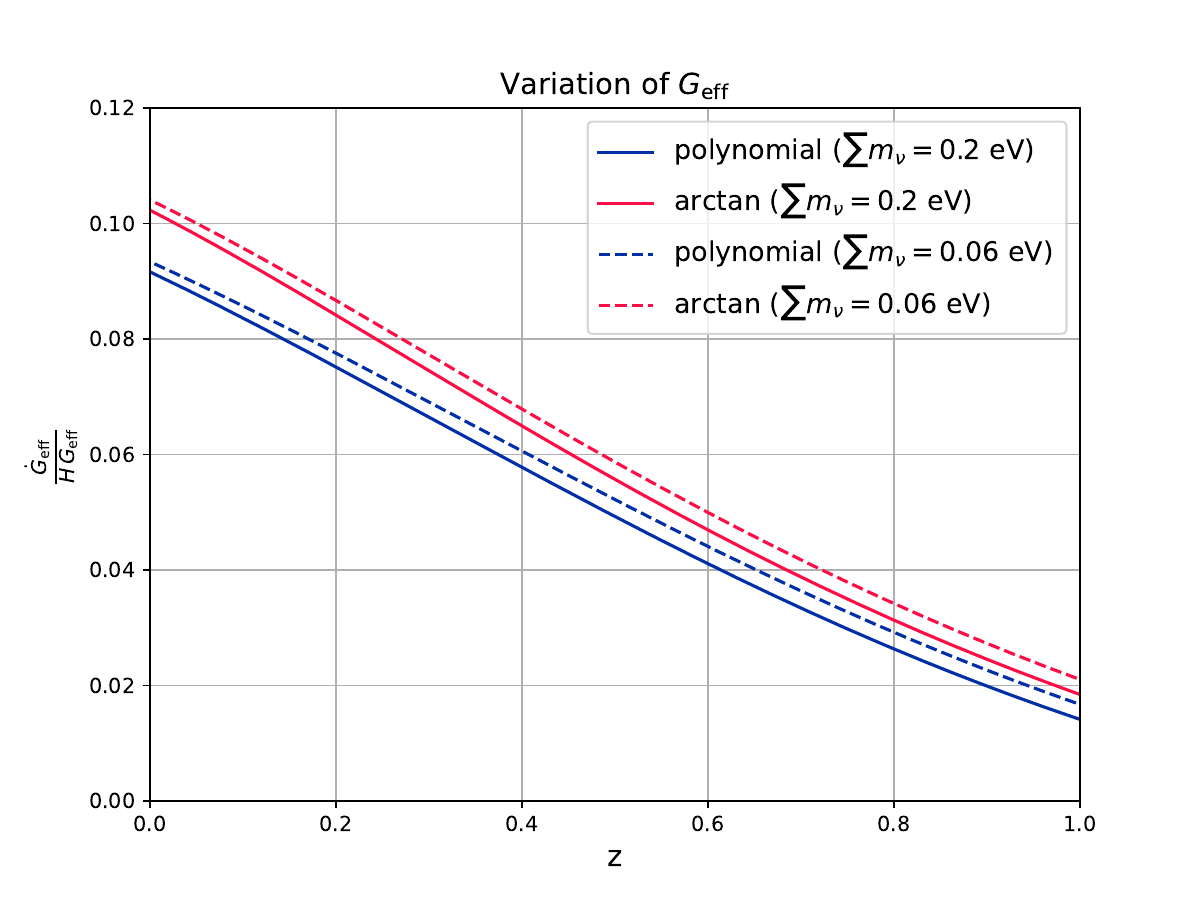} 
\caption{The variation of $G_{\rm eff}$ for the models considered.}
\label{fig:varg}
\end{figure}

\subsection{The coupled runaway dilaton}
In appendix \ref{app:dil}, we consider a third and simpler model where the K-essence part is removed and canonical scalar fields are taken into account. This is a coupled dilaton model which is found to be compatible with cosmological data in the same way as Kinessence. Hence in   the context of future large scale surveys we have found three models, i.e. the coupled dilaton, cubic and $\arctan$ Kinessence models, which pass present cosmological tests and are only distinguishable at short scales in the Solar System. Only one, i.e. the $\arctan$ Kinessence model, can describe phenomena from the Solar System to cosmological scales. In this paper, we have not considered the non-linear regime of cosmology. This could also lead to differences that could become observationally testable in the future. This would require N-body simulations left for subsequent work.

\begin{figure}[!h]
\centering 
\includegraphics[width=0.9\linewidth]
{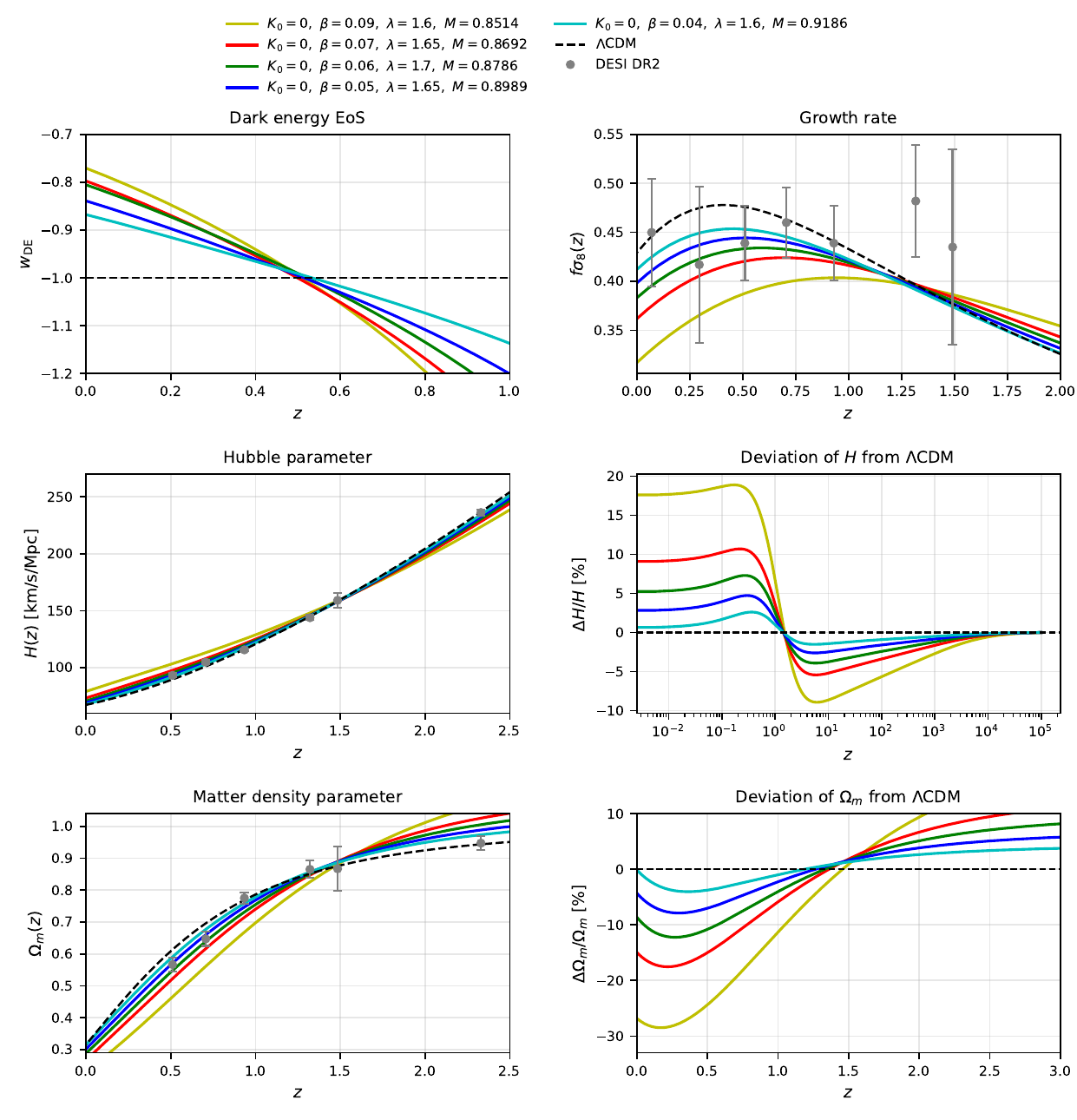} 
\caption{The equation of state, the growth rate, the Hubble rate, the variation of the Hubble rate with respect to $\Lambda$CDM, $f\sigma_8$, $\Omega_m$ and the relative deviation of $\Omega_m$ with respect to $\Lambda$CDM for a model with a canonical kinetic term ($K_0=0$) {and two massive neutrinos of total mass 0.2 eV}.}
\label{fig:K0a}
\end{figure}

\begin{figure}[!h]
\centering 
\includegraphics[width=0.9\linewidth]
{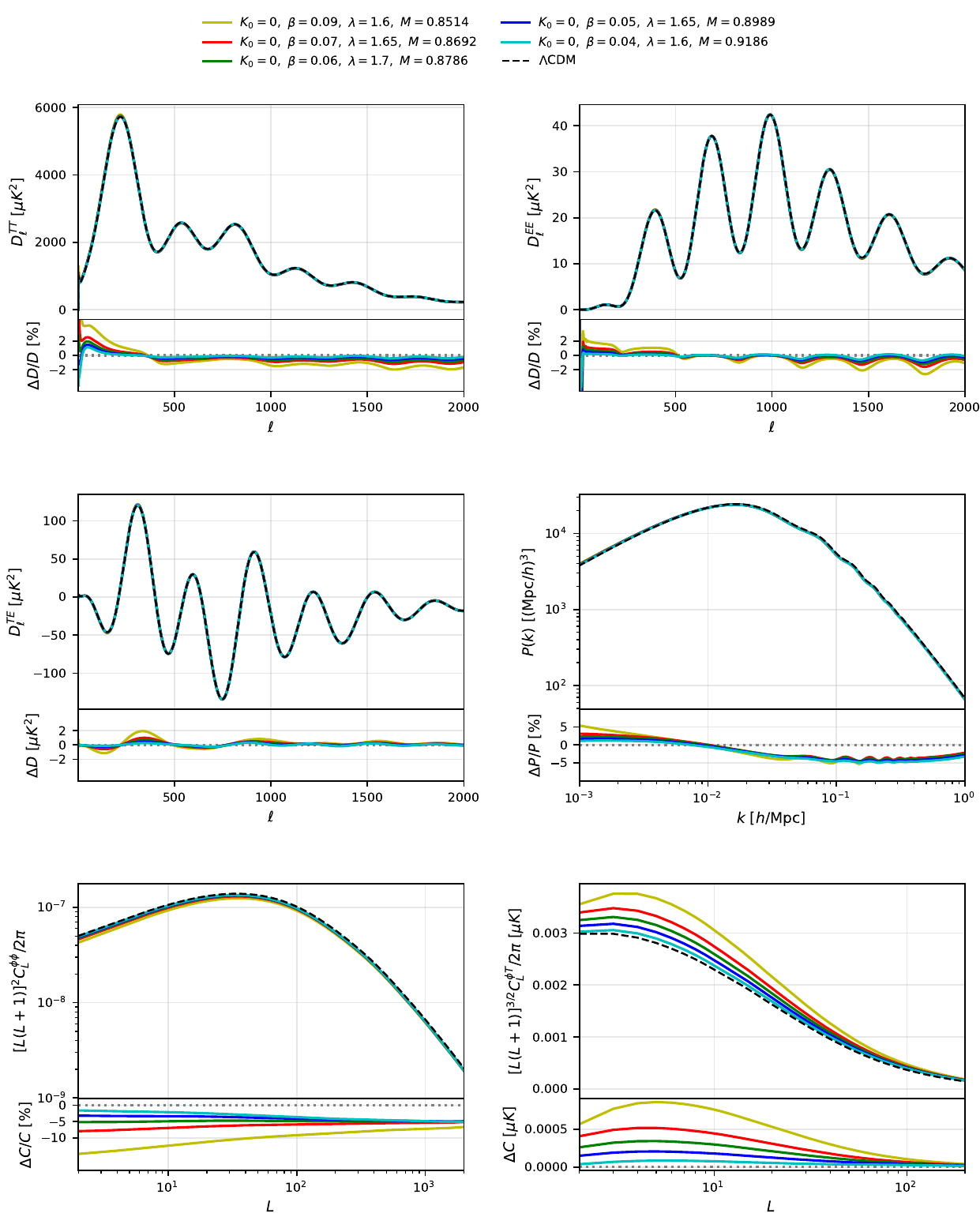} 
\caption{The temperature, E-polarisation TE cross-correlation angular power spectrum, the matter power spectrum and the CMB lensing potentials $C_L^{\phi\phi}$ and $C_L^{\phi T}$ for a runaway model with a canonical kinetic term ($K_0=0$) and coupled to matter {and two massive neutrinos of total mass 0.2 eV}.}
\label{fig:K0b}
\end{figure}

\begin{figure}[!h]
\centering 
\includegraphics[width=0.9\linewidth]
{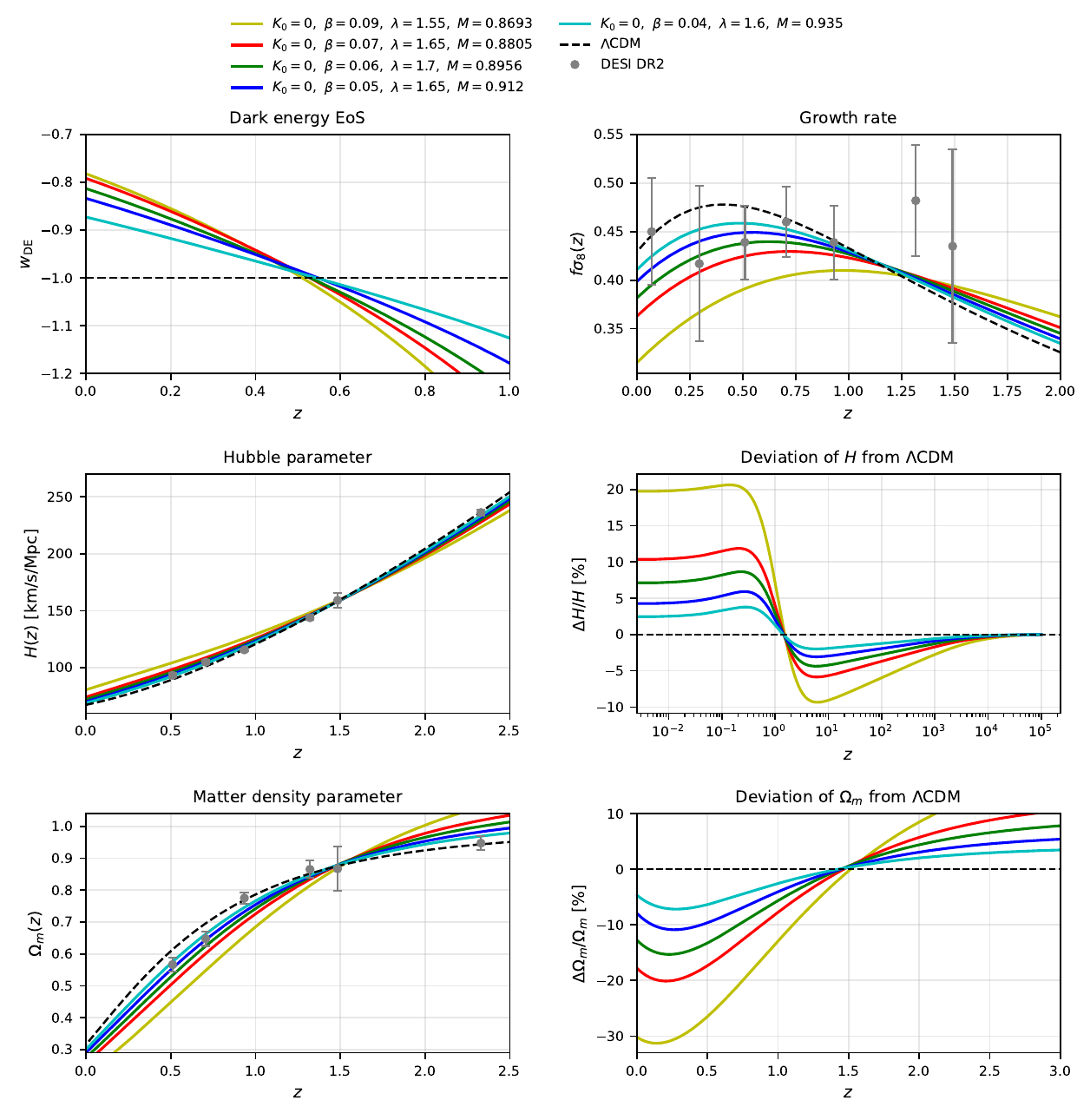} 
\caption{The equation of state, the growth rate, the Hubble rate, the variation of the Hubble rate with respect to $\Lambda$CDM, $f\sigma_8$, $\Omega_m$ and the relative deviation of $\Omega_m$ with respect to $\Lambda$CDM for a model with a canonical kinetic term ($K_0=0$) and a single massive neutrino of mass 0.06.}
\label{fig:K0a_006}
\end{figure}

\begin{figure}[!h]
\centering 
\includegraphics[width=0.9\linewidth]
{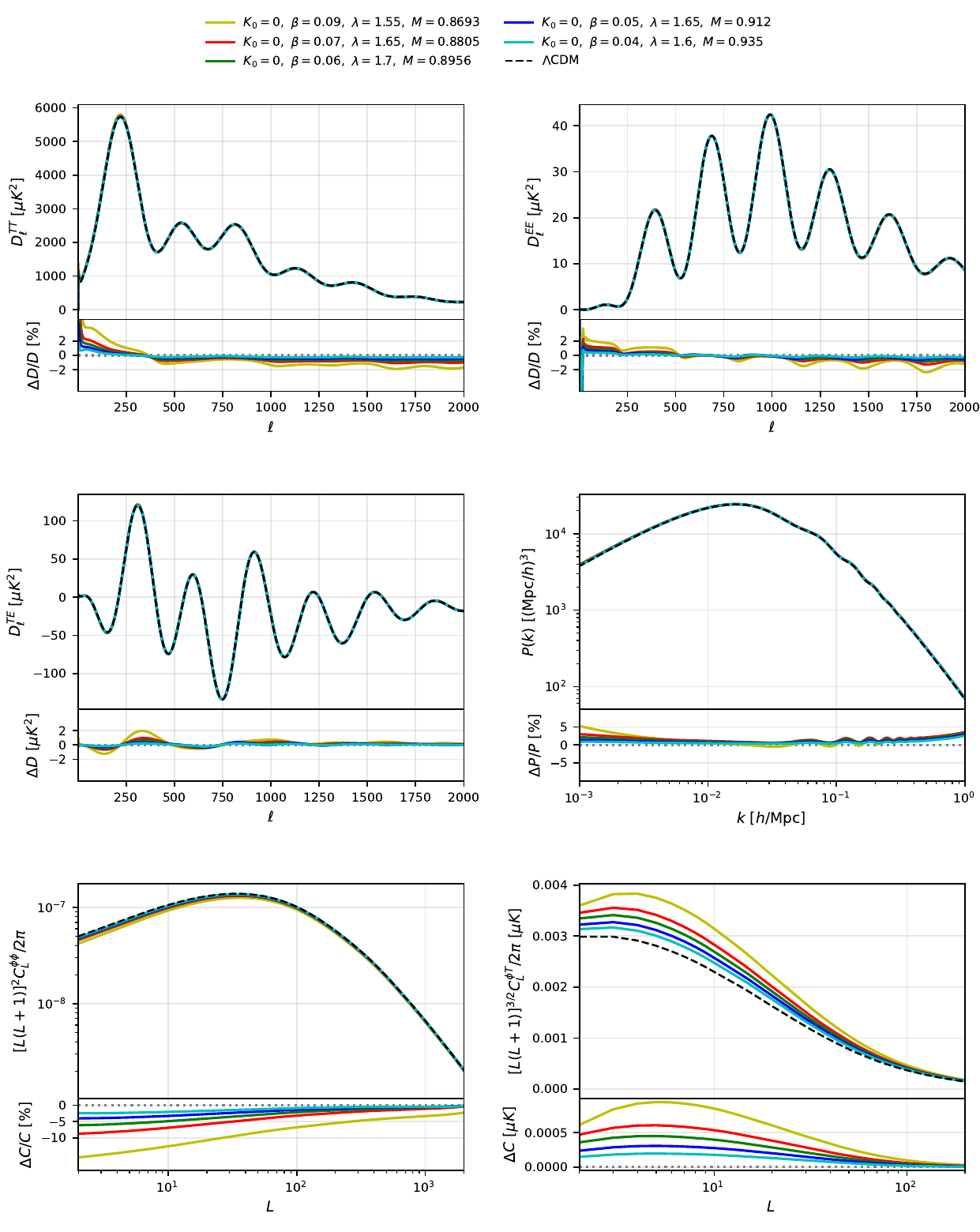} 
\caption{The temperature, E-polarisation TE cross-correlation angular power spectrum, the matter power spectrum and the CMB lensing potentials $C_L^{\phi\phi}$ and $C_L^{\phi T}$ for a runaway model with a canonical kinetic term ($K_0=0$) and coupled to matter and a single massive neutrino of mass 0.06.}
\label{fig:K0b_006}
\end{figure}

\subsection{The variation of Newton's constant}
{Finally let us discuss the constraint from $\dot G_{\rm eff}/G_{\rm eff}$ \cite{Williams:2004qba,Muller:2007zzb,Biskupek:2020fem,BeltranJimenez:2015sgd,Brax:2018zvh}. The K-mouflage mechanism does not screen the cosmological variation of $\phi$ \cite{Brax:2018zvh}. This implies that the time variation of the effective Newton constant $G_{\rm eff}$ is not protected by the large field gradient locally.  The time variation is typically constrained to be less than $0.02 H_0$ in the recent past from the Lunar Laser Ranging (LLR) measurements \cite{Williams:2004qba} and even a stronger $10^{-4} H_0$ \cite{Biskupek:2020fem} from more recent analyses. We have for the time variation
\be 
\frac{\dot G_{\rm eff}}{G_{\rm eff}}= \frac{2\beta}{M_{\rm Pl}} \dot \phi.
\label{eq:varg}
\ee
In the regime where the potential is negligible and matter dominates the Klein-Gordon equation gives $\dot \phi \sim - \frac{4\Omega_{m0}}{3}\beta M_{\rm Pl}H$ therefore  implying that
\be 
\left| \frac{\dot G_{\rm eff}}{G_{\rm eff}}\right| \sim \frac{8\Omega_{m0}}{3}\beta^2 H.
\ee
When the potential is present  and dominates the late time dynamics with an equation of state $\omega_0$, we have for  the variation
\be 
\frac{\dot G_{\rm eff}}{G_{\rm eff}}=2\sqrt{3(1+\omega_0)\Omega_\Lambda}\beta H_0.
\ee
When the field follows the tracking solution \cite{Copeland:1997et} with $1+\omega_0= \frac{\lambda^2}{3}$, we find the estimate
\be 
\frac{\dot G_{\rm eff}}{G_{\rm eff}}\sim 2\sqrt{\Omega_\Lambda}\lambda \beta H_0.
\ee
This selects low values of $\beta$ and $\lambda$. For kinessence where both the potential and the coupling to matter are relevant, the variation interpolates between a variation in $\beta^2 H_0$ and $\beta\lambda H_0$. 
}

{In Fig. \ref{fig:varg} we represent the variation of $G_{\rm eff}$ divided by $H$ for the models considered as obtained  from Eq. \eqref{eq:varg} as a function of  the redshift. We note that the LLR constraint is not satisfied as the ratio $\frac{\dot G_{\rm eff}}{G_{\rm eff}H_0} \sim 0.1$, significantly higher than 0.02 \cite{Babichev:2011iz} or $10^{-4}$ \cite{Biskupek:2020fem}. In order to decrease this ratio in our models, it is necessary to reduce the magnitude of $\beta$, but this is not possible while keeping the observed equation of state compatible with the latest DESI measurements. Indeed, a reduction in  $\beta$ leads to a decrease in the magnitude of the slope of the equation of state $w(z)$, which in turn requires a reduction of $\lambda$ to keep the phantom crossing redshift correct, but this in turn shifts the value of $w(z=0)$ towards $-1$, i.e. in tension with the DESI measurements.  One can envisage a solution to this issue by replacing our potential with an Albrecht-Skordis potential \cite{Albrecht:1999rm}, which exhibits an attractor-type behaviour corresponding to the local minimum of the field. A potential of the form
\be
f(\phi)=U(\phi) e^{-\lambda\phi/M_{\rm Pl}}
\ee
where $U(\phi)\sim 1$ away from the minimum $\phi=\phi_{\rm min}$ and  $f(\phi) \sim 1+ \frac{\gamma}{2 M^2_{\rm Pl}}(\phi -\phi_{\rm min})^2$ in its vicinity  would lead to oscillations. As long as $\phi$ reaches the point where the field stops before decreasing towards the minimum for $z= 0$, the variation of Newton's constant could be made to vanish at $z=0$. Assuming that the evolution of the field $\phi$ tracks an attractor when $U(\phi)\sim 1$, the convergence towards $\dot\phi\sim 0$ for $z=0$  would not require a fine-tuning of the initial conditions, see appendix \ref{app:C} for more detail. Of course, 
this would still lead to an equation of state close to $-1$ at very small redshift. It would have to be seen if this can be made compatible with the DESI results. As the modification of the equation of state would happen for very small redshift this would have no impact on the CMB physics as presented in this paper and on the crossing of the equation state below $-1$. Such a replacement is left for a subsequent publication in conjunction with a MCMC analysis of the data.}

\section{Conclusions}
\label{sec:conc}
We have introduced the Kinessence models, obtained by combining a K-essence model and a quintessence potential, with a Yukawa coupling to matter. The interplay between these features produces a distinctive phenomenology: the quintessence potential term dominates at low redshift, leading to $w_{DE}>-1$, while the K-essence kinetic term coupled to matter drives $w_{DE}<-1$ at intermediate redshifts leading to a phantom crossing at $z \approx 0.5$. This model can thus explain the recent DESI measurements. 

We have investigated the dynamics of such models at the background and perturbative level in the Jordan frame, showing how the equation of state crosses the phantom divide and affects both the matter content of the Universe and the Hubble rate at low redshift. We have implemented the models numerically and considered two scenarios for the kinetic function, a polynomial and an arctan function. The latter satisfies all the Solar System constraints whilst the former fails to describe the precise trajectory of the Moon around the Earth. We have analysed the effective equation of state and the matter fraction for the background, as well as the effects of the models on the Hubble evolution. We have employed the $\mathcal{H}$-EFTCAMB code to evaluate the CMB spectra and the linear power spectrum.

Our work shows that Kinessence models are able to describe the qualitative nature of the DESI measurements successfully, while maintaining a CMB power spectrum compatible with the \textit{Planck} observations. The coupling to matter $\beta$ must be significantly away from zero potentially leading to gravitational issues in the Solar System. This is resolved by the K-mouflage mechanism on small scales. This distinguishes Kinessence models from coupled dilatons with a coupling to matter as the latter cannot screen the induced fifth force in the Solar System. Our study emphasises the necessary link between cosmological models and their short-distance behaviour. A description on all scales is necessary to validate Kinessence on large scales compared to models with canonical kinetic terms for instance.

The fact that a coupling between matter and dark energy is necessary in these models could lead to interesting signatures on cluster scales for instance where K-mouflage screening is less efficient. This is left for future work. 
A full Markov-Chain-Monte-Carlo analysis of the models is left for a subsequent paper too.  This will make more precise the possibility that CMB-S4/Simons-Observatory-class lensing reconstruction, targeting sub-percent precision in that $\ell$ range, could help distinguish between  Kinessence models and $\Lambda$CDM \cite{CMB-S4:2016ple,LiteBIRD:2023iiy}. {This will also specify how much reduction of growth is allowed and give clues on the neutrino mass. {Finally the tight constraint on the variation of Newton's constant requires to alter the shape of the potential to force the time variation of the field to vanish at very small redshift. This should have no impact on the cosmology of the model for BAO and the CMB. On the other hand, this introduces a new coincidence problem: why would the scalar field be stopped so close to the present time? The compatibility between the cosmological data and the LLR results using Albrecht-Skordis potentials will be analysed in future work. } }

\acknowledgments
{We would like to thank J. Hallam, V. Poulin, M. Tristram, L. Vacher and A. Smith for crucial remarks. AL has been supported by the Romanian Ministry of Research, Innovation and Digitalization under the
Romanian National Core Program LAPLAS VII-contract no. 30N/2023. TC would like to thank CEA Saclay for hospitality. }

\appendix

\section{The coupled dilaton}
\label{app:dil}
A non-trivial $K$ function is only required to satisfy the Solar System constraints, as discussed in \ref{subsec:stability_screening}. Models with $K_0=0$ can successfully pass cosmological tests. These models are simple dilaton runaway scenarios with a coupling to matter. They are well-known to have an effective potential in the Einstein frame with a minimum. This is not enough for the chameleon screening mechanism to be efficient enough to satisfy Solar System bounds \cite{Brax:2010gi}. Similarly, the mass of the dilaton at the minimum of the potential is not large enough to attract the dilaton cosmologically towards the minimum. 
{If the minimum of the potential were an attractor, the ratio $\delta=\frac{\beta^2 H_0^2}{m_0^2}$ where $m_0$ is the cosmological mass of the dilaton in the Einstein frame would be bounded as $0.047\lesssim \delta \lesssim 0.093$ \cite{Brax:2025ahm}. We find that this ratio is slightly outside this interval apart for the large value of $\beta=0.09$. We consider $(\beta,\lambda, \delta)= (0.09,1.6,0.059), (0.07,1.65,0.045), (0.06,1.7,0.038), (0.05,1.65,0.033), (0.04,1.6,0.027)$.}

Such models only depend on three parameters, $\beta$, $\lambda$ and $M$. By fixing the first two, we note that $M$ is no longer free, as this parameter can be employed to fix $\sigma_8$ and hence the amplitude of the matter power spectrum. We therefore proceed  by fixing the value of $\beta$, and then choosing $\lambda$ and $M$ such that the expected crossing redshift  and matter power spectrum amplitudes are obtained. We show that the expected equation of state as well as the other DESI measurements can be satisfied, together with percent-level deviations of the CMB temperature and polarisation spectra with respect to $\Lambda$CDM.

This appendix presents the plots corresponding to the variation of the parameters defining the coupled dilaton models. We show that a good agreement with data can be achieved.
In Figs. \ref{fig:K0a}-\ref{fig:K0b} we have represented the most reasonable  fit for the observable and the best values obtained. We observe that the fit for growth rate is improved as $\beta$ is decreased. The value of the dark energy equation of state at $z=0$ is, however, decreasing at the same time, degrading the fit. The deviations of the Hubble parameter and $\Omega_m$ with respect to $\Lambda$CDM are also reduced. The CMB and matter spectra present small modifications with respect to $\Lambda$CDM. {The values $\sigma_8$ are around 0.79 for all models, with difference only at the next decimal place. The decrease of the value of $\sigma_8$ with respect to $\Lambda$CDM is due to the presence of massive neutrinos.}

In Figs. \ref{fig:K0a_006}-\ref{fig:K0b_006} we present the corresponding results for the coupled dilation model with a single massive neutrino of mass 0.06 eV. In this case, the values of $\sigma_8$ are 0.81, with differences at  next decimal point.

All in all, we find that the dilaton model with no K-essence part reproduces reasonably current data. Hence, we have found three models whose chief difference comes from their small-scale  properties. Without K-mouflage, the dilaton models do not pass the Cassini test. When introducing cubic K-mouflage, the Cassini bound is satisfied whilst the moon periastron is violated. Only the $\arctan$ models satisfies the constraints. It would be extremely interesting to see if the three models differ in the non-linear regime of structure formation.

\section{Variation of the parameters}
\label{app:plots}
 {We  consider several variations of Kinessence, and we present for each case, from top to bottom and left to right: the dark energy equation of state, the Hubble deviation with respect to $\Lambda$CDM, the growth rate $f\sigma_8$, the evolution of $\Omega_m$, all with respect to redshift, the CMB temperature, polarisation and mixed TE spectra, and the matter power spectrum. In Figs. \ref{fig:varnu_poly} and \ref{fig:varnu_arctan} we investigate the effects of the neutrino mass for the polynomial and arctan models respectively. Figs. \ref{fig:varbeta_poly} and \ref{fig:varbeta_arctan} focus on the $\beta$ variation, Figs. \ref{fig:varlambda_poly} and \ref{fig:varlambda_arctan} on the $\lambda$ variation, Figs. \ref{fig:varM_poly} and \ref{fig:varM_arctan} on the $M_{\rm num}$ variation. In Figs. \ref{fig:varK0_poly} and \ref{fig:varm_poly} we show how physical quantities vary when $K_0$ and $m$ change in the polynomial model, and in Figs. \ref{fig:varKs_arctan} and \ref{fig:varxs_arctan} we present analogous results for $K_*$ and $\chi_*$ in the arctan model. The physical description of these variations is discussed in the main text.}

\begin{figure}[!h]
\centering
\includegraphics[width=\linewidth]
{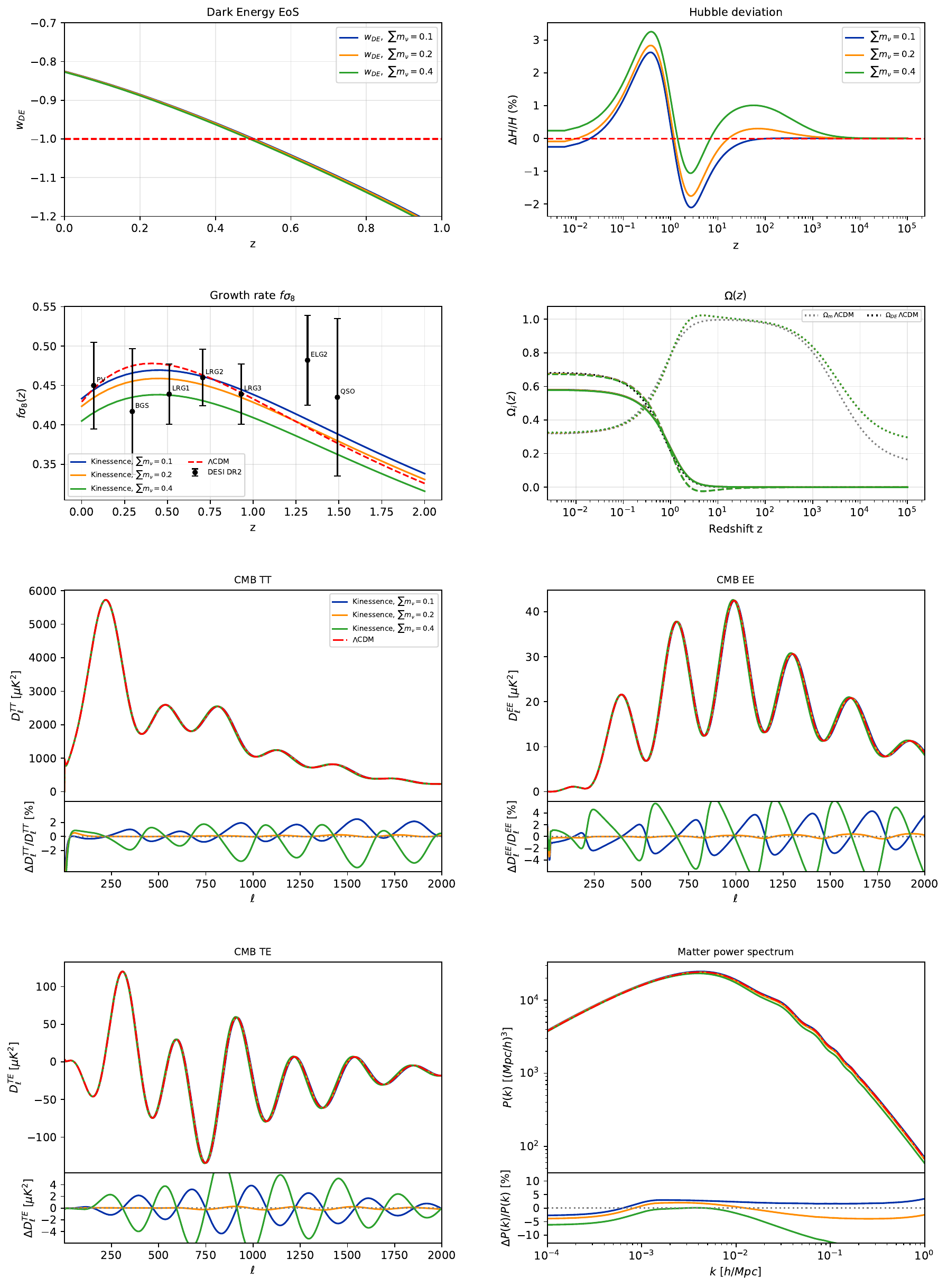} 
\caption{Variations of equation of state, the Hubble rate, $f\sigma_8$, the energy fractions, and the CMB and power spectra as a function of the neutrino mass for the polynomial model.}
\label{fig:varnu_poly}
\end{figure}

\begin{figure}[!h]
\centering 
\includegraphics[width=\linewidth]
{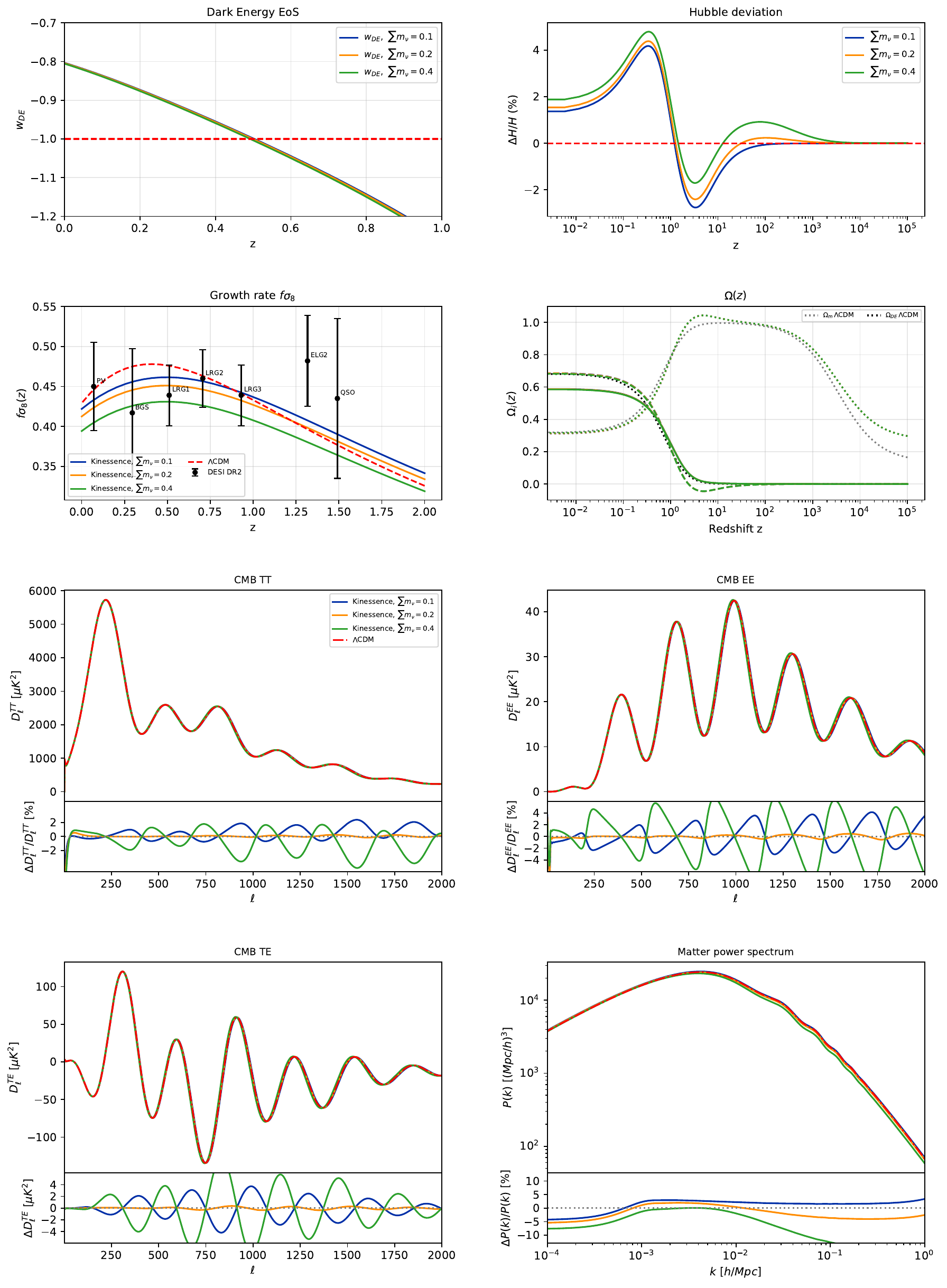} 
\caption{Variations of equation of state, the Hubble rate, $f\sigma_8$, the energy fractions, and the CMB and power spectra as a function of the neutrino mass for the arctan model.}
\label{fig:varnu_arctan}
\end{figure}

\begin{figure}[!h]
\centering
\includegraphics[width=\linewidth]{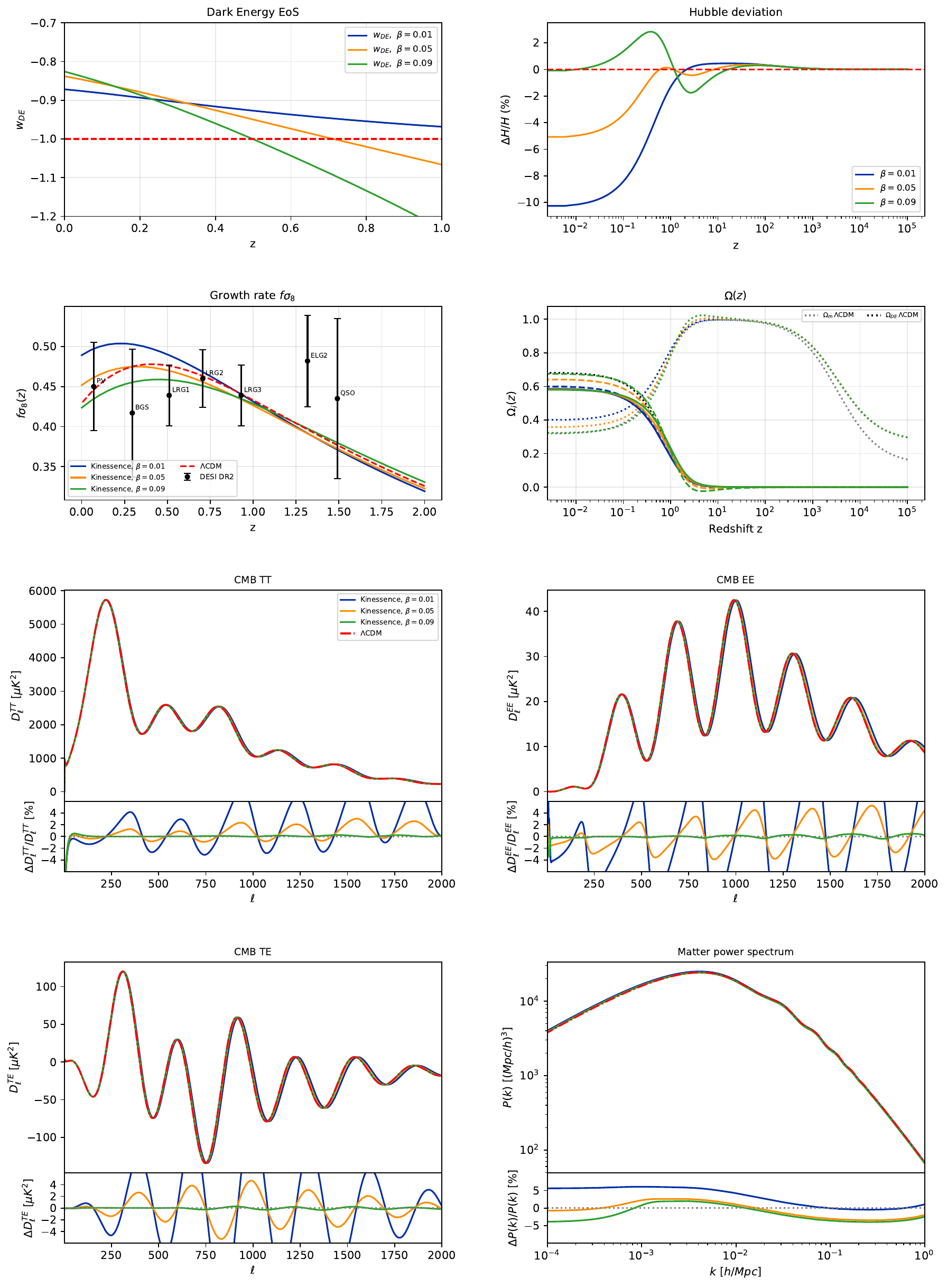}
\caption{Variations of the equation of state, the Hubble rate, $f\sigma_8$, the energy fractions, and the CMB and power spectra as a function of  $\beta$ for the polynomial model.}
\label{fig:varbeta_poly}
\end{figure}

\begin{figure}[!h]
\centering
\includegraphics[width=\linewidth]{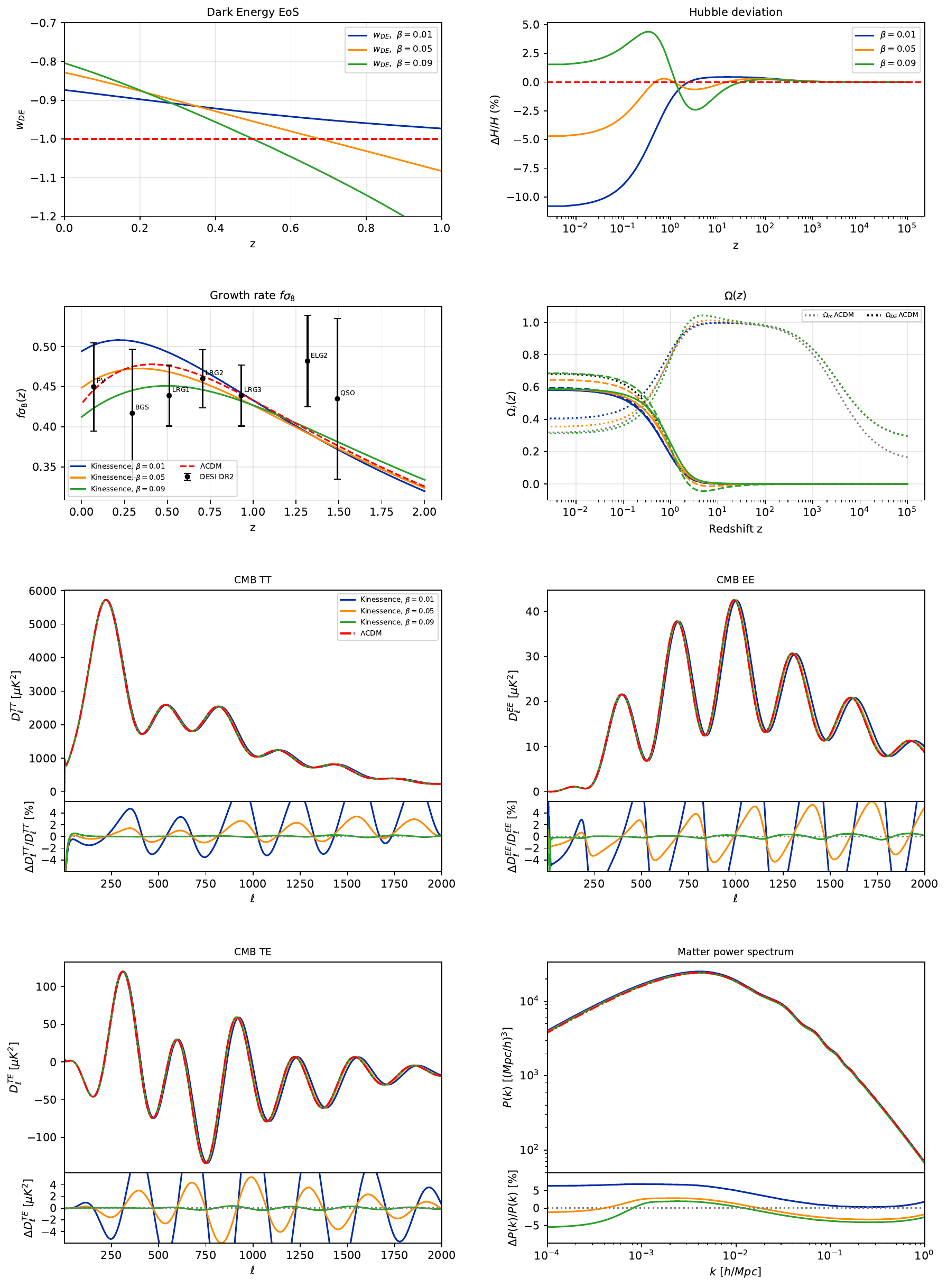}
\caption{Variations of the equation of state, the Hubble rate, $f\sigma_8$, the energy fractions, and the CMB and power spectra as a function of $\beta$ for the arctan model.}
\label{fig:varbeta_arctan}
\end{figure}

\begin{figure}[!h]
\centering
\includegraphics[width=\linewidth]{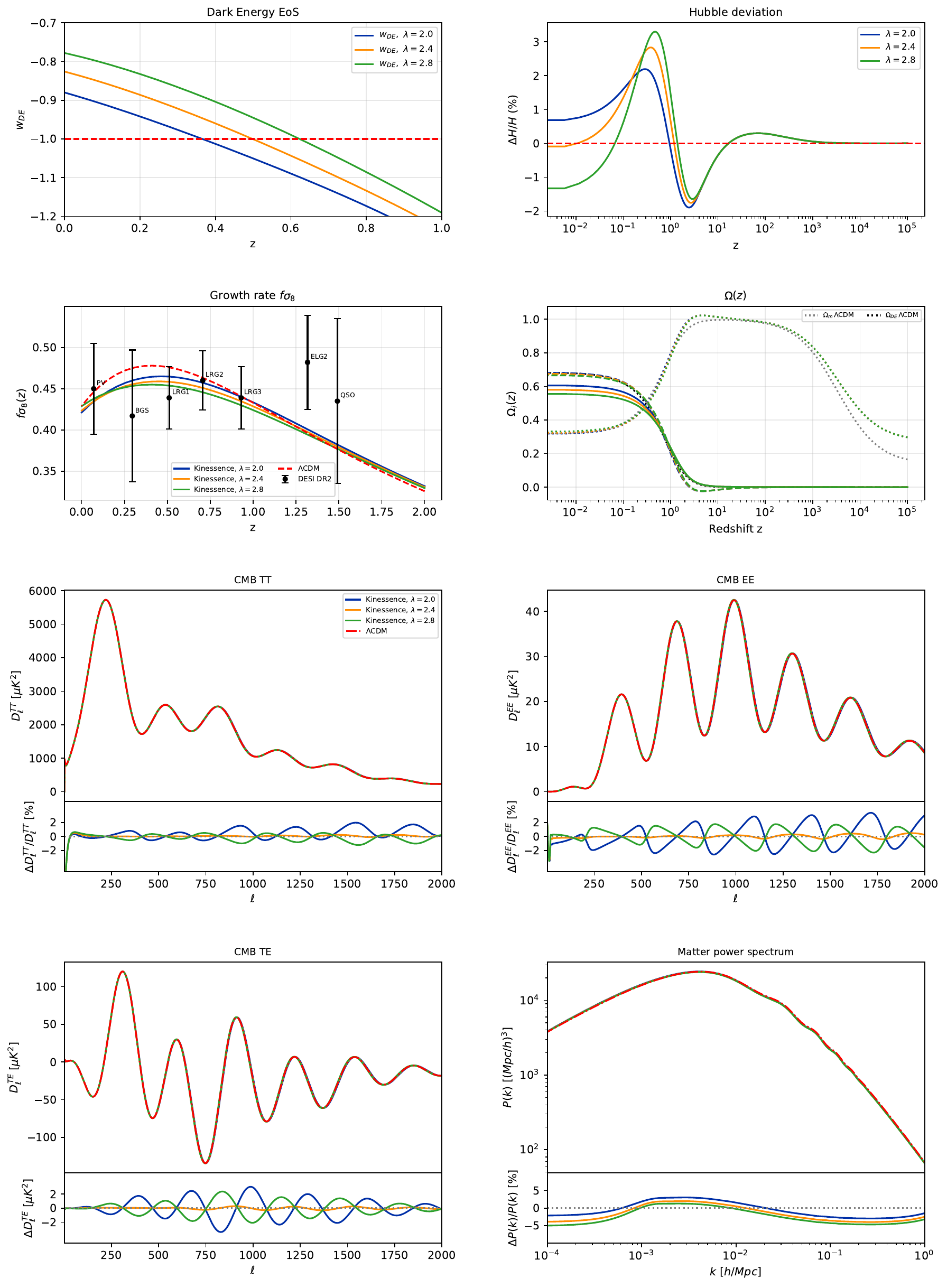}
\caption{Variations of the equation of state, the Hubble rate, $f\sigma_8$, the energy fractions, and the CMB and power spectra as a function of  $\lambda$ for the polynomial model.}
\label{fig:varlambda_poly}
\end{figure}

\begin{figure}[!h]
\centering
\includegraphics[width=\linewidth]{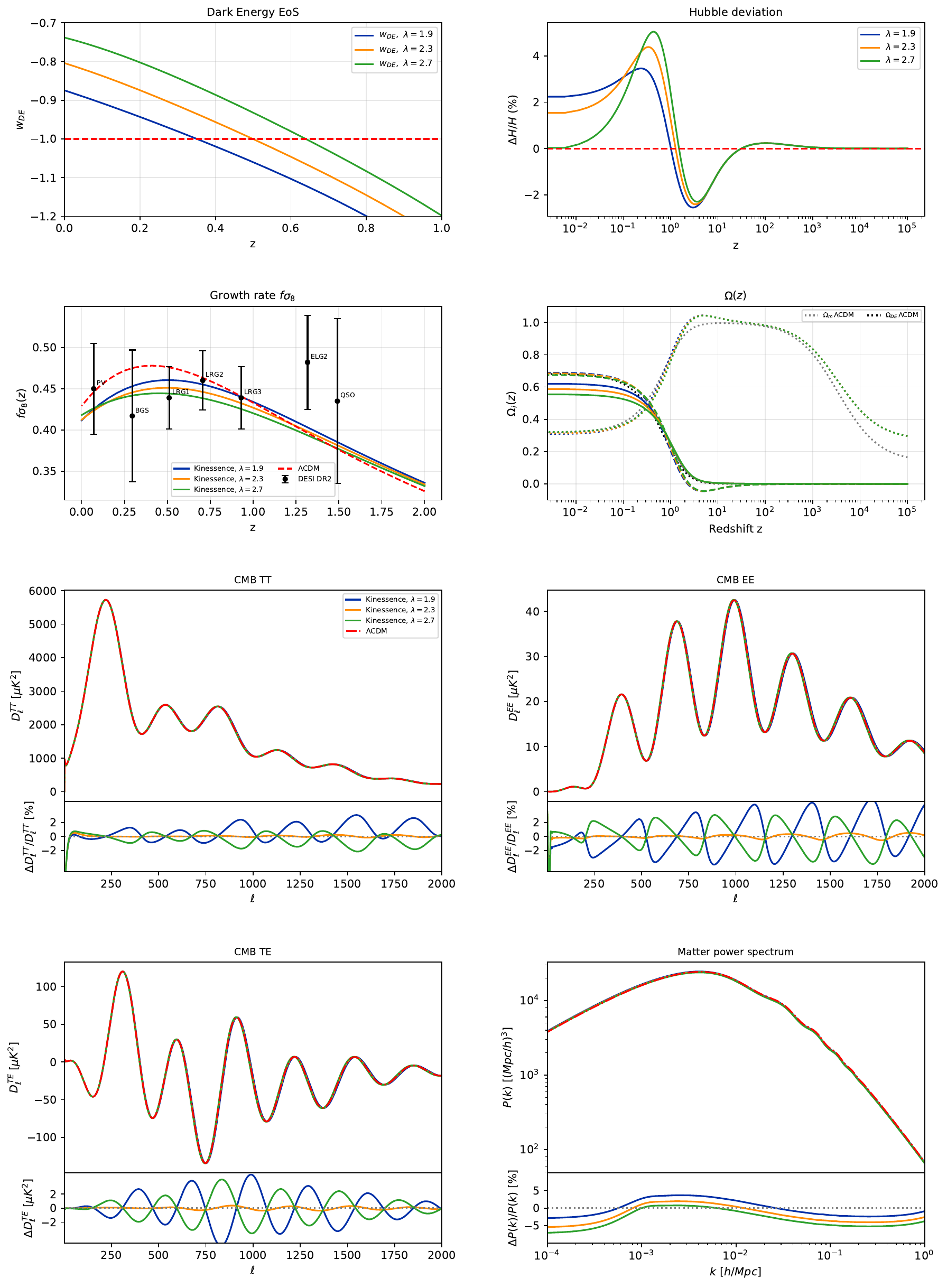}
\caption{Variations of the equation of state, the Hubble rate, $f\sigma_8$, the energy fractions, and the CMB and power spectra as a function of $\lambda$ for the arctan model.}
\label{fig:varlambda_arctan}
\end{figure}

\begin{figure}[!h]
\centering
\includegraphics[width=\linewidth]{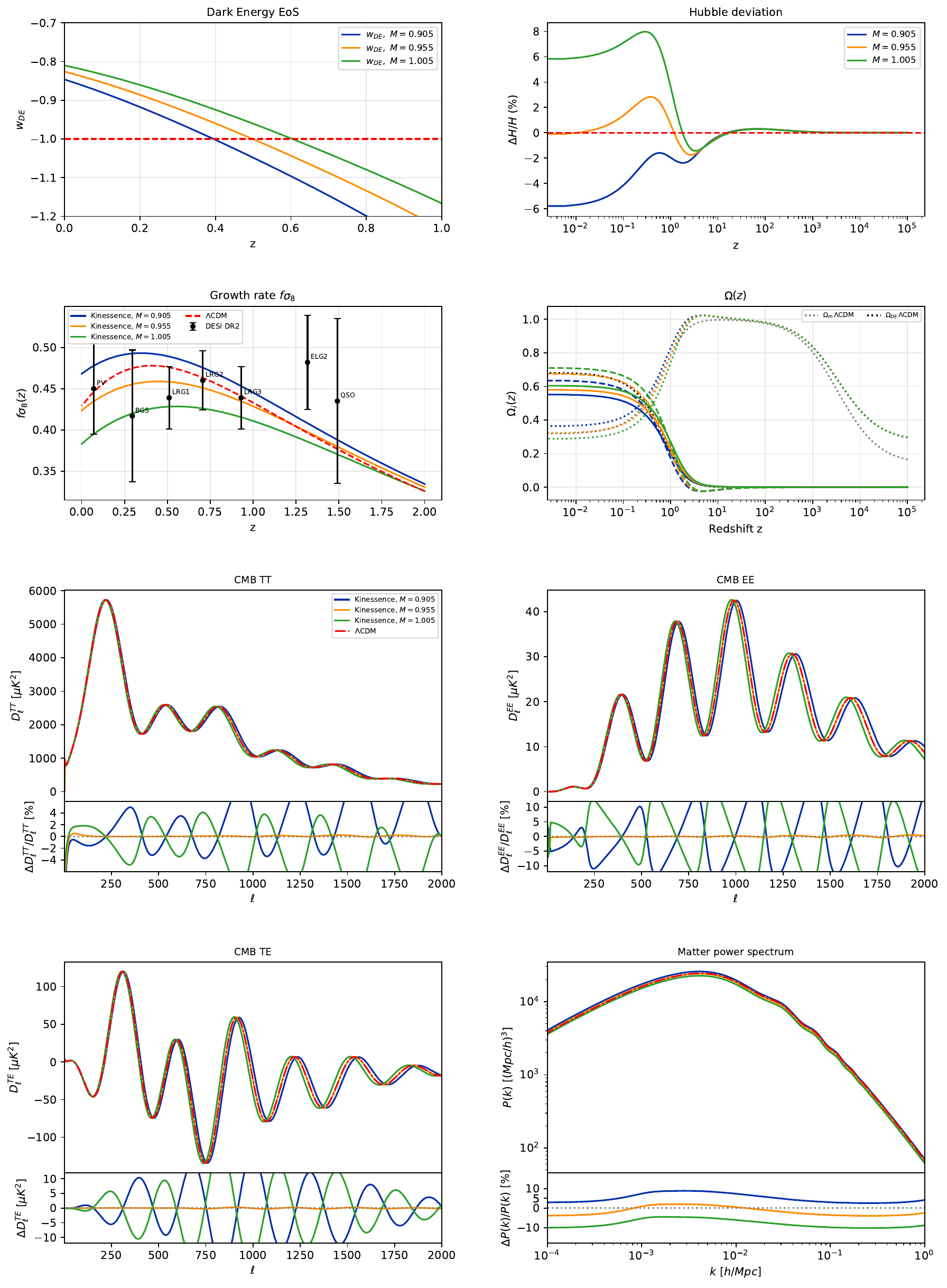}
\caption{Variations of the equation of state, the Hubble rate, $f\sigma_8$, the energy fractions, and the CMB and power spectra as a function of $M_{\rm num}$ for the polynomial model.}
\label{fig:varM_poly}
\end{figure}

\begin{figure}[!h]
\centering
\includegraphics[width=\linewidth]{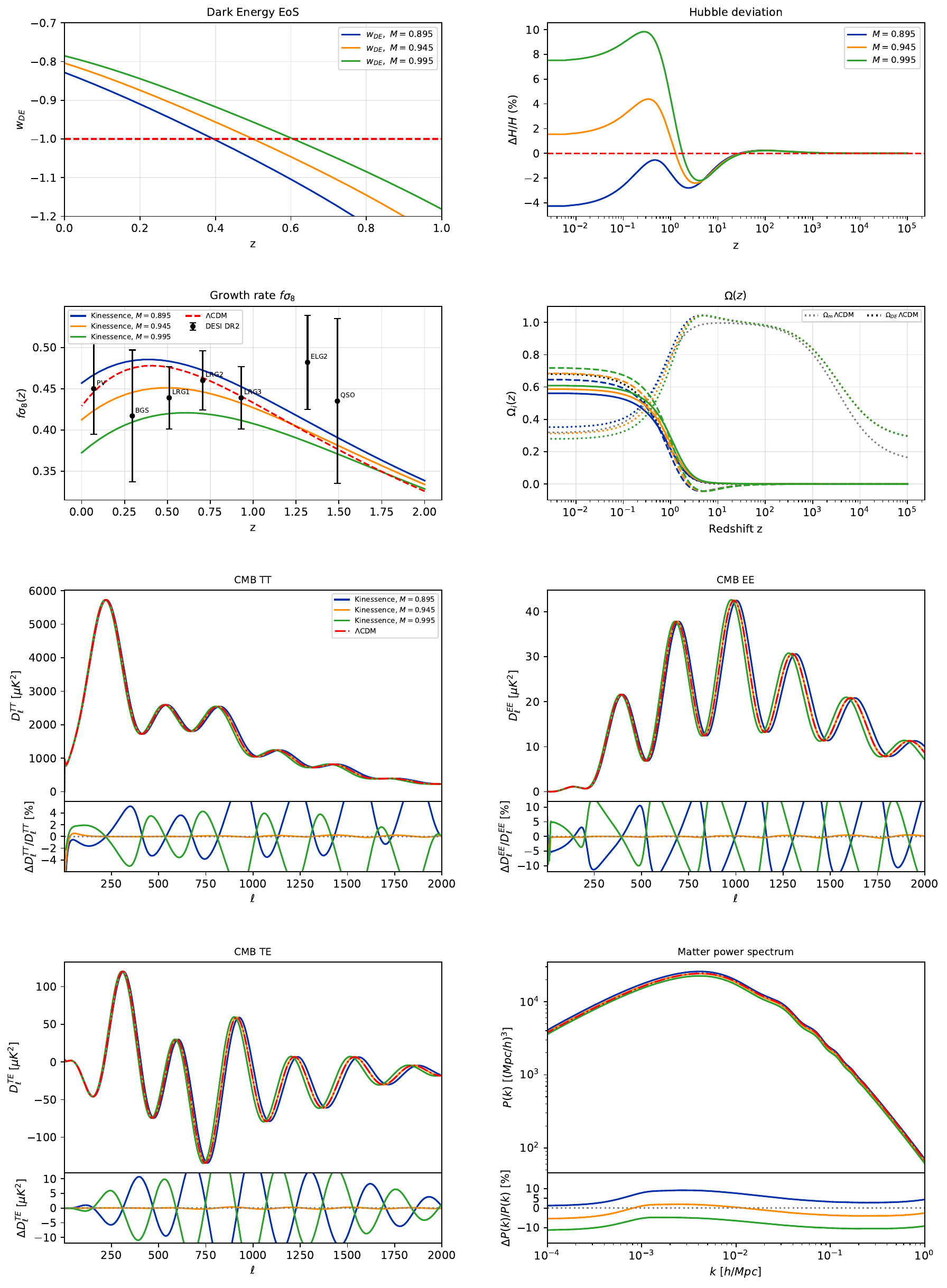}
\caption{Variations of the equation of state, the Hubble rate, $f\sigma_8$, the energy fractions, and the CMB and power spectra as a function of $M_{\rm num}$ for the arctan model.}
\label{fig:varM_arctan}
\end{figure}

\begin{figure}[!h]
\centering
\includegraphics[width=\linewidth]{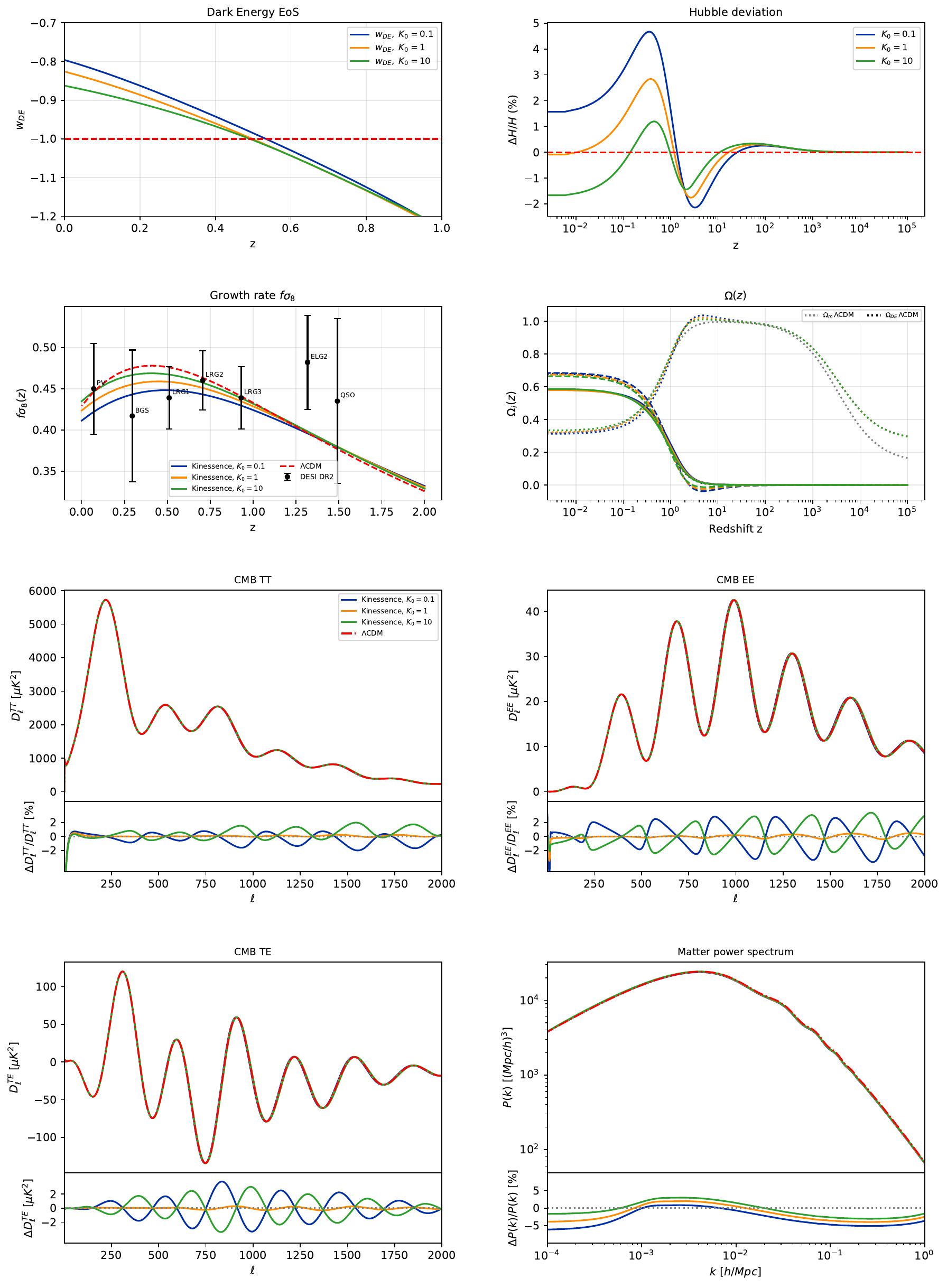}
\caption{Variations of the equation of state, the Hubble rate, $f\sigma_8$, the energy fractions, and the CMB and power spectra as a function of $K_0$ for the polynomial model.}
\label{fig:varK0_poly}
\end{figure}

\begin{figure}[!h]
\centering
\includegraphics[width=\linewidth]{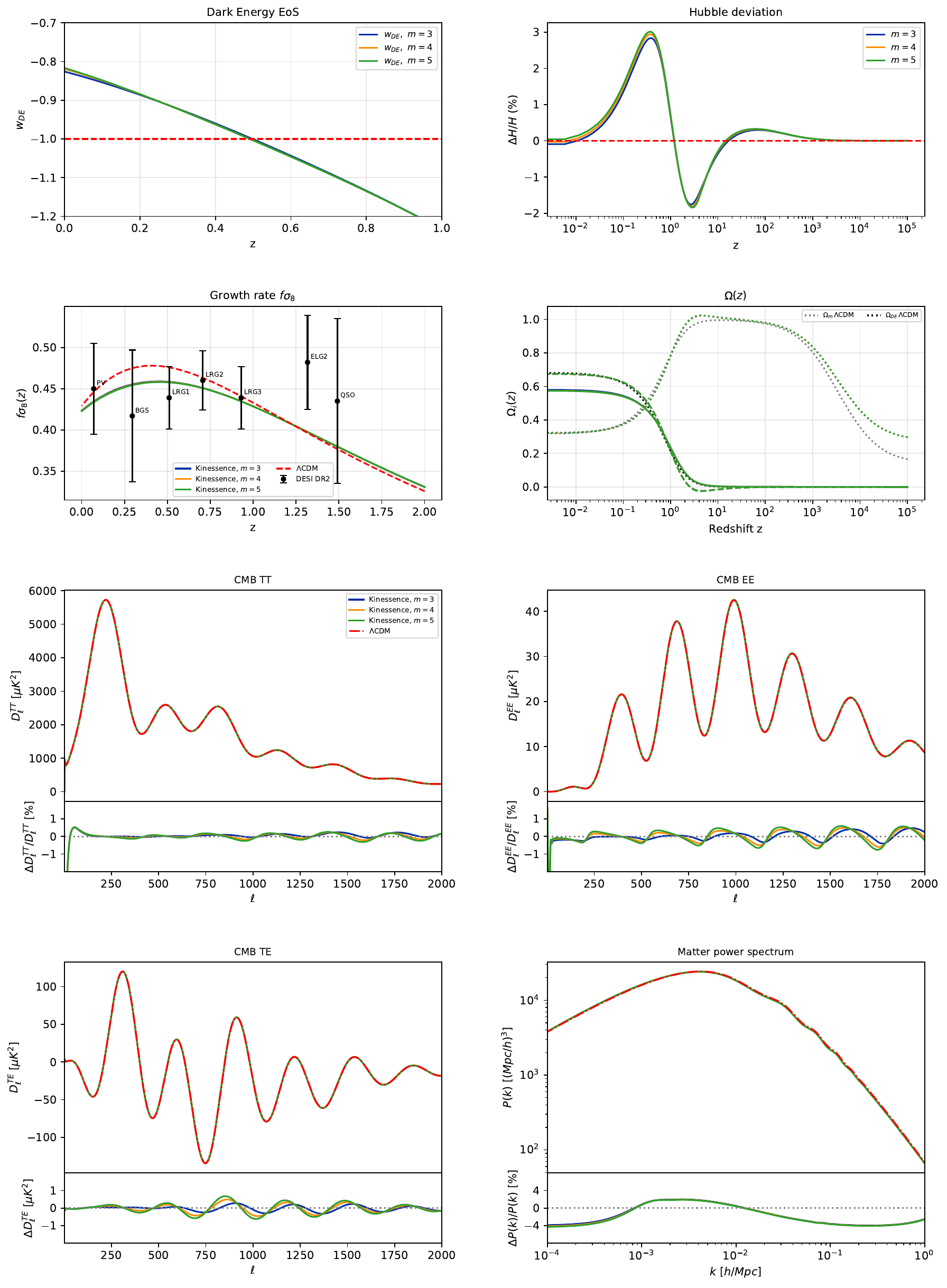}
\caption{Variations of the equation of state, the Hubble rate, $f\sigma_8$, the energy fractions, and the CMB and power spectra as a function of $m$ for the polynomial model.}
\label{fig:varm_poly}
\end{figure}

\begin{figure}[!h]
\centering
\includegraphics[width=\linewidth]{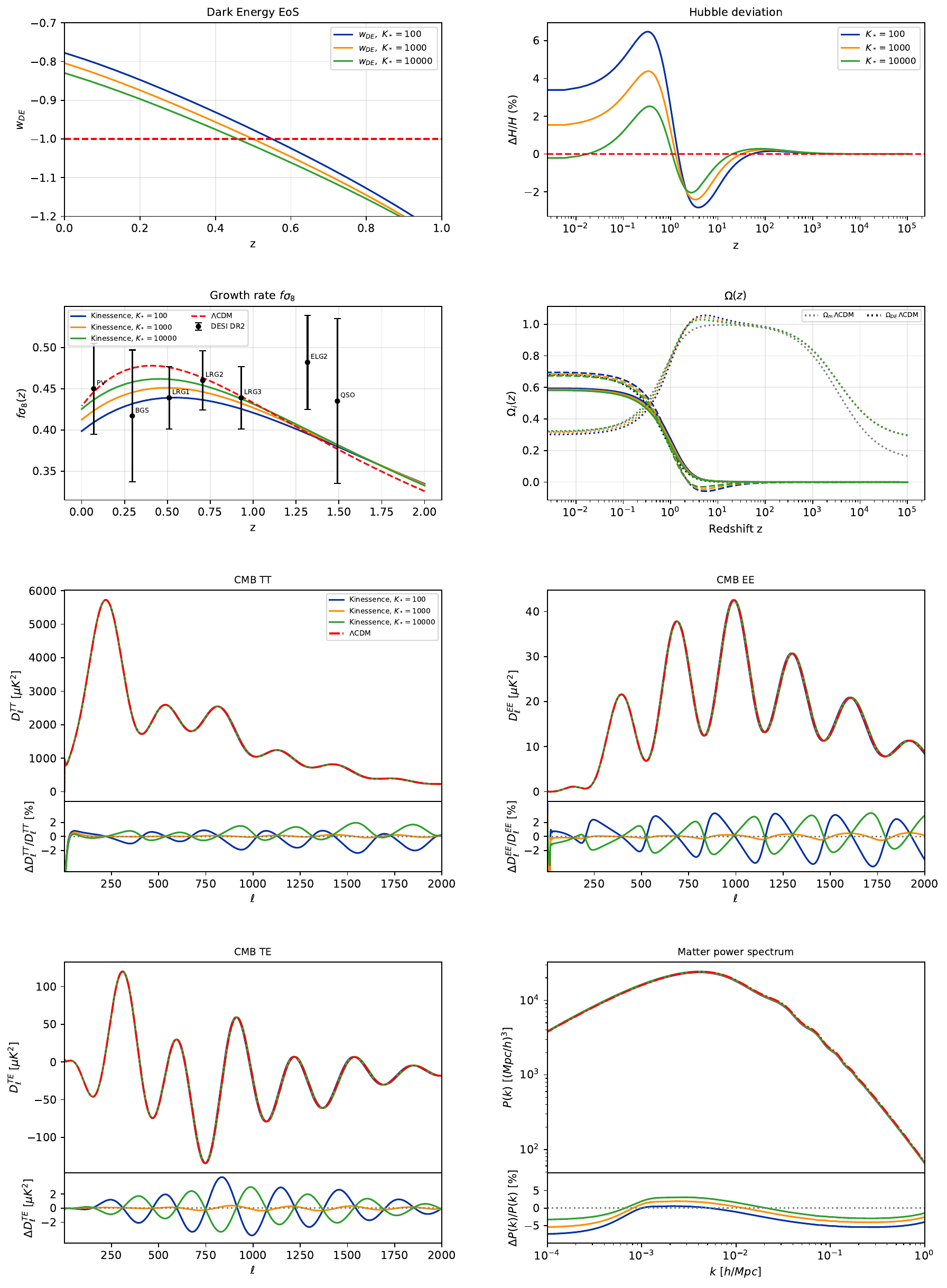}
\caption{Variations of the equation of state, the Hubble rate, $f\sigma_8$, the energy fractions, and the CMB and power spectra as a function of $K_*$ for the arctan model.}
\label{fig:varKs_arctan}
\end{figure}

\begin{figure}[!h]
\centering
\includegraphics[width=\linewidth]{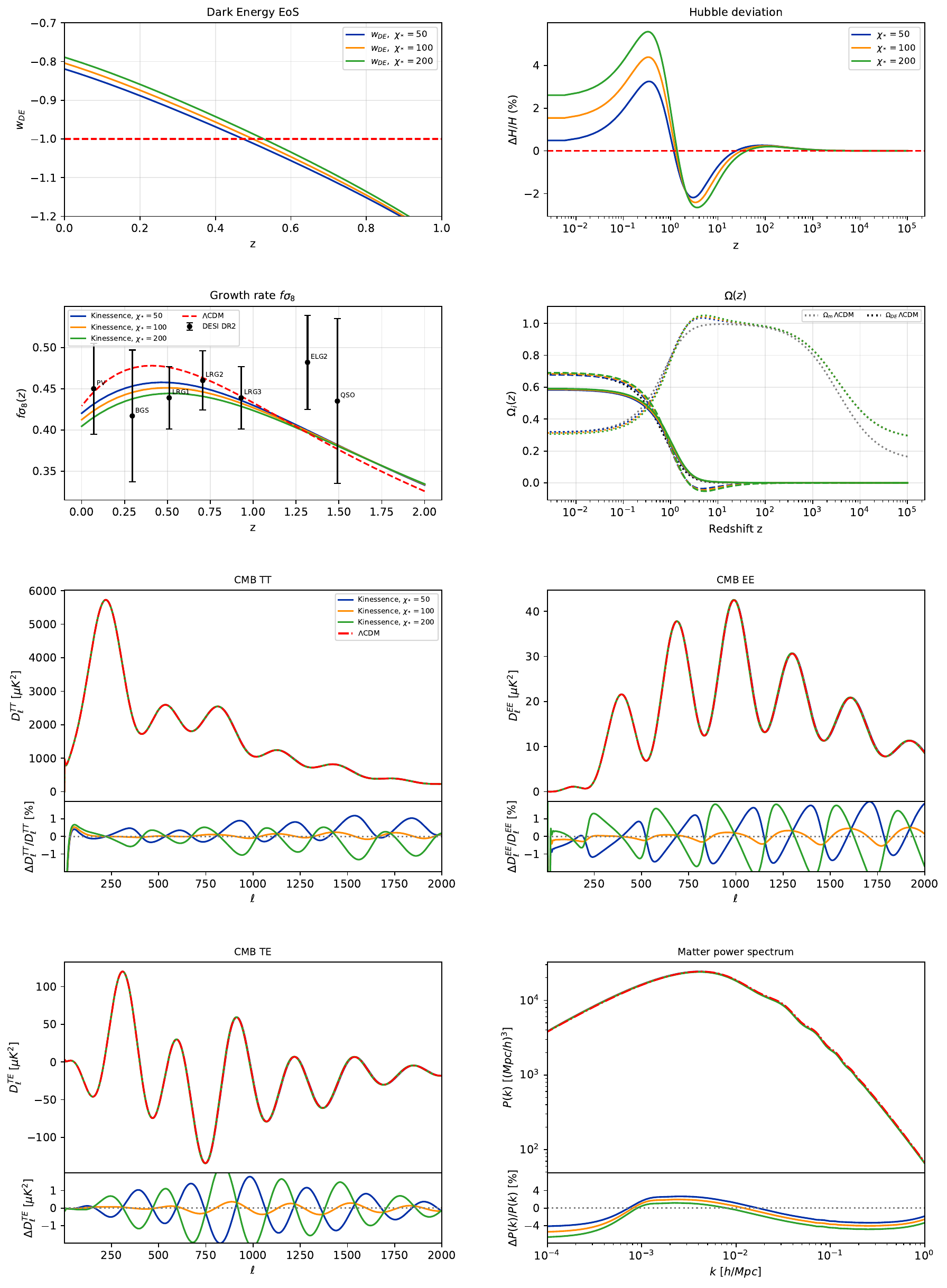}
\caption{Variations of the equation of state, the Hubble rate, $f\sigma_8$, the energy fractions, and the CMB and power spectra as a function of $\chi_*$ for the arctan model.}
\label{fig:varxs_arctan}
\end{figure}
\section{The LLR constraint and the Albrecht-Skordis potential}
\label{app:C}
The convergence to a small $\dot G_{\rm eff} $ at $z=0$ can be analysed in a simplified setting where $K(\chi) \sim \chi$ at small $z$.
As long as matter dominates over dark energy the scalar field satisfies
\be 
\frac{d}{dt}(a^3 \dot \phi)= - \frac{\beta}{M_{\rm Pl}} a^3\rho.
\ee
In the matter era where $H^2\sim \frac{\rho}{3M_{\rm Pl}^2}$ and assuming that $A(\phi)\sim 1$ with  $\rho a^3 = \rho_0$, this implies that
\be 
\phi=\phi_{\rm ini}-\frac{4\Omega_{m0}}{3} \beta M_{\rm Pl}\ln \frac{t}{t_{\rm ini}}.
\label{convq}
\ee
This is valid until $\rho \sim M^4 f(\phi)$ where the potential starts dominating. Assuming this takes place in the range of field values  $\phi\lesssim \phi_\star$ for which  $f(\phi)\sim e^{-\lambda \phi/M_{\rm Pl}}$  then the field converges to the attractor  defined by
\be 
\phi = \phi_{\rm DE}+ \frac{2M_{\rm Pl}}{\lambda}\ln \frac{t}{t_{\rm DE}}
\label{attrac}
\ee
where the Hubble rate is 
\be 
H= \frac{2}{\lambda^2 t}
\ee
and $V(\phi_{\rm DE}) = \frac{6 M^2_{\rm Pl}}{t^2_{\rm DE}\lambda^2}(\frac{2}{\lambda^2}-\frac{1}{3})$ requires $\lambda \le \sqrt 6$. This behaviour is only valid until the field reaches the vicinity of the minimum of the potential where 
\be V(\phi)\sim {V_{\rm DE}} + \Omega^2 (\phi-\phi_{\rm min})^2.
\ee
When the field oscillates around the minimum of the potential we have $V(\phi)\sim V_{\rm DE}$ and the field behaves as
\be 
\phi -\phi_{\rm min}= e^{-\frac{3}{2}H_{\rm DE}(t-t_\star)}(A \cos \Omega (t-t_\star) + B \sin \Omega (t-t_\star))
\ee
where $A= \phi_\star -\phi_{\rm min}$ and $\Omega B= \dot \phi_\star +\frac{3}{2} H_{\rm DE} (\phi_\star -\phi_{\rm min})$. The field has a vanishing derivative when
\be 
\tan \Omega (t_0-t_\star)= \frac{\Omega B- \frac{3H_{\rm DE}}{2}A}{\Omega A+ \frac{3H_{\rm DE}B}{2}}
\ee
where $t=t_0$ corresponds to $z=0$. This is a constraint relating $\Omega$ and $\phi_\star -\phi_{\rm min}$ to $\dot \phi_\star$. As long as the field has converged to the attractor (\ref{attrac}), the value of $\dot \phi_\star$ does not depend on the initial conditions of the field. This implies that the value of $\Omega$ is independent of the initial conditions. 
This can be seen using further approximations. We have $\phi_\star \sim \phi_{\rm min}$ when the potential around the minimum is steep enough. Using $H_{\rm DE}\sim H_0$ this implies that 
\be 
\tan z_\star \frac{\Omega}{H_0}\sim \frac{2}{3} \frac{\Omega}{H_0}
\ee
where $z_\star= H_0 (t_0 -t_\star)$ is the redshift when the oscillation start, i.e. when the potential leaves a pure exponential decrease.
This implies that
\be 
\frac{\Omega}{H_0}\sim \frac{\pi}{2z_\star}.
\ee
Assuming that $z_\star \sim 0.01$ corresponding to the redshift of the closest SNIA supernovae, the oscillations around the minimum take place in a fraction of the age of the Universe. For such a large value of $\Omega$, the convergence towards $\dot \phi=0$ is independent of the initial conditions and only depends on the redshift when the potential changes shape. This is only dependent on the attractor behaviour (\ref{attrac}).

In practice, the assumption that $K(\chi)\sim \chi$ at small redshift fails as 
(\ref{convq}) implies that $\chi \sim \beta^2 \frac{H^2 M^2_{\rm Pl}}{M^4}$ which is not small. The tuning of the potential close to its minimum in order to reach small values of $\dot G_{\rm eff}/G_{\rm eff}$ must be performed numerically and is left to future work. 

\bibliography{Bibliography}

\end{document}